\documentclass[useAMS,usenatbib]{mnras}
\usepackage{amsmath,graphicx,bm,amssymb,aas_macros,url,array,amssymb,breqn,aas_macros,graphicx,comment,float}
\def\BCCP{$^3$}
\def\BCCPtxt{$^3$Dept. of Astronomy and Berkeley Center for Cosmological Physics, University of California, Berkeley, Berkeley, California 94720, USA }

\def\ASU{$^9$}
\def\ASUtxt{$^9$Arizona State University, School of Earth and Space Exploration, Tempe, AZ 85287, USA}

\def\UW{$^8$}
\def\UWtxt{$^8$University of Washington, Department of Physics, Seattle, WA 98195, USA}

\def\UWeScience{$^{17}$}
\def\UWeSciencetxt{$^{17}$University of Washington, eScience Institute, Seattle, WA 98195, USA}

\def\SKASA{$^{10}$}
\def\SKASAtxt{$^{10}$Square Kilometre Array South Africa (SKA SA), Park Road, Pinelands 7405, South Africa}

\def\RU{$^{11}$}
\def\RUtxt{$^{11}$Department of Physics and Electronics, Rhodes University, Grahamstown 6140, South Africa}

\def\CfA{$^4$}
\def\CfAtxt{$^4$Harvard-Smithsonian Center for Astrophysics, Cambridge, MA 02138, USA}

\def\ANU{$^{12}$}
\def\ANUtxt{$^{12}$Australian National University, Research School of Astronomy and Astrophysics, Canberra, ACT 2611, Australia}

\def\CAASTRO{$^{7}$}
\def\CAASTROtxt{$^{7}$ARC Centre of Excellence for All-sky Astrophysics (CAASTRO)}

\def\Haystack{$^{13}$}
\def\Haystacktxt{$^{13}$MIT Haystack Observatory, Westford, MA 01886, USA}

\def\MIT{$^2$}
\def\MITtxt{$^2$MIT Kavli Institute for Astrophysics and Space Research, Cambridge, MA 02139, USA}

\def\MITP{$^1$}
\def\MITPtxt{$^1$ Department of Physics, Massachusetts Institute of Technology, Cambridge, MA, 02139 USA}

\def\Curtin{$^{14}$}
\def\Curtintxt{$^{14}$International Centre for Radio Astronomy Research, Curtin University, Perth, WA 6845, Australia}

\def\Victoria{$^{18}$}
\def\Victoriatxt{$^{18}$Victoria University of Wellington, School of Chemical \& Physical Sciences, Wellington 6140, New Zealand}

\def\UWisc{$^{19}$}
\def\UWisctxt{$^{19}$University of Wisconsin--Milwaukee, Department of Physics, Milwaukee, WI 53201, USA}

\def\UMichigan{$^{20}$}
\def\UMichigantxt{$^{20}$University of Michigan, Department of Atmospheric, Oceanic and Space Sciences, Ann Arbor, MI 48109, USA}

\def\UMelbourne{$^{21}$}
\def\UMelbournetxt{$^{21}$The University of Melbourne, School of Physics, Parkville, VIC 3010, Australia}

\def\USydney{$^{15}$}
\def\USydneytxt{$^{15}$Sydney Institute for Astronomy, School of Physics, The University of Sydney, NSW 2006, Australia}

\def\CASS{$^{22}$}
\def\CASStxt{$^{22}$CSIRO Astronomy and Space Science (CASS), PO Box 76, Epping, NSW 1710, Australia}

\def\Tata{$^{23}$}
\def\Tatatxt{$^{23}$National Centre for Radio Astrophysics, Tata Institute for Fundamental Research, Pune 411007, India}

\def\RRI{$^{24}$}
\def\RRItxt{$^{24}$Raman Research Institute, Bangalore 560080, India}

\def\Brown{$^{25}$}
\def\Browntxt{$^{25}$Brown University, Department of Physics, Providence, RI 02912, USA}

\def\NRAO{$^{26}$}
\def\NRAOtxt{$^{26}$National Radio Astronomy Observatory, Charlottesville and Greenbank, USA}

\def\UWA{$^{27}$}
\def\UWAtxt{$^{27}$International Centre for Radio Astronomy Research, University of Western Australia, Crawley, WA 6009, Australia}

\def\ASTRON{$^6$}
\def\ASTRONtxt{$^6$Netherlands Institute for Radio Astronomy (ASTRON), PO Box 2, 7990 AA Dwingeloo, The Netherlands}

\def\Dunlap{$^{16}$}
\def\Dunlaptxt{$^{16}$Dunlap Institute for Astronomy and Astrophysics, University of Toronto, ON M5S 3H4, Canada}

\def\SNS{$^5$}
\def\SNStxt{$^5$Scuola Normale Superiore, Piazza dei Cavalieri 7, I-56126 Pisa, Italy}

\newcommand{\be}{\begin{equation}}
\newcommand{\ee}{\end{equation}}

\def\Let@{\def\\{\notag\math@cr}}

\pdfoutput=1

\title[First Limits on the 21 cm EoX Power Spectrum]{First Limits on the 21\,cm  Power Spectrum during the Epoch of X-ray heating.}

\author[A Ewall-Wice et al.]{A.~Ewall-Wice,\MITP$^,$\MIT \thanks{E-mail: aaronew@mit.edu}~
Joshua~S.~Dillon,\MITP$^,$\MIT $^,$\BCCP~
J.~N.~Hewitt,\MITP$^,$\MIT~
A.~Loeb,\CfA~
\newauthor
A.~Mesinger,\SNS~
A.~R.~Neben,\MITP$^,$\MIT~
A.~R.~Offringa,\ASTRON$^,$\CAASTRO~
M.~Tegmark,\MITP$^,$\MIT~
N.~Barry,\UW~
\newauthor
A.~P.~Beardsley,\ASU~
G.~Bernardi,\SKASA$^,$\RU$^,$\CfA~
Judd~D.~Bowman,\ASU~
F.~Briggs,\ANU$^,$\CAASTRO~
\newauthor
R.~J.~Cappallo,\Haystack~
P.~Carroll,\UW~
B.~E.~Corey,\Haystack~
A.~de~Oliveira-Costa,\MIT~
D.~Emrich,\Curtin~
\newauthor 
L.~Feng,\MITP$^,$\MIT~
B.~M.~Gaensler,\USydney$^,$\CAASTRO$^,$\Dunlap~
R.~Goeke,\MIT~
L.~J.~Greenhill,\CfA~
B.~J.~Hazelton,\UW$^,$\UWeScience~
\newauthor
N.~Hurley-Walker,\Curtin~
M.~Johnston-Hollitt,\Victoria~
Daniel~C.~Jacobs,\ASU~
D.~L.~Kaplan,\UWisc~
\newauthor
J.~C.~Kasper,\UMichigan$^,$\CfA~
HS Kim,\UMelbourne$^,$\CAASTRO~
E.~Kratzenberg,\Haystack~
E.~Lenc,\USydney$^,$\CAASTRO~
J.~Line,\UMelbourne$^,$\CAASTRO~
\newauthor
C.~J.~Lonsdale,\Haystack~
M.~J.~Lynch,\Curtin~
B.~McKinley,\UMelbourne$^,$\CAASTRO
S.~R.~McWhirter,\Haystack~
\newauthor
D.~A.~Mitchell,\CASS$^,$\CAASTRO~
M.~F.~Morales,\UW~
E.~Morgan,\MIT~
Nithyanandan~Thyagarajan,\ASU~
\newauthor
D.~Oberoi,\Tata~ 
S.~M.~Ord,\Curtin$^,$\CAASTRO~
S. Paul,\RRI~
B.~Pindor,\UMelbourne$^,$\CAASTRO~
J.~C.~Pober,\Brown$^,$\UW~
T.~Prabu,\RRI~
\newauthor
P.~Procopio,\UMelbourne$^,$\CAASTRO~
J.~Riding,\UMelbourne$^,$\CAASTRO~
A.~E.~E.~Rogers,\Haystack~ 
A.~Roshi,\NRAO~ 
\newauthor
N.~Udaya~Shankar,\RRI~
Shiv~K.~Sethi,\RRI~
K.~S.~Srivani,\RRI~
R.~Subrahmanyan,\RRI$^,$\CAASTRO~
\newauthor
I.~S.~Sullivan,\UW~
S.~J.~Tingay,\Curtin$^,$\CAASTRO~
C.~M.~Trott,\Curtin$^,$\CAASTRO
M.~Waterson,\Curtin$^,$\ANU~
\newauthor
R.~B.~Wayth,\Curtin$^,$\CAASTRO~
R.~L.~Webster,\UMelbourne$^,$\CAASTRO~
A.~R.~Whitney,\Haystack~
A.~Williams\Curtin~ 
\newauthor
C.~L.~Williams,\MITP$^,$\MIT~
C.~Wu,\UWA~
J.~S.~B.~Wyithe\UMelbourne$^,$\CAASTRO~
\\
\MITPtxt \\
\MITtxt \\
\BCCPtxt \\
\CfAtxt \\
\SNStxt \\
\ASTRONtxt \\
\CAASTROtxt \\
\UWtxt \\
\ASUtxt \\
\SKASAtxt \\
\RUtxt \\
\ANUtxt \\
\Haystacktxt \\
\Curtintxt \\
\USydneytxt \\
\Dunlaptxt \\
\UWeSciencetxt \\
\Victoriatxt \\
\UWisctxt \\
\UMichigantxt \\
\UMelbournetxt \\
\CASStxt \\
\Tatatxt \\
\RRItxt \\
\Browntxt \\
\NRAOtxt \\
\UWAtxt
}

\begin{document}
\label{firstpage}
\maketitle
\voffset=-.6in
\clearpage

\begin{abstract}
	We present first results from radio observations with the Murchison Widefield Array seeking to constrain the power spectrum of 21\,cm brightness temperature fluctuations between the redshifts of 11.6 and 17.9 (113 and 75\,MHz). Three hours of observations were conducted over two nights with significantly different levels of ionospheric activity. We use these data to assess the impact of systematic errors at low frequency, including the ionosphere and radio-frequency interference, on a power spectrum measurement. We find that after the 1-3 hours of integration presented here, our measurements at the Murchison Radio Observatory are not limited by RFI, even within the FM band, and that the ionosphere does not appear to affect the level of power in the modes that we expect to be sensitive to cosmology. Power spectrum detections, inconsistent with noise, due to fine spectral structure imprinted on the foregrounds by reflections in the signal-chain, occupy the spatial Fourier modes where we would otherwise be most sensitive to the cosmological signal. We are able to reduce this contamination using calibration solutions derived from autocorrelations so that we achieve an sensitivity of $10^4$\,mK on comoving scales $k\lesssim 0.5\,h$Mpc$^{-1}$. This represents the first upper limits on the $21$\,cm power spectrum fluctuations at redshifts $12\lesssim z \lesssim 18$ but is still limited by calibration systematics. While calibration improvements may allow us to further remove this contamination, our results emphasize that future experiments should consider carefully the existence of and their ability to calibrate out any spectral structure within the EoR window.  
\end{abstract}
\begin{keywords}
dark ages, reionization, first stars -- techniques: interferometric -- radio lines: general -- X-rays: galaxies
\end{keywords}

\section{Introduction}
Mapping the 21\,cm transition of neutral hydrogen at high-redshift promises to revolutionize our knowledge on the first generations of stars and galaxies and to provide a unique probe of the ``Dark Ages" preceding this first generation of luminous objects (see \citet{Barkana:2001, Furlanetto:2006Review, Morales:2010} for reviews ). Planned instruments such as the Square Kilometre Array (SKA) \citep{Koopmans:2015} and the Hydrogen Epoch of Reionization Array (HERA) \citep{Pober:2014} are expected to elucidate the formation of the first luminous structures and to place strict constraints on the properties of the sources that reionized the intergalactic medium \citep{Pober:2014,Greig:2015}. A number of experiments including the Giant Metrewave Telescope (GMRT) \citep{Pagica:2013}, the Low-Frequency Array (LOFAR) \citep{VanHaarlem:2013}, the Murchison Widefield Array (MWA) \citep{Bowman:2013,Tingay:2013a}, and the Precision Array for Probing the Epoch of Reionization (PAPER)  \citep{Parsons:2010} are already underway to explore the challenges of separating the faint cosmological signal from bright foregrounds and to attempt a first detection of the power spectrum of the cosmological 21\,cm emission line. 

Thus far, these experiments have targeted redshifts between $6$ and $12$. During this Epoch of Reionization (EoR), ultraviolet photons from the first generations of luminous sources transformed the intergalactic medium (IGM) from predominantly neutral to ionized. Over the past several years, deep integrations have placed significant upper limits on the power spectrum during reionization \citep{Pagica:2013,Dillon:2014,Parsons:2014,Ali:2015,Dillon:2015b,Trott:2016a}. The best upper limits of 502\,mK$^2$ at $z\sim 8.4$ \citep{Ali:2015} have begun to rule out scenarios where the neutral IGM experiences little or no heating \citep{Pober:2015}. Integrations at comparatively high redshifts have also been carried out. \citet{Dillon:2014} put an upper limit on the power spectrum at $z=11.7$ using the 32-tile MWA pathfinder. A much deeper integration at redshift 10.3 was performed with PAPER's 32-element configuration \citep{Jacobs:2015}, though it was limited by residual foregrounds at the edge of the instrumental bandpass. 

While observations of the power spectrum during the EoR alone will shed light on the sources and astrophysics that drove reionization, it is only the final milestone in the evolution of the neutral IGM. Before reionization, the gas was heated, most likely by the first generations of high mass X-ray binaries (HMXB) \citep{Mirabel:2011} and/or hot interstellar medium (ISM) \citep{Pacucci:2014}. Brightness temperature fluctuations from inhomogenous heating at these early times can yield power spectrum amplitudes that are over an order of magnitude larger than those expected during reionization \citep{Pritchard:2007,Mesinger:2013}. Even before the X-ray heating, fluctuations in the brightness temperature were likely sourced by fluctuations in the Lyman-$\alpha$ flux field from the first stars \citep{Barkana:2005b,Pritchard:2006}. 

The ultimate goal of 21\,cm cosmology is a three dimensional map of the entire IGM between  $z\approx 200$ and reionization since, at least in principle, the 21\,cm line is a cosmological observable accessible all the way back through the dark ages to the decoupling of the spin temperature from the cosmic microwave background (CMB) \citep{Furlanetto:2006Review}. Even though the ionosphere obscures extraterrestrial radio signals below about $30$\,MHz \citep{Jester:2009}, it is expedient to use ground based experiments to cover as great a redshift span as possible. Because foreground amplitudes and ionospheric effects grow progressively at lower frequency, the most reasonable next step after reionization in our march into the dark ages is the Epoch of X-ray heating (EoX). The exact redshift range for the EoX depends on the astrophysical model (see for example \citet{Mesinger:2013,Mesinger:2014,Pacucci:2014}), but a reasonable range, targeted in this work, is z=11.6 (113\,MHz) to z=17.9 (75\,MHz) .

First experimental 21\,cm constraints on the thermal history of the IGM come from \citet{Pober:2015} who used upper limits on the power spectrum at reionization redshifts to rule out inefficient heating histories. These constraints arise from the fact that the observable brightness temperature difference from the CMB depends on the spin temperature as
\begin{equation}\label{eq:dTb}
\Delta T_b \propto \left( 1 - \frac{T_\gamma}{T_s} \right),
\end{equation}
where $T_\gamma$ is the temperature of the CMB and $T_s$ is the spin temperature of the gas which is expected to be closely coupled to the gas kinetic temperature before substantial heating takes place \citep{Furlanetto:2006Review}. For a cold IGM, $1-T_\gamma/T_s$ is large and negative leading to large amplitude contrasts between neutral and ionized regions. 

However, assuming that the number of X-rays per baryon involved in star formation is the same as what is observed in nearby star forming galaxies \citep{Mineo:2012a}, the HI spin temperature is expected to be heated well above the CMB by the time reionization begins, saturating the effect of heating on equation \ref{eq:dTb} \citep{Furlanetto:2006Global}. Hence, direct measurements of the 21\,cm line during the EoX will be necessary if we want to learn about the detailed properties of the thermal history and the astrophysical phenomena that influence it. Recent work has shown that if X-ray heating proceeds inefficiently, the current generation of interferometers will be sensitive enough to detect the power spectrum sourced by spin temperature fluctuations at $z \approx 12$ \citep{Christian:2013}. Next generation of 21\,cm observatories will detect the heating power spectrum for a wide range of heating scenarios out to redshifts as high as 20 \citep{Mesinger:2014} and place percent level constraints on the properties of the earliest X-ray sources \citep{EwallWice:2015}. While pre-reionization measurements are expected to shed light on the first stellar mass black holes or the hot ISM, they may also offer us insights into other astrophysical processes. It is possible for dark matter annihilation \citep{Valdes:2013} and the existence of warm dark matter \citep{Sitwell:2014} to create observable impacts on the IGM thermal history. Finally, the IGM is especially cool and optically thick during the beginning of the heating process, making it ideal for 21\,cm forest \citep{Loeb:2002,Carilli:2002,Furlanetto:2006Forest,Mack:2012,Ciardi:2013} studies should any radio loud sources exist at those redshifts. It is also possible to constrain the source population itself by detecting its signature in the 21\,cm power spectrum \citep{EwallWice:2014}. 

Complementary observations of the sky-averaged (the ``global") 21\,cm signal with a single dipole can also explore the reionization and pre-reionization epochs and experiments such as  EDGES \citep{Bowman:2010}, LEDA \citep{Greenhill:2012}, DARE \citep{Burns:2012}, SARAS \citep{Patra:2013}, SciHI \citep{Voytek:2014}, and BIGHORNS \citep{Sokolowski:2015} are beginning to take data. While demanding much greater sensitivity than global signal experiments, power spectrum measurements with an interferometer probe fine frequency scales while foregrounds occupy a limited region of Fourier-space, known as the ``wedge" \citep{Datta:2010,Parsons:2012,Morales:2012,Vedantham:2012,Trott:2012,Hazelton:2013,Thyagarajan:2013,Liu:2014a,Liu:2014b,Thyagarajan:2015a,Thyagarajan:2015b}. The region of Fourier space outside of the wedge, in principle free of foregrounds and therefore having greater sensitivity to brightness temperature fluctuations, is known as the ``EoR window" (henceforth ``window"). foreground modeling and calibration errors that are smooth in frequency should have limited impact within the window.

In this paper we assess the levels of systematic errors that are especially severe at the lower EoX frequencies (relative to those typical of EoR studies) including the ionosphere, radio-frequency interference (RFI), and the enhanced noise and foregrounds from a sky that is both intrinsically brighter at lower frequencies and observed with a larger primary beam. In \S~\ref{sec:obs} we describe the MWA, our observations, and data reduction. In \S~\ref{sec:sys} we address the systematic errors that are especially challenging below EoR frequencies and our efforts to mitigate them. The limiting systematic error that we encounter is fine frequency structure in the instrumental bandpass due to standing wave reflections in the cables between the MWA's beamformers and receivers. After making a reasonable assumption about the relationship between our autocorrelations and the gain amplitudes, we achieve notable improvement in calibration but we are still left with significant foreground contamination.

Power spectrum upper limits are derived (\S~\ref{sec:results}) which are broadly consistent with thermal noise except in several regions of Fourier space corresponding to the light travel time delays of the reflections. We expect that refined calibration techniques employing better foreground models \citetext{Caroll et al., in preparation} and redundant baselines \citep{Wieringa:1992,Liu:2010,Zheng:2014} can improve the removal of this contamination. In order to avoid signal loss and the introduction of spurious spectral structure, we have been conservative in the number of free parameters allowed in our gain solutions; increasing these may also resolve this problem. Reduced cable lengths expected in upcoming experiments such as HERA and the SKA will ameliorate the problem of reflections.

\section{Observing and Initial Data Reduction}\label{sec:obs}
	We begin our discussion with an overview of our observations and our data reduction procedure. Our analysis yields two different image products with 112\,s cadence: high resolution continuum images created from bandwidth multifrequency synthesis (MFS) (with $\approx 6'$ resolution) where baselines across all fine frequencies are combined into a single image, and naturally weighted multifrequency data cubes, where each fine frequency channel is imaged separately and  integrated over three hours. We use the MFS images to evaluate ionospheric conditions, and we use the multifrequency data cubes in our power spectrum analysis. We note that with 112\,s averaging, we are performing significant averaging over fine time-scale ionospheric effects which for the MWA baselines have a typical coherence time of $\approx 10-44$\,s \citep{Vedantham:2015a} (henceforth V15a).  After outlining the instrument and our observing strategy (\S~\ref{ssec:obs}), we discuss our initial calibration procedure (\S~\ref{ssec:initCal}) and finish with the production MFS images (\S~\ref{ssec:mfs}) and data cubes (\S~\ref{ssec:cubes}) which serve as the input to our power spectrum pipeline. 
		
	\subsection{Observations with the MWA}\label{ssec:obs}
	
	The MWA \citep{Lonsdale:2009,Tingay:2013a} is a 128 antenna interferometer located at the Murchison Radio Observatory (MRO) in Western Australia (26.70$^{\circ}$S, 116.67$^{\circ}$ E) with an analog bandpass of 80-300\,MHz. Each correlated antenna tile consists of 16 dual polarization dipole elements arranged in a four-by-four grid. The phased output of the dipoles on each tile is summed together in an analog beam-former and delivered to one of sixteen different receiver units where 30.72\,MHz of bandwidth is digitized before correlation in an on-site building. We refer the reader to \citet{Prabu:2015} and \citet{Ord:2015} for a detailed discussion of the MWA's receivers and correlator, respectively. The instrument is designed to achieve a diverse set of science goals \citep{Bowman:2013} including a first detection of the 21\,cm power spectrum during the EoR, detecting and monitoring transients, pulsars \citep{Tremblay:2015}, solar and heliospheric science \citep{Tingay:2013b}, and a low-frequency survey of the sky below DEC=$+25^{\circ}$ \citep{Wayth:2015}. 
		
\begin{figure}
	\includegraphics[width=.48\textwidth]{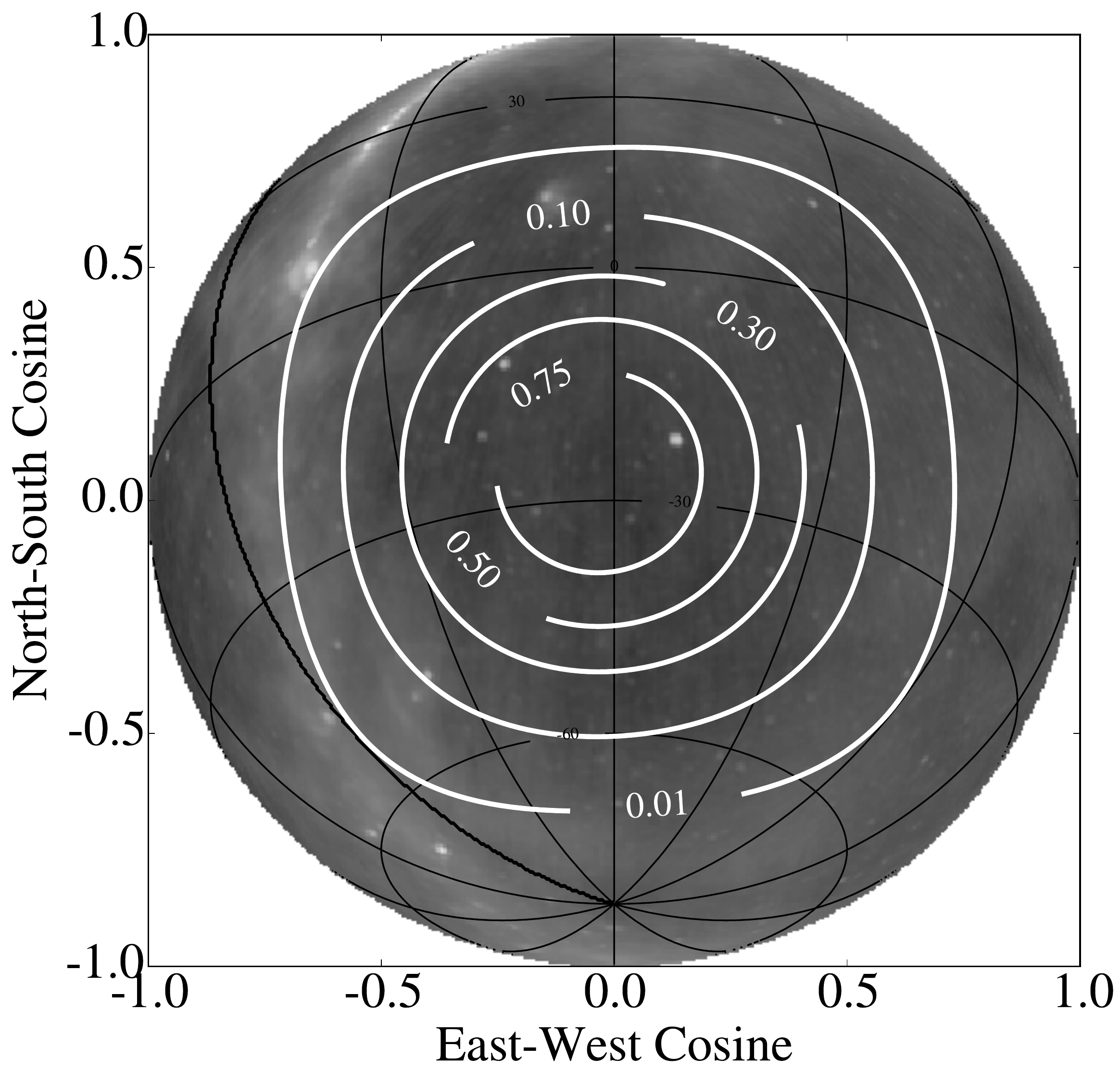}
	\caption{A radio map at 408\,MHz \citep{Haslam:1982}  sin-projected over the region of the sky observed in this paper. Cyan through magenta contours indicate the total fraction of observation time weighted by our primary beam gain for our three hours of observation at 83~MHz. Red contours indicate R.A.-decl. lines. Observation tracked the position  (R.A.(J2000) = $4^h0^m0^s$, decl.(J2000)=$-30^\circ 0' 0''$) on a region of the sky with relatively little galactic contamination and dominated by the resolved sources Fornax A and Pictor A. The galactic anticentre and bright diffuse sources, such as the Gum Nebula, are below 1$\%$ bore-site gain. }
	\label{fig:field}
	\end{figure}

\begin{figure}
	\includegraphics[width=.48 \textwidth]{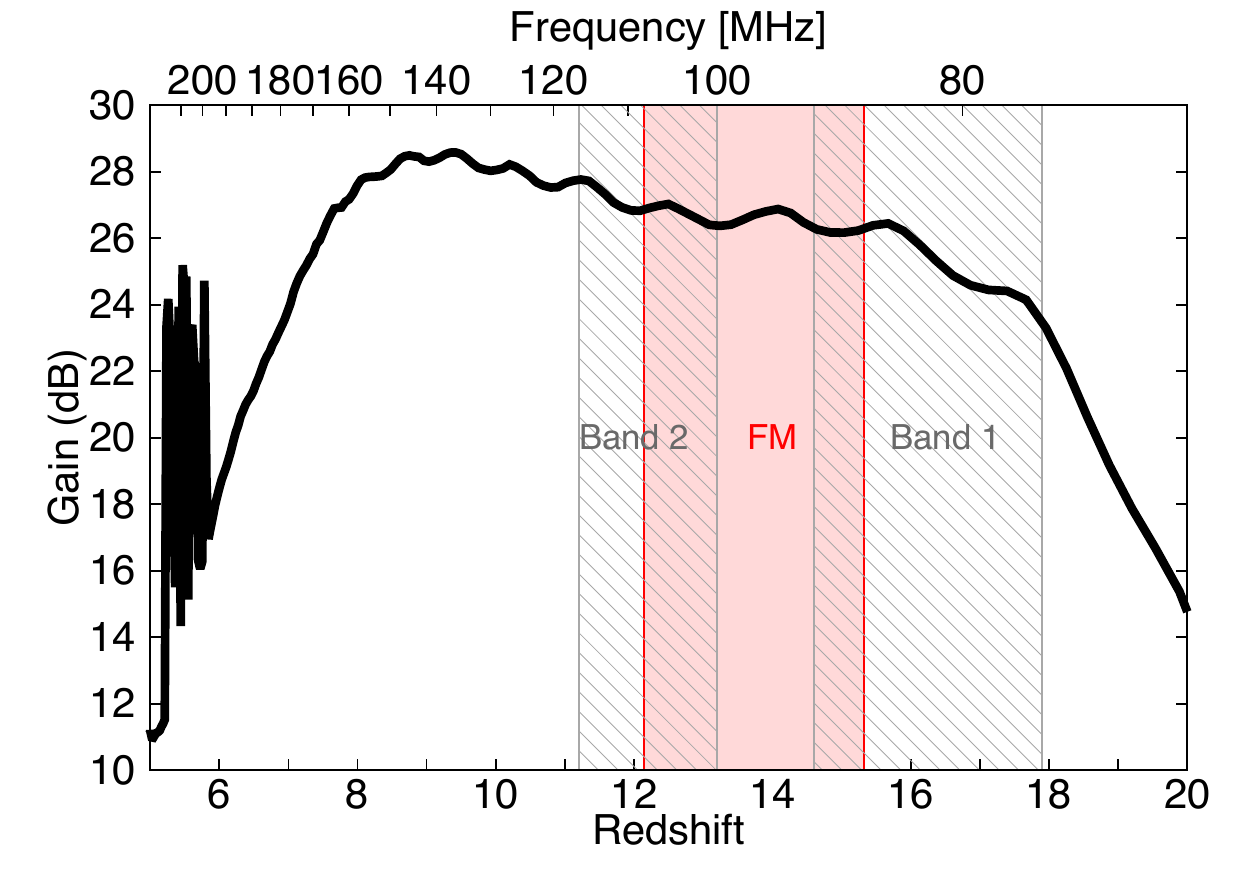}
	\caption{The autocorrelation spectrum of a single tile, showing the MWA bandpass, is plotted here (solid black line) along with the frequency ranges over which our data was taken (gray striped rectangles). Observations were performed simultaneously on two non-contiguous bands located on either side of the 88-108 MHz FM band (red shaded region) to both assess conditions within the FM band and to preserve some usable bandwidth should it have proven overly contaminated by RFI. }
	\label{fig:bandpass}
	\end{figure}

	Observations of a field centered at R.A.(J2000) = $4^h0^m0^s$ and decl.(J2000)=$-30^\circ 0' 0''$  were carried out for 4.13 hours over two nights on September 5th and 6th, 2013, with primary beams of the 128 MWA antenna elements (``tiles") formed at five different altitude/azimuth pointings each night to track the field. After flagging for RFI and anomalous behavior that we will discuss in detail in \S~\ref{ssec:rfi}, our total observation time for our power spectrum upper limit is 3.08 hours. In Fig.~\ref{fig:field} we show the relative integration time on the sky weighted by the primary beam over all observations. We observed with 40\,kHz spectral resolution simultaneously over two contiguous bands; a 16.64\,MHz interval between 75.52\,MHz and 90.88\,MHz (Band 1) and a 14.08\,MHz band between 98.84\,MHz and 112.64\,MHz (Band 2). Both of these sub-bands overlap with the FM band (88-108\,MHz). In Fig.~\ref{fig:bandpass} we show our observed bands superimposed on the autocorrelation spectrum of a single MWA tile. Observations in Band 1 took place right on the edge of the analogue cutoff of the MWA, making its shape relatively complicated to model in calibration (our calibration is direction independent and described in \S~\ref{ssec:initCal} and \S~\ref{ssec:autocal}). Band 2, while in a flatter region of the bandpass, has a larger overlap with the FM band. Observations were divided into 112\,s snapshot with data averaged into 0.5\,s integration intervals by the correlator.

	\subsection{RFI flagging and Initial Calibration}\label{ssec:initCal}
	
	The data were first flagged for RFI contamination. An optimized version of {\tt aoflagger} \citep{Offringa:2012}, called {\tt cotter} \citep{Offringa:2015}, was run on each snapshot with automatic RFI identification performed only on the visibility cross correlations. Additional flags were applied to the center and 40\,kHz edges of each 1.28\,MHz resolution coarse spectral channel to remove digital artifacts arising from the two stage channelization scheme used in the MWA. After flagging, visibilities were averaged in time from 0.5 to 2\,s and in frequency from 40 to 80\,kHz. The averaged visibilities were than converted to Common Astronomy Software Applications Package ({\tt CASA}) measurement sets \citep{McMullin:2007} which served as the inputs to all ensuing steps in our pipeline. The percentage of all data flagged by {\tt cotter} was approximately 0.75\% for Band 1 and $2\%$ for Band 2. As described in \S~\ref{ssec:rfi}, we also implement additional (and highly conservative) flagging of observations by inspecting autocorrelations which leads us to discard $25\%$ of our data.  
	Our initial calibration was divided into three steps: A preliminary complex antenna gain solution using an approximate sky model, one iteration of self calibration, and polynomial fitting to reduce noise and limit fine frequency scale systematics that might arise from our incomplete sky model.
		
	For the first step, our sky model combined a list of point sources with images of the two bright resolved sources in our field: Fornax A and Pictor A. For the point source model we included the 200 brightest sources at our frequencies based on extrapolated  power law fits to data from the Coolgoora survey \citep{Slee:1995}, the Molongolo Reference Catalogue \citep{Large:1981} and more recent measurements by PAPER at 145\,MHz \citep{Jacobs:2011}. Fornax A is the brightest extended source in our field and is highly resolved, so we modeled it with a VLA image taken by \citet{Fomalant:1989} at 1.4\,GHz, scaled to match the flux density and spectral index measured by \citet{Bernardi:2013} at 180\,MHz, and extrapolated to our band with a spectral index of $-0.88$. For Pictor A, we used a VLA image at 333\,MHz by \citet{Perley:1997} and extrapolated to our band with a spectral index of $-0.71$. The model components for our initial calibration extended down to an apparent flux density of $\approx 5$\,Jy which is comparable to the flux uncertainties in the brightest sources in the initial catalog. Due to the high uncertainty in the sky at our frequencies, this model was updated by a round of self calibration which we describe shortly.
	
\begin{figure*}
	\includegraphics[width=\textwidth]{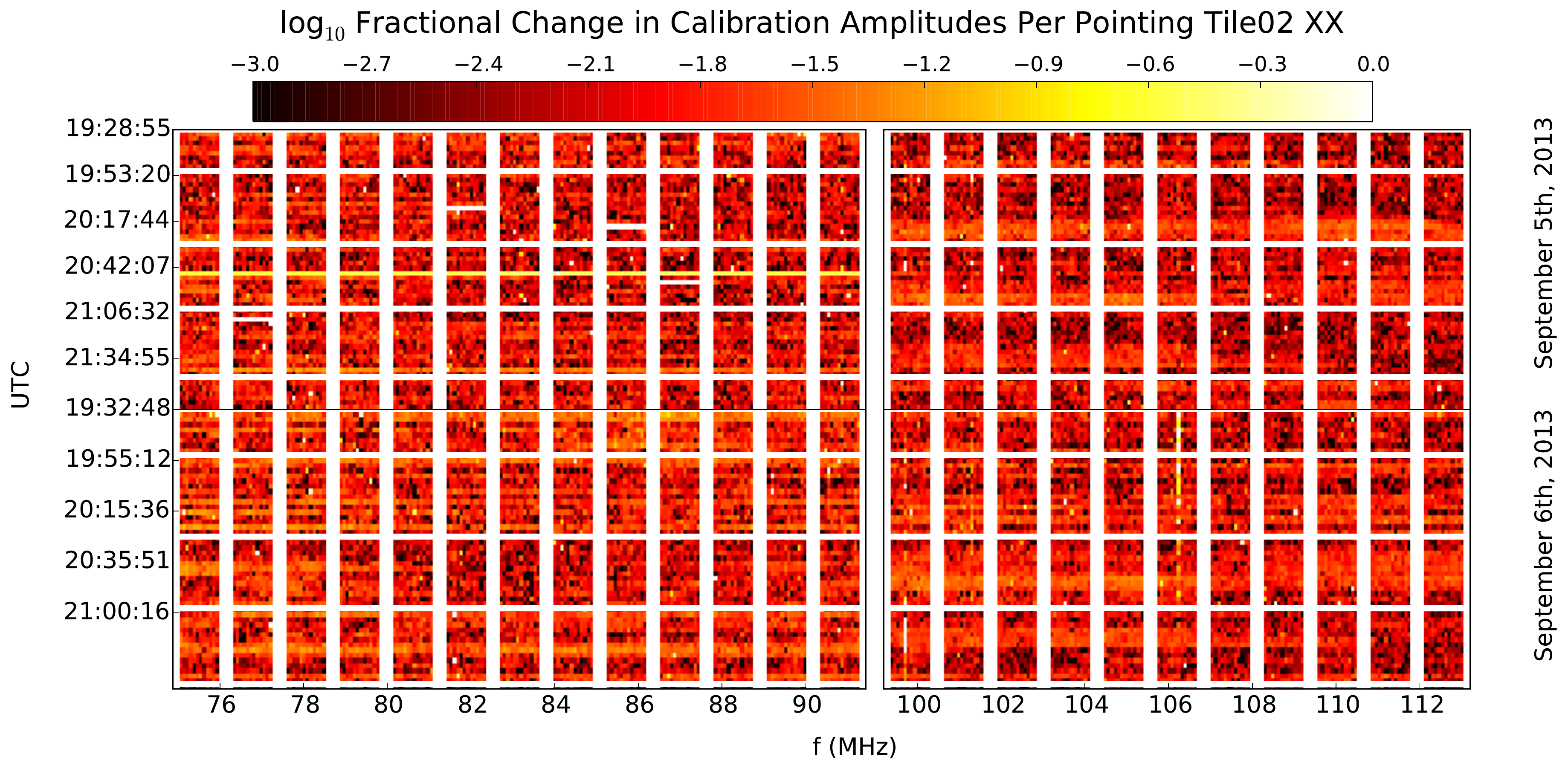}
	\caption{A false-color plot of the fractional change in our calibration amplitudes over each pointing (in which beamformer delay settings are fixed). Pointing changes are marked by the solid white horizontal lines. We see that the calibration amplitudes vary within a pointing on the order of several percent with little systematic variation.  There are several coarse channel scale jumps on September 5th that correspond to observations in which the number of sources identified within a snapshot image were reduced (see Fig.~\ref{fig:counts}). We found that these jumps corresponded to excess flagging from {\tt cotter}, indicating high levels of RFI or other bad data, and dropped them from our analysis. Vertical lines from RFI are visible at $98$~MHz and $107$~MHz, especially on September 6th. }
	\label{fig:cal}
	\end{figure*}

	The {\tt CASA} {\tt bandpass} function was used to obtain a first set of best fit calibration gains, averaging over 32 fine channels for each solution. Since our starting model was uncertain at the 10\% level, an iteration of self calibration was performed by MFS imaging and deconvolving $10^4$ components with {\tt WSClean} \citep{Offringa:2014}. The CLEAN components were used as a model for a second run of {\tt bandpass} where we relaxed channel averaging so that the complex gain for each 80~kHz fine channel was found independently. In Fig.~\ref{fig:cal} we show the fractional change in calibration amplitude for one antenna tile over time intervals in which beamformer settings were held constant (pointings). Over our two nights we find that variations are on the order of a few per cent and dominated by uniform amplitude jumps across the entire frequency range that are strongly correlated between all antennas. Observations of the autocorrelations do not show such behavior within each pointing so these gain jumps must arise from the calibration routine itself. Possible sources of  these jumps could be variations in the the overall amplitudes in self-calibration which occurs if the cleaning step does not recover all of the flux density on large scales, unmodeled sources moving through the sidelobes of the primary beam, or the result of a varying sky in the presence of beam modeling errors. Since these features do not introduce fine frequency structure, we do not think they are an impediment to the power spectrum analysis in this work. 

	We attempted to apply calibration solutions in which each frequency channel was allowed to vary freely but we found significant power was introduced on fine scales, contaminating the EoR window. This could be due to unmodeled sources adding spectral structure from long to short baselines or insufficient SNR on the solutions themselves. There is also possibility for signal loss given the large number of degrees of freedom. The degree to which calibration can remove signal and how unmodeled sources with enhanced spectral structure at long baselines can be mixed into short baselines are open questions that are beyond the scope of this analysis and are being investigated by \citet{Barry:2016,Trott:2016b,EwallWice:2016b}. With these concerns in mind, we erred on the side of caution and fit each j$^{th}$ antenna gain with the product of three smooth functions in frequency, f,
	\begin{equation}\label{eq:smooth}
		g_j(f) = P_j(f) R_{j,7m}(f)B(f),
	\end{equation}
	where $P_j(f)$ is a third order polynomial in amplitude and first order polynomial in phase; B(f) is a median bandpass that accounts for the course band shape determined by taking the median of calibration amplitudes across all tiles and polarizations; and $R_{j,7m}(f)$ is a reflection function that accounts for standing wave ripples in the bandpass arising from impedance mismatches in the 7-meter low noise amplifier (LNA) to beamformer cable connections.  It is straightforward to show (see Appendix \ref{app:reflections}) that a cable with length $L(j)$ introduces a multiplicative reflection term to the overall gain, 
	\begin{equation}\label{eq:reflection}
	R_j(f) = \frac{1}{1 - r_j e^{i (2 \pi f \tau_{L(j)} + \phi_j)} } ,
	\end{equation}
	where $r_j e^{i \phi_j}$ is a complex coefficient that is a function of the impedance of the cable and its connections at either end and $\tau_{L(j)}$ is the time for a signal to travel from one end of the cable and back. We note that this multiplicative term is derived from the infinite sum of reflections occurring from each $n^{th}$ round trip of the reflected wave and hence accounts for higher order reflections, not just the first round trip contribution. 
	
	When we formed power spectra from visibilities calibrated by this initial method, we found that our beamformer to receiver cables introduced spectral structure at the $\lesssim 1 \%$ level into our instrumental bandpass which were not removed by this smooth model (we return to this issue in \S~\ref{ssec:reflections}). The effect of this spectral structure on the MFS images, used to measure ionospheric refraction in \S~\ref{ssec:ion}, was negligible and we therefore employed this calibration for their production. A more refined calibration procedure, which we describe in \S~\ref{ssec:autocal}, was used for the data cubes and our power spectra.

\subsection{MFS Imaging and Flux Scale Corrections}\label{ssec:mfs}

	To form  MFS images, we averaged the antenna phases over each night and held them constant for every snapshot calibration solution to average over time-variability introduced by the ionosphere. This was done in order to ensure that short time-scale snapshot-to-snapshot variations were not due to time variations in the calibration solutions caused by the ionosphere. A multifrequency synthesis image was created from each snapshot, band, and polarization. From each XX and YY polarization snapshot which we call $I_{XX}$ and $I_{YY}$ respectively, we created a Stokes I snapshot corrected for the primary beam using an analytic dipole model of the MWA primary beam (which we denote as $B_{XX}$ and $B_{YY}$). 
	\begin{equation}\label{eq:stokesI}
	I_I(\theta,\phi) = \frac{I_{XX}(\theta,\phi) B_{XX}(\theta,\phi) + I_{YY} (\theta,\phi) B_{YY}(\theta,\phi)}{B_{XX}^2(\theta,\phi) + B^2_{YY}(\theta,\phi)}. 
	\end{equation}

	Sources were identified using the \texttt{Aegean} source finder \citep{Hancock:2012}. An overall flux scale for each stokes I snapshot was set following the technique, described in \citet{Jacobs:2013}, in which the flux scale for all sources is simultaneously fit to catalog flux densities using a Markov chain Monte Carlo method. We used the ten highest signal-to-noise point sources in each field and catalog flux densities interpolated between 74~MHz measurements from the Very Large Array Low-Frequency Sky Survey Redux \citep{Lane:2014} and 80 and 160 MHz measurements from the Culgoora catalog \citep{Slee:1995}. Since a flux scale error has no frequency dependence and the errors themselves evolve slowly in time near the center of the primary beam, we do not think that such mismodeling will result in frequency dependent errors. The dominant uncertainty in our flux scale is the systematic uncertainty in the model source fluxes themselves which are on the order of $\approx 20\%$\citep{Jacobs:2013} while uncertainties in the beam model contribute at the several percent level \citep{Neben:2015}.  On September 6th, we observed systematically smaller source counts than on September 5th (Fig.~\ref{fig:counts}). We attribute this difference to greater ionospheric turbulence on September 6th; and discuss this result further in \S~\ref{ssec:ion}. 
	
\begin{figure*}
	\includegraphics[width=\textwidth]{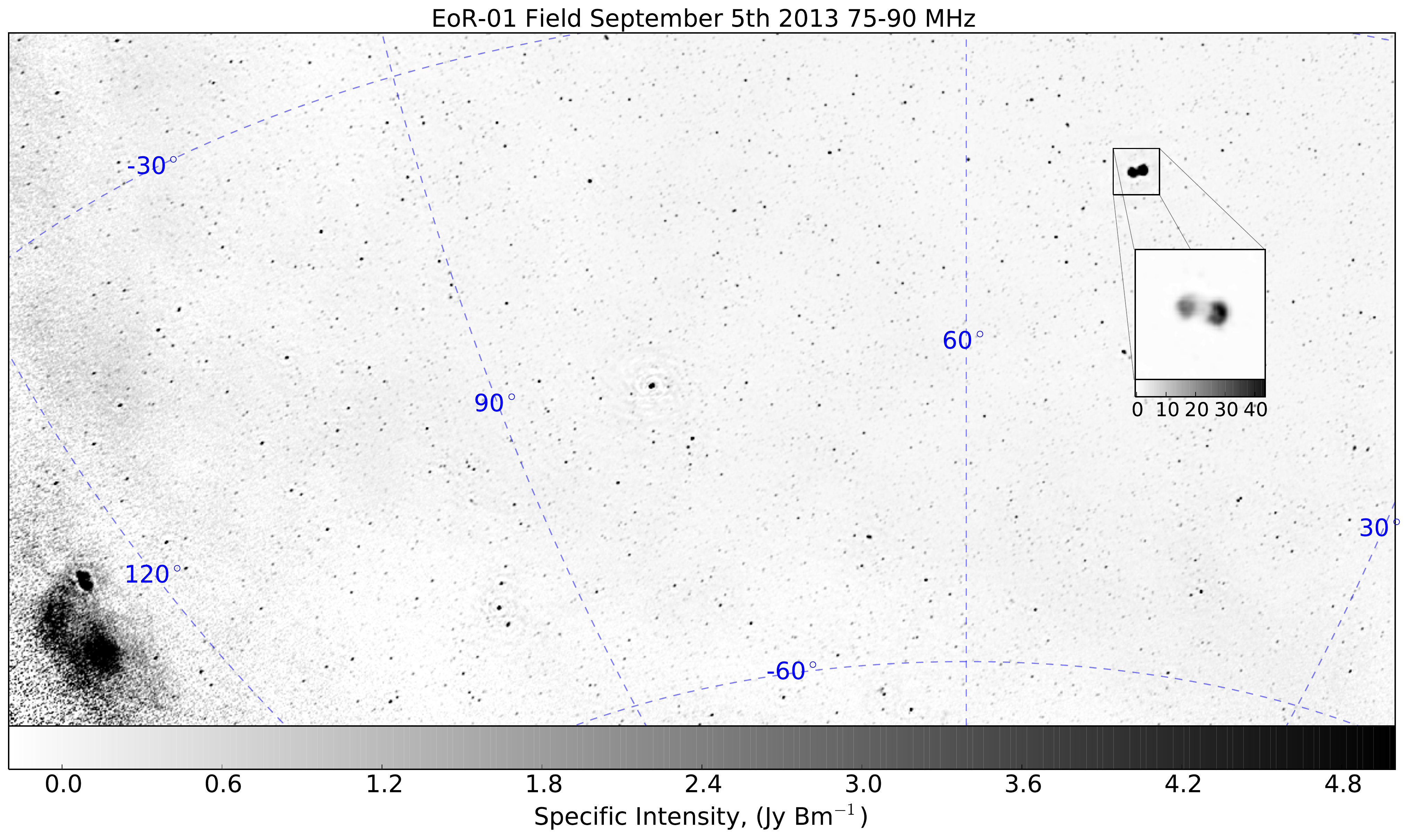}
	\caption{A deep image of the MWA ``EoR-01" field centered at (R.A.(J2000) = $4^h0^m0^s$ and decl.(J2000)=$-30^\circ 0' 0''$), derived by stacking restored multifrequency synthesis Stokes XX and YY images produced by {\tt WSClean} on Band 1. The dominant source in our field is Fornax A (detailed in the inset) whose structure is well recovered in imaging. Pictor A is also present at the center of the image (at $\sim 30 \%$ beam) along with the diffuse Puppis and Vela supernova remnants on the left.}
	\label{fig:mfs}
	\end{figure*}

	In Fig.~\ref{fig:mfs} we show a deep, primary beam corrected, integration of a portion of our field formed from a MFS image of Band 1. Known sources are well reproduced. The diffuse structures of the Vela and Puppis supernova remnants are clearly visible along with the fine scale structure of Fornax A.

Similar to previous upper limits on the power spectrum \citep{Dillon:2014,Parsons:2014,Dillon:2015b,Ali:2015,Trott:2016a}, our analysis does not consider the cross polarization products from the interferometer, which would require an additional calibration step to solve for the arbitrary phase difference between the $X$ and $Y$ polarized arrays \citep{Cotton:2012,Moore:2015}.

\subsection{Data Cubes for Power Spectrum Analysis}\label{ssec:cubes}

The inputs for our power spectrum analysis are multifrequency data cubes further calibrated with autocorrelation data from each tile. We describe the autocorrelation calibration technique in \S~\ref{ssec:autocal} where we address systematic errors revealed by our first look at data cubes and power spectra. 
In this section we describe our procedure for building the data cubes. While the MWA's long baselines, extending out to 2864\,m,  are useful for gauging ionospheric conditions and other systematics, the $uv$ plane is completely filled out to $\lesssim 160$\,m after $\approx 3$ hours of rotation synthesis. Since forming image power spectra using data with incomplete uv coverage leads to unwanted mode-mixing and spectral artifacts \citep{Hazelton:2013}, we threw away the sparse regions of the uv plane and reprocessed the data for our power spectrum analysis at much lower angular resolution than our MFS images. For each 112\,s snapshot, we divided the data into even and odd time step visibility sets formed from every other two second integration step, which we later cross multiply to form power spectra without noise bias \citep{Dillon:2014}. A naturally weighted, dirty snapshot cube was produced for each set with 80 kHz spectral resolution, 80 pixels on a side and $1.0^\circ$ ($0.75^\circ$) resolution for Band 1 (2).  
To form a power spectrum we will have to uniformly weight the sum of these cubes, hence the array point spread function (PSF), ${\bf s}_{j,even/odd}$, which is the 2d Fourier transform of the number of samples within each $uv$ cell; and the analytic primary beam matrix, ${\bf B}_j$, were also saved for each snapshot, polarization and interleaved time step. The power spectrum formalism of \citep{Dillon:2013} assumes the flat sky approximation and requires a sufficiently small field of view to be computationally tractable. Hence, before creating a uniformly weighted data cube by stacking the naturally weighted images in the $uv$ plane and dividing by their cumulative sampling function, we cropped each snapshot from $80 \times 80$ to $24 \times 24$ pixels.

We created even and odd, uniformly weighted temperature cubes  by summing all naturally weighted snapshots across both the XX and YY polarizations and dividing by the sum of their cumulative sampling function in the $uv$ plane,
\begin{equation}\label{eq:Uniform}
\boldsymbol{\widehat{x}}_{even/odd}  = \frac{\lambda^2}{2 k_b \Omega_{pix}} \boldsymbol{\mathcal{F}}_{\bf 2}^{-1} \left[ \frac{\sum_j \boldsymbol{\mathcal{F}}_{\bf 2}{\bf B}_j  {\bf \widehat{x}}_{j,even/odd} }{\sum_j \boldsymbol{\mathcal{F}}_{\bf 2} {\bf B}^2_j {\bf s}_j} \right],
\end{equation}
where the division of the two sums is to be understood as element-wise division. $\boldsymbol{\mathcal{F}}_{\bf 2}$ denotes the two dimensional Fourier transform matrix.  Indexing by angle cosines, $\ell,m$, their duals, $u$ and $v$, and frequency, $f$, we may write $\boldsymbol{\mathcal{F}}_{\bf 2}$ and its inverse as,
\begin{align}
\left[ \boldsymbol{\mathcal{F}}_{\bf 2} \right]_{u v f'  \ell m f}&= \Omega_{pix}  e^{-2 \pi i (\ell u + m v)} \delta_{f' f}\\
\left[\boldsymbol{\mathcal{F}}^{\bf -1}_{\bf 2} \right]_{\ell m f u v f'}& = \frac{1}{N_\ell N_m} \Omega_{pix} e^{2 \pi i( \ell u + m v)}  \delta_{f f'},
\end{align}
where $\Omega_{pix}$ is the solid angle of each pixel. Note that we are dividing by the convolution of the sampling function with the beam squared in the $uv$ plane so we do not have to worry about dividing by the beam nulls in image space. The prefactor at the front of equation \ref{eq:Uniform} converts from brightness to temperature units.

 By dividing by the point spread function in the $uv$ plane, equation~\ref{eq:Uniform} is essentially the application of uniform weighting to our data with some additional factors of the primary beam which warrant explanation. An additional factor of the primary beam was included in the sum to upweight regions of the field to which the MWA has the greatest gain, and is equivalent to optimal mapmaking \citep{Tegmark:1997b, Dillon:2015a}. Other 21\,cm pipelines, notably Fast Holographic Deconvolution \citep{Sullivan:2012} perform a similar upweighting by gridding visibilities with the primary beam while the fringe rate filtering procedure  \citep{Liu:2015} weights without gridding or imaging at all. 
 
 The impact of multiplying by the two factors of the beam in \ref{eq:Uniform} has the effect of convolving the true visibilities the $uv$ space beam convolved with its complex conjugate. Since our $uv$ cells are quiet large, this is well approximated by multiplication of the visibilities by the convolution of the $uv$ beam with its complex conjugate. We therefore also include the factor ${\bf B}^2$ in the denominator to correctly normalize the data in the $uv$ plane.

 Our ${\bf \widehat{x}}$ estimate of I in equation \ref{eq:Uniform} is similar to the Stokes-visibility I approximated in previous power spectrum analsyses \citep{Dillon:2014,Parsons:2014,Dillon:2015b,Ali:2015}. Visibility stokes I is susceptible to leakage from visibility stokes $Q \equiv \frac{1}{2} (XX - YY)$ due to beam ellipticity which has the potential to introduce fine frequency structure, caused by Faraday rotation, into the EoR window  \citep{Jelic:2010,Geil:2011,Moore:2013,Jelic:2014,Moore:2015,Asad:2015}. We can obtain an upper limit on polarization leakage in our power spectrum estimate by considering the best upper limits to date of the polarized $Q$,$U$, and $V$ visibility power spectra that might leak into our I visibilities, measured by \citet{Moore:2015} over the large field of view of PAPER. In this analysis the authors place limits of $\approx 5 \times 10^4$\,mK$^2$ on Q,U, and V at $\approx 120$\,MHz. Scaling by the frequency dependence of the sky, the polarization power spectrum levels at our frequencies should be below $ \approx 5 \times 10^4\,\text{mK}^2 (80 \,\text{MHz}/120\,\text{MHz})^5 \approx 3.7 \times 10^5\, \text{mK}^2$. The leakage from Q/U to I is given by equations~15 and 16 in \citet{Moore:2015}, and is equal to the product of the polarized power spectrum and the ratio between the integral of the differences of the $X$ and $Y$ polarized beams squared and the integral of the sum of the polarized beams squared. Using a short dipole model of our primary beam, we find that this ratio for the MWA beam is $\approx 5 \times 10^{-3}$. We therefore estimate an upper limit on the stokes Q,U to I leakage in our power spectra to be $\approx 5 \times 10^{-3} \times 3.7 \times 10^5$\,mK$^2 \approx 1.5 \times 10^3$\,mK$^2$. This is slightly larger than the EoX power spectrum which is anticipated to be several hundred mK$^2$ \citep{Pritchard:2007,Santos:2008,Baek:2010,Mesinger:2013} so it is still possible that polarized leakage may limit a detection unless direction-dependent polarization corrections are applied. However, this number is an upper limit and the actual leakage is probably lower. As of now, the most sensitive limits on the EoR power spectrum formed from $XX$ and $YY$ visibilities \citep{Ali:2015} limit (Q,U) $\to$ I leakage from the similarly elliptical PAPER beam to below the level of $\approx 500$\,mK$^2$ between $k_\parallel \approx 0.2-0.5$\,$h$Mpc$^{-1}$ at 150\,MHz while \citet{Asad:2015} predict stokes polarized power spectrum from observations of the 3C196 field at $\approx 142$\,MHz to be at the level of only $10^2-10^{3}$\,mK$^2$ at $k_\perp \lesssim 0.1$\,$h$Mpc$^{-1}$ (see their figure 12, panel a). Scaling this up by $(140\,\text{MHz}/80\,\text{MHz})^5 \approx 16$ to account for the increasing sky temperature and applying our ellipticity factor of $5 \times 10^{-3}$ gives a polarization leakage of $\approx 8-80\,$mK$^2$ which is still several times smaller than the predicted amplitude of the EoX power spectrum. 

In order to reduce artifacts from aliasing at the edges of the coarse channels due to the two-stage polyphase filter bank, we flagged 240\,kHz at each side. Finally, $uv$ cells with poor sampling were flagged since sampling and noise in these cells can change rapidly with frequency, leading to fine frequency artifacts \citep{Hazelton:2013}. 

To check our flux scale, we estimate the level of thermal noise and the system temperature  from the difference of our even/odd interleaved cubes, ${\bf \widehat{x}}_{even} - {\bf \widehat{x}}_{odd}$. Since the PSF is virtually identical between 2 second time steps, each pixel of ${\bf \widehat{x}}_{even}-{\bf \widehat{x}}_{odd}$ in the ($u,v,f$) basis has zero mean and a variance of \citep{Thompson:1986}

\begin{equation}\label{eq:var}
\left \langle | [ \boldsymbol{\mathcal{F}}_2{\bf \widehat{x} } ]_{u v f} |^2 \right \rangle = \frac{\lambda^4 T^2_{sys}(f)}{ 2 A^2_e(f) t(u,v,f) df},
\end{equation}
where $df$ is the fine channel frequency width, $T_{sys}$ is the system temperature, and $t(u,v,f)$ is the total integration time in the $uv$ cell equal to the sampling function in the $uv$ plane multiplied by the 2 second integration time step $dt$,
\begin{equation}
t(u,v,f) = \sum_j \left[ \boldsymbol{\mathcal{F}}_{\bf 2}{\bf B}_j^2  {\bf s}_j \right]_{u v f} dt.
\end{equation}
 $A_e(f)$ is the effective area of the MWA tile at frequency $f$ computed from an analytic dipole model. We may determine $T_{sys}$ by taking the ratio of the variance across $uv$ cells at each frequency in ${\bf \widehat{x}_2-\widehat{x}_1}$ and the average across $uv$ cells of our model variance predicted by equation~(\ref{eq:var}) without the $T_{sys}$ factor. We find that $T_{sys}(f)=2091$K (1139 K) at 83 (106) MHz with an error of $\approx 20$\% which is dominated by the systematic uncertainty in the fluxes of the sources used to set our flux scale. Assuming a spectral index of $-2.6$ \citep{Rogers:2008,Fixsen:2011}, these values imply a system temperature of $T_{sys} \approx 470$K at $150$\,MHz, consistent with what is found at higher frequencies in \citep{Dillon:2015b}. 

\begin{figure*}
\includegraphics[width=\textwidth]{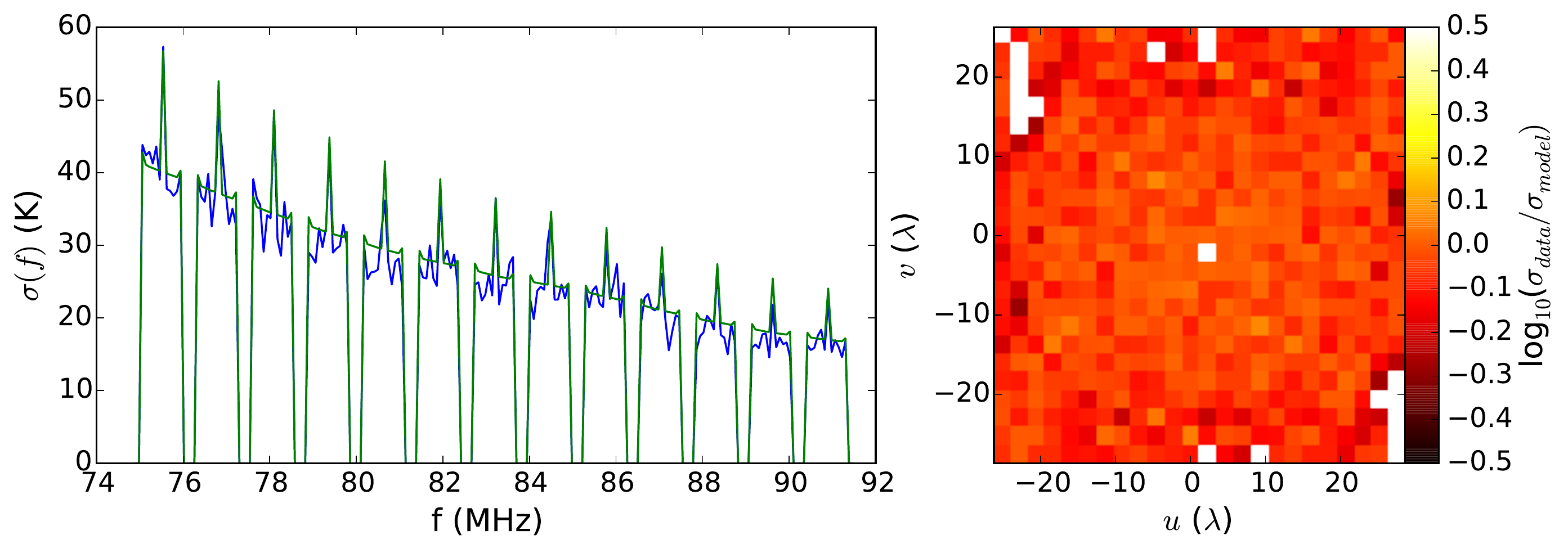}
\caption{Left: The standard deviation over $uv$ cells as a function of frequency for an even/odd time step difference cube after three hours of integration. The blue line is derived from data, while the green line is a model with a system temperature of 470\,K at 150\,MHz and a spectral index of $-2.6$. Spikes in the standard deviation are present at the center of each coarse channel since the center channel has one half of the data due to flagging the center channel which is contaminated by a digital artifact before averaging from 40 to 80\,kHz. Right: The ratio of variance taken over frequency in each $uv$ cell and our variance model using the same system temperature as on the left. The ratio between our model and observed variance is close to unity across the $uv$ plane. White cells indicate $uv$ voxels that were flagged at all frequencies due to poor sampling. All data in this figure is from Band 1.}
\label{fig:noise}
\end{figure*}

In Fig.~\ref{fig:noise} we show the standard deviation across all $uv$ cells at each frequency in our Band 1 difference cube along with the square root of the mean of our model variances at each frequency assuming $T_{sys} \propto f^{-2.6}$ and normalized to the center channel. We see that they are in good agreement. We also find that the standard deviation across frequency in each $uv$ cell is consistent with the square root of the mean across frequency of our model variances.

An interesting question is whether or not our determination of $T_{\text{sys}}$ is contaminated by ionospheric scintillation noise. For baselines within the Fresnel radius, $r_F=\sqrt{\lambda h/(2\pi)}$, of an ionospheric plasma screen of height $h$ (which is the case for the MWA core), V15a determine the coherence time for scintillation noise to be set by the length of time it takes for overhead plasma, traveling at velocity $v$, to cross the Fresnel radius, $2 r_F/v$. For typical plasma velocities of $\approx 100-500$\,km\,s$^{-1}$ and $h \approx 600$\,km, consistent with measurements \citep{Loi:2015a}, we obtain coherence times between 10 and 44 seconds at $83$\,MHz, which is signifcantly greater than the two second interleaving of our data cubes. Hence, ionospheric fluctuations are likely subtracted away in our differencing on two second intervals. For a much lower altitude of $100$\,km, the correlation time is still $\approx 4$\,seconds at $83$\,MHz. Even if there was still some variation between the time slices, we can put an upper limit on the variation relative to thermal noise by comparing the amplitude of scintillation noise we might expect given our primary beam and the parameters of the phase power spectrum of the ionospheric fluctuations that we determine in \S~\ref{ssec:ion}. We find that the level of scintillation noise on a single visibility in a two second snapshot (appendix~\ref{app:Scintillation}) is only $\lesssim 2$\,\% the thermal noise level. Further suppression of the scintillations comes from the fact that we approximate $T_{\text{sys}}$ at each frequency by taking the standard deviation across the $uv$ plane within the Fresnel zone, in which the noise is expected to be coherent and would not contribute to such a standard deviation. The coherence in frequency of the inosopheric fluctuations (V15a,b) would also suppress their contribution to the standard deviation across frequency in each $uv$ cell. For these reasons, we expect scintillation noise to have a very small contribution to our determination of $T_{\text{sys}}$ at or below the $1$\% level. 

\section{Addressing the Challenges of Low Frequency Observing}\label{sec:sys}

A number of systematics associated with observing the EoR become dramatically more challenging as one moves to higher redshift. Because the Epoch of X-ray heating (EoX) spans the FM band, we expect enhanced RFI contamination. The ionosphere's influence on electromagnetic wave propagation increases with wavelength, though its smooth evolution in frequency should cause its impact on source mis-subtraction and calibration to be contained within the wedge \citep{Trott:2015,Vedantham:2015b} (henceforth V15b). Moving down in frequency, the larger primary beam extends foreground emission to higher delays and hence larger $k_\|$ while the foregrounds increase rapidly in brightness temperature as $\approx f^{-2.6}$ \citep{Rogers:2008,Fixsen:2011}. Finally, spectral structure in our gains at fixed delays move down in $k_\|$ at higher redshift and, due to the increased primary beam width, occupy a greater extent in k-space as well\citep{Thyagarajan:2015a}.  In this section we determine the impact (if any) of each these obstacles on our power spectrum analysis and describe our strategies for mitigating them. We deal with RFI, ionosphere, and spectral structure in \S~\ref{ssec:rfi},\S~\ref{ssec:ion}, and \S~\ref{ssec:reflections}/\ref{ssec:autocal} respectively. 

\subsection{Radio-Frequency Interference}\label{ssec:rfi}

\begin{figure*}
	\includegraphics[width=\textwidth]{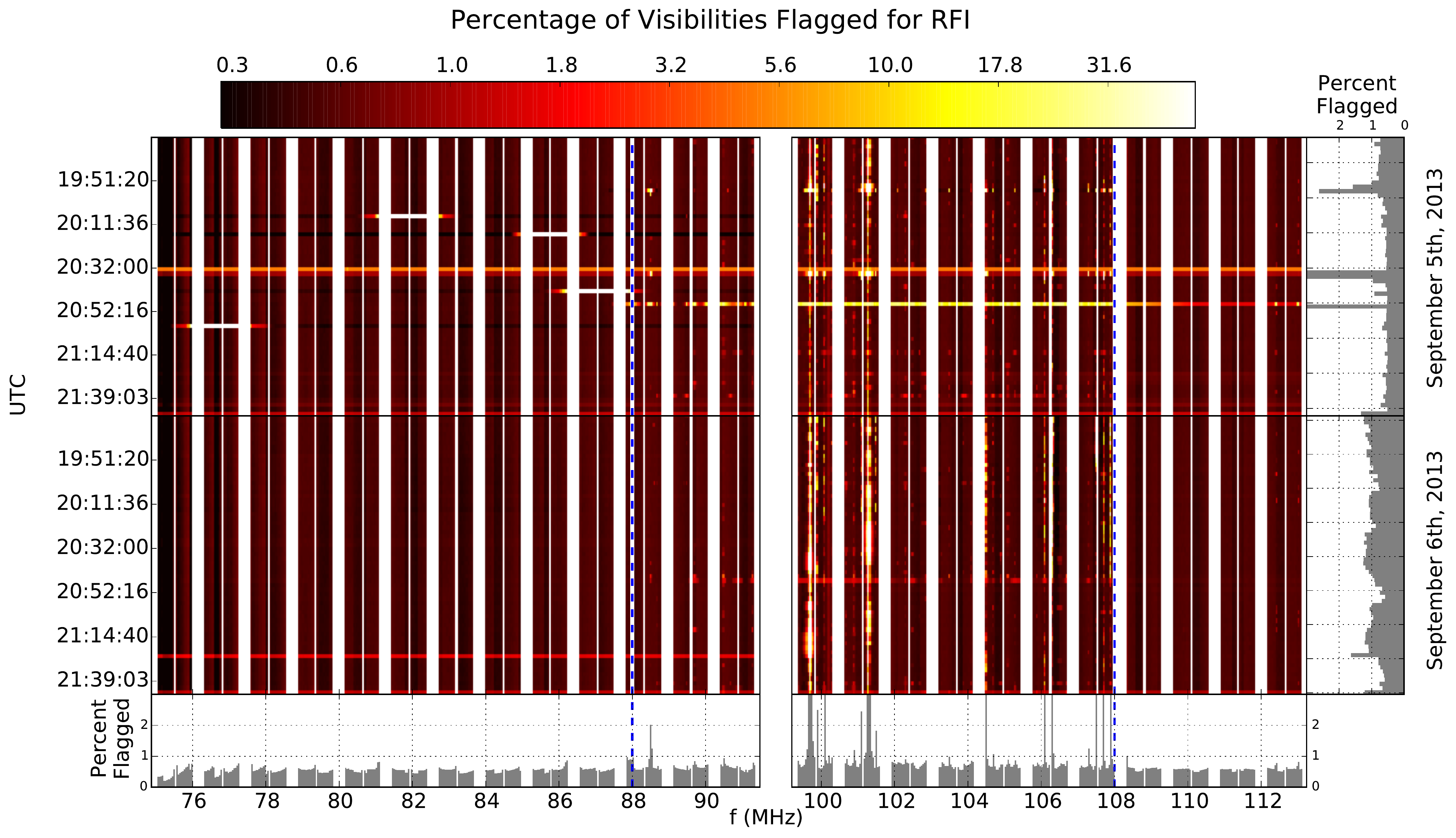}
	\caption{The percentage of visibilities flagged for RFI by {\tt cotter} as a function of time and frequency. White regions indicate missing data including the coarse band edges and blue-dashed vertical lines indicate the edge of the FM band. While Band 1 remains predominantly clear with a few sparse events within the lower end of the FM band, contamination is significantly greater over Band 2. Even in the FM band, RFI events are either isolated in frequency or time, allowing us to flag them. A handful of observations in Band 1 on September 5th are missing entire coarse channels which we also discard. Bar plots on the bottom and right show the averages of the RFI flagging fraction over time and frequency respectively.}
	\label{fig:cotter}
	\end{figure*}

	As explained above, automated RFI detection and flagging was performed using {\tt cotter} on the 0.5\,s, 40\,kHz resolution cross correlations before time and frequency averaging. To illustrate the time-frequency structure of RFI contamination, we plot the fraction of visibilities flagged at each fine frequency channel and 112\,s snapshot interval in Fig.~\ref{fig:cotter}. One can see that the majority of Band 1 is clear of RFI. Even within the region overlapping with the FM, events are sparse in time and frequency. In Band 2 we see significantly greater interference, especially in the two lowest coarse channels. There are clearly a greater number of events contained within the FM band however they only appear intermittently with the exception of a handful of 40\,kHz fine-channels. The existence of intermittent FM signals, even in a radio quite site such as the MRO is possibly due to signals from over the horizon, reflected off of the bottom of the ionosphere 
	
	It is possible for interference that is present for extended periods of time but weak enough to remain below the 112\,s noise floor over which RFI is flagged to contaminate our data. RFI can also make it past flagging through calibration solutions which are derived from the autocorrelations on which {\tt cotter} does not perform flagging (\S  \ref{ssec:autocal}). When we first created integrated data cubes there were clear signs in Band 2 that some low level RFI contamination remained in the form of ripples in the power spectrum and spikes in the frequency domain of our gridded visibility cubes.  We identified and discarded observations that appeared to contain increased flagging for a wide range of frequencies and completely flagged any channels that contained spikes in our final data cubes.
			
	The lower two coarse channels in Band 2 were contaminated for a wide range of times (Fig.~\ref{fig:cotter}) and were thus excluded from our power spectrum cubes entirely. In addition, we flagged a total of 9, 80\,kHz channels (7 \% of our data) which appeared to be contaminated by RFI at contiguous intervals over a significant number of observations; 104.08, 104.48, 106.08, 106.24, 106.32, 107.44, 107.60, and 107.84\,MHz. The rest of the FM band appears clean after routine {\tt cotter} flagging is applied. After these channels are discarded, we see no evidence that our three hour power spectrum results are limited by RFI (see the end of \S~\ref{ssec:ps2d} for further discussion). 	

\subsection{Ionospheric Contamination}\label{ssec:ion}

 The refraction induced by gradients in the total electron content (TEC) of the ionosphere scales as $\lambda^2$ and is therefore expected to be more severe at our frequencies than those associated with the EoR. In this section we quantify the severity of ionospheric conditions over our observations by measuring the differential refraction of point sources relative to known catalog positions. We find that the ionospheric gradients change considerably over the duration of our observations, despite nearly constant and mild solar weather indicators (Table~\ref{table:bulk}). However, when we form power spectra from data with nearly a factor of two difference in the observed gradients, we see no effect on the power spectrum within the EoR window (Fig.~\ref{fig:ionCompare}).  The analysis presented here is meant to assess the level of refraction and, in \S~\ref{ssec:ionCompare}, its impact on the power spectrum. Readers who are interested in a more detailed analysis of TEC gradients over the MRO and an interpretation of their physical origin or their impact on time domain astrophysics should consult \citet{Loi:2015a,Loi:2015b}. %

Radiation passing through a plasma of electrons with spatial density $N_e(r)$ acquires a phase given by \citep{Rybicki:1979}
\begin{equation}
\phi \approx \frac{e^2}{c m_e f} \int N_e(r) dr,
\end{equation}
where $c$ is the speed of light, $e$ is the electron charge, and $m_e$ is the electron mass. If we assume the ionosphere is a flat screen of plasma at height $h$ and that the TEC changes linearly over scales comparable to the separation between antenna pairs in an interferometer, ${\bf x_i}-{\bf x_j} = {\bf b_{ij}}$, the visibilities formed by cross multiplying and averaging the electric field measured by two antennas is given by (see V15a for a derivation)
\begin{align}\label{eq:ionvis}
V_{ij} &= \left \langle E(\boldsymbol{x}_i) E^*(\boldsymbol{x}_j) \right \rangle \nonumber \\ &\approx \int d^2 \boldsymbol{s} I(\boldsymbol{s}) e^{i \boldsymbol{\nabla} \phi ( \boldsymbol{s}) \cdot \boldsymbol{b}_{ij}} e^{-2 \pi i \boldsymbol{s} \cdot \boldsymbol{b}_{ij}f/c},
\end{align}
This is the standard equation for radio interferometric visibilities \citep{Thompson:1986} where the angular positions ${\bf s}$ of the sources with intensity $I({\bf s})$ have been modified to be 
\begin{equation}
\boldsymbol{s} \to \boldsymbol{s'} = \boldsymbol{s} + \frac{c}{2 \pi f} \boldsymbol{\nabla} \phi (\boldsymbol{s}).
\end{equation}
where ${\bf \nabla} \phi({\bf s})$ is the gradient of the phase screen with respect to the E-W and N-S directions. Hence one can measure gradients in the TEC by observing offsets in the positions of point sources. Note that the gradients themselves are a function of position $\nabla \phi ({\bf s})$. While we focus on refractive effects as a proxy for ionospheric gradients, we also note that a significant number of the MWA baselines are within the Fresnel radius at these frequencies and experience significant fluctuations in the source amplitudes as well, but we do not attempt a detailed analysis of these amplitude fluctuations in this work. 

\citet{Cohen:2009} have observed ionospheric TEC gradients with the VLA using the differential refraction statistic and \citet{Helmboldt:2012} have measured the 2d power spectrum of spatial and temporal fluctuations in TEC over the VLA. A similar power spectrum analysis exploiting the MWA's much larger field of view was recently carried out by \citet{Loi:2015a,Loi:2015b}. Because the ionosphere is curved and the derivation of equation~(\ref{eq:ionvis}) relies on the Fresnel approximation this model is only strictly valid for small fields of view so we only measure source shifts within 15 degrees of the phase center.  To obtain a global picture of ionospheric conditions, we turn to the differential refraction statistic employed in \citet{Cohen:2009} which we now briefly describe.

For an ensemble of source pairs with an angular separation of $\theta$, the one dimensional differential refraction statistic, $D(\theta)$, is defined as 
\begin{equation}\label{eq:diff}
D(\theta)= \left \langle |\boldsymbol{\Delta} \boldsymbol{\theta}_1 - \boldsymbol{\Delta} \boldsymbol{\theta}_2|^2 \right \rangle,
\end{equation}
where ${\bf \Delta}\boldsymbol{\theta}_{1/2}$ is the measured offset of source (1/2) from its known catalog position.

If $ \boldsymbol{\Delta} \boldsymbol{\theta} = {\bf \Delta} \boldsymbol{\theta}_1 - {\bf \Delta} \boldsymbol{\theta}_2  = (\Delta \alpha, \Delta \delta)$ is a two-dimensional vector with each component distributed with standard deviation $\sigma$, than the probability density function of the amplitude square is given by an exponential distribution. 
 We compute an estimate of $D(\theta)$, $\widehat{D}(\theta)$, within each angular separation bin by fitting a histogram of the lower 80\% of source separations to an exponential distribution in order to eliminate potentially spurious outliers. $D(\theta)$, as will be explained, is directly related to the power spectrum of phase fluctuations whose properties we will determine below.

Each $\boldsymbol{\Delta \theta}$ is the sum of an ionospheric offset and a noise term arising from random errors in determining the position of the source, 
\begin{equation}
\boldsymbol{\Delta}\boldsymbol{ \theta}_{1/2} = \boldsymbol{\Delta}{\bf I}_{1/2} + \boldsymbol{\Delta}{\bf N}_{1/2},
\end{equation}
where $\boldsymbol{\Delta} {\bf I}$ is the contribution to position offset due to TEC gradients and $\boldsymbol{\Delta} {\bf N}$ is the contribution from position errors. 
Expanding equation~(\ref{eq:diff}),
\begin{align}\label{eq:diffexp}
D(\theta) &=\left \langle | \boldsymbol{\Delta}{\bf I_1} |^2 \right \rangle + \left \langle|\boldsymbol{ \Delta} {\bf I_2}|^2 \right \rangle - 2 \left \langle \boldsymbol{ \Delta} {\bf I_1 } \cdot \boldsymbol{ \Delta }{\bf I_2} \right \rangle \nonumber \\ & + \left \langle | \boldsymbol{ \Delta}{\bf N_1}|^2 \right \rangle + \left \langle |\boldsymbol{\Delta}{\bf N_2}|^2 \right \rangle - 2 \langle \boldsymbol{\Delta}{\bf N}_{\bf 1} \cdot \boldsymbol{\Delta}{\bf N}_{\bf 2} \rangle.
\end{align}
 For separations greater than several times the width of the synthesized beam, the background noise is uncorrelated. In our analysis we only consider separations that are greater than $\approx 8.4'$ while our synthesized beam has a diameter of 4.2'. In this regime, we can ignore the cross term in equation~(\ref{eq:diffexp}). Furthermore, it is roughly stationary over the center of the primary beam lobe so that the noise terms add a $\theta$ independent offset to equation~(\ref{eq:diffexp}). If $\theta$ is small enough that both sources fall behind a single isoplanatic ionosphere patch but large enough such that the synthesized beams do not significantly overlap, $\left \langle | \boldsymbol{\Delta} {\bf I_1}|^2 \right \rangle = \left \langle| \boldsymbol{\Delta} {\bf I_2}|^2 \right \rangle = \left \langle \boldsymbol{\Delta}{\bf I_1} \cdot \boldsymbol{\Delta}{\bf I_2} \right \rangle$ so the ionospheric terms in equation~(\ref{eq:diffexp}) cancel out and we are left with only the noise bias terms. We may therefor determine the noise bias from smallest non-zero separation bin and subtract it. Our estimate of the square root of the structure function of the ionospheric fluctuations is
\begin{equation}\label{eq:structure}
\widehat{R}(\theta) = \sqrt{ \widehat{D}(\theta) -\widehat{D}(\epsilon) },
\end{equation}
where $\epsilon$ is the median angular separation of our smallest bin which is $30'$. 

\begin{figure*}

\includegraphics[width=\textwidth]{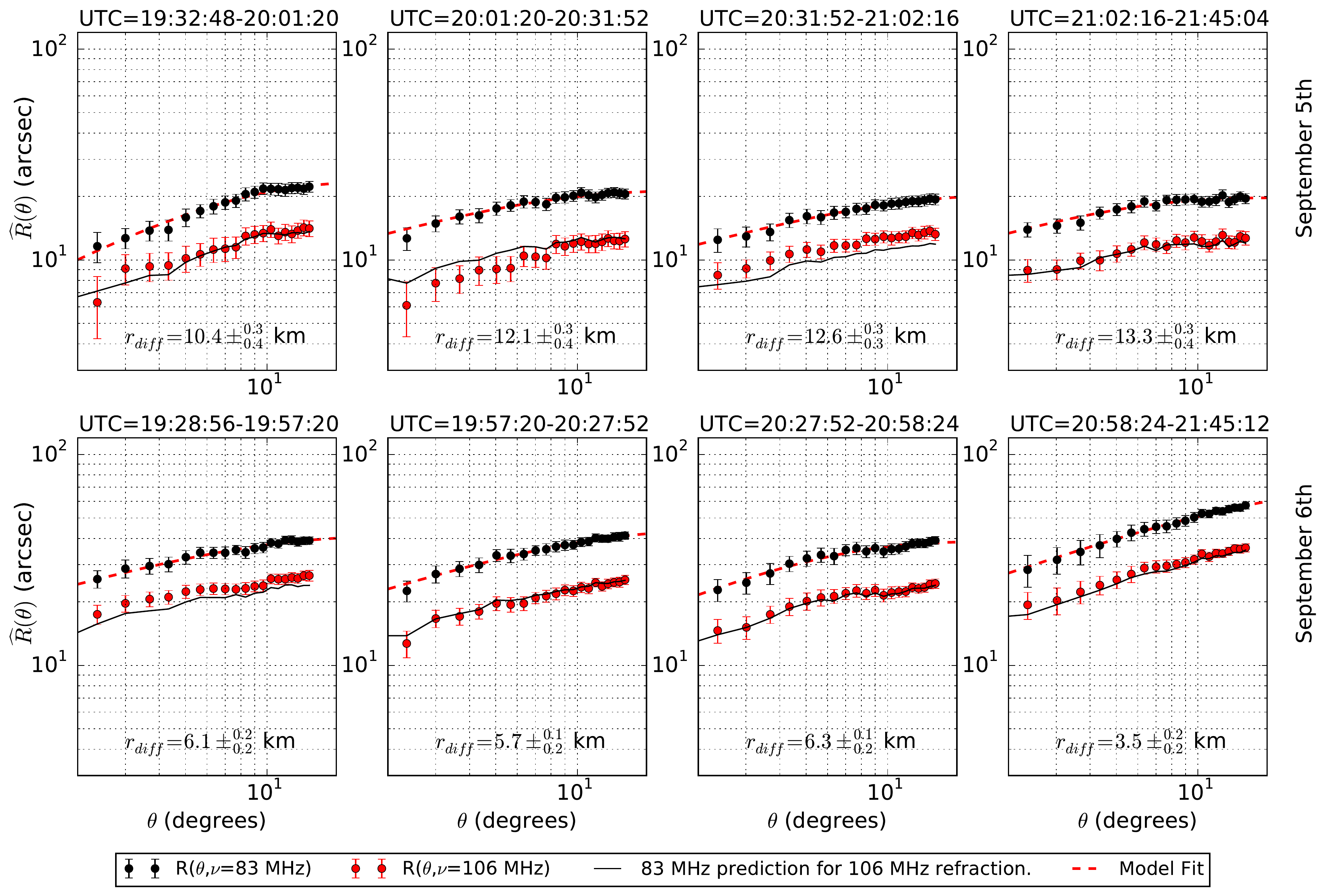}
\caption{Differential refraction derived from source pairs within 30 minute bins on September 5th, 2013 (top row) and September 6th, 2013 (bottom row). Band 1 (black points) scaled by the ratio of the band center frequencies square (solid black line) very nicely predicts the differential refraction in Band 2 (red points), indicating that the refraction we are measuring here is indeed due to ionospheric fluctuations. The magnitude of ionospheric activity differs significantly between September 5th and 6th and peaks over the last observations taken on the 6th. We also show fits to an isotropic power spectrum model of differential refraction at 83\,MHz (dashed black line) and print the inferred diffractive scale. }
\label{fig:refraction}
\end{figure*}

In Fig.~\ref{fig:refraction} we show the differential refraction computed from all differential source pair separations over 30 minutes on September 5th and 6th, 2013 for both of our observing bands. We see that the level of fluctuations recorded in both bands scales as $\lambda^2$ indicating that it indeed originates from ionospheric effects. \citet{Kassim:1993} use a similar comparison to confirm ionospheric refraction as a source of variation in visibility phases. On September 6th, the levels of refraction are approximately a factor of two greater than those observed on September 5th, peaking at the end of the night.

\begin{figure}
\includegraphics[width=.5\textwidth]{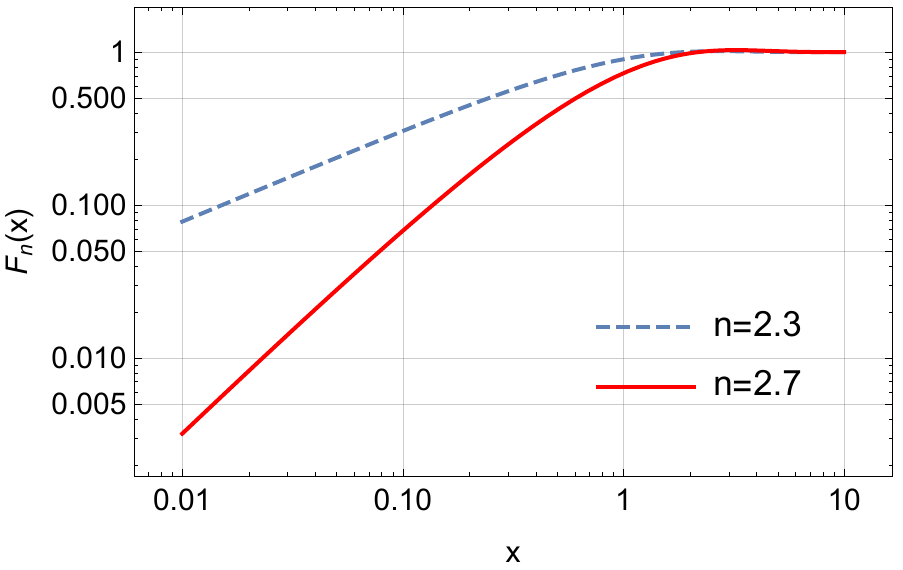}
\caption{The dimensionless integral, $F_n(x)$ normalized to unity at $F_n(10)$, given in equation~\ref{eq:Fn}. For small values of $x$, $F_n(X)$ is well approximated by a power law, but flattens out towards $x=1$. Hence, the structure function of observed source offsets levels out at the outer energy injection scale of the turbulence. }
\label{fig:Fn}
\end{figure}

We can relate our differential refraction measurements to the underlying power spectrum of phase fluctuations. In appendix~\ref{app:Refraction}), we derive the relationship
\begin{equation}\label{eq:diffPowerSpectrum}
D(\theta) = \frac{2}{2 \pi} \left( \frac{c}{2 \pi f}\right)^2 \int dk k^3 (1 - J_0(k h \theta)) P(k)
\end{equation}
where $h$ is the height of the plasma screen, ${\bf k}$ is the 2d wave vector, and $J_0$ is the bessel function of the first kind. We parameterize $P(k)$ as a generalization of the form given in V15a which describes fluctuations that level out at some outer scale $r_0 = 2 \pi/k_0$. 
\begin{equation}\label{eq:IonPowerSpectrum}
P(k) = \phi_0^2 \frac{4 \pi(n-1)}{k_0^2} \left[ \left(\frac{k}{k_0} \right)^2 + 1 \right]^{-n}.
\end{equation}
Substituting this form of the power spectrum into equation~\ref{eq:diffPowerSpectrum}, we obtain
\begin{equation}\label{eq:structurePS}
D(\theta) = 4 (n-1) \left( \frac{c}{2 \pi f} \right)^2 \phi_0^2 k_0^2 F_n(k_0 h \theta),
\end{equation}
where $F_n(x)$ is a dimensionless integral. 
\begin{equation}\label{eq:Fn}
F_n(x) = \int dq q^3 (1-J_0(qx)) [q^2 + 1]^{-n}.
\end{equation}
In Fig.~\ref{fig:Fn} we show several examples of $F_n(x)$, noting that it exhibits power law behavior for small values of $x$ and levels off towards $x=1$. Assuming a plasma height of $h=600$\,km and fitting our structure functions to equation~\ref{eq:structurePS}, we obtain values for the power spectrum normalization, $\phi_0$, the spectral index, $n$, and the outer energy injection scale, $r_0$. We show our fits to our $83$\,MHz band 1 as dashed black lines in Fig.~\ref{fig:refraction}. Fitted values for $r_0$ were on the order of several hundred kilometers, $n$ ranged between $2.3$ and $2.7$, and $\phi_0$ between $4$ rad on the quietest times and $45$ rad during the most severe refraction at the end of Sept 6th. The reported $n$ values are somewhat steeper than $n=11/6\approx 1.8$ for Kolmogorov turbulence. However, the spectral index of ionospheric fluctuatons has been found to vary significantly \citep{Rufenach:1972,Cohen:2009}. We check the slopes of $F_n(x)$ for small $x$ for our fitted indices and find that they lie within the slopes of the power law fits derived from various time ranges by \citet{Cohen:2009}. We note that there are small systematic departures from the smooth behavior described by $F_n(x)$ in most of our intervals. These are possibly due to anisotropies in the phase fluctuation fields and departure from turbulent behavior over the short $30$-minute time-scales due to transient phenomena such as traveling ionospheric disturbances. 

Given our fitted values, we can also compute the diffractive scale of the ionospheric fluctuations $r_{\text{diff}}$ which gives the scale at which the structure function of the phase field, $\phi$ reaches unity. Phase fluctuations that have large amplitudes and deocorrelate rapidly with separation give a smaller diffractive scale, allowing it to serve as a single number indicator of the severity of the fluctuations. We determine $r_{\text{diff}}$ for each 30-minute interval by computing the structure function from each fitted power spectrum and numerically solving for $r_{\text{diff}}$. We obtain error bars by computing $r_{\text{diff}}$ for $1000$ instances of $(r_0,n,\phi_0)$ drawn from a multivariate gaussian distribution whose covariance is the estimate of the fitted parameter covariances given by the Levenberg-Marquardt method as implemented in {\tt scipy}\footnote{\url{https://github.com/scipy/scipy}}. We use the $16$ and $84$\% percentile values of the resulting distributions to obtain $1\sigma$ upper and lower bounds. On September 5th, the median $r_{\text{diff}}$, at $83$\,MHz, ranged between $\approx 11-13$\,km while on September 6th, it ranged between $\approx 4-6$\,km. We report these values and their errors in Fig.~\ref{fig:refraction}. We also show the values of $r_{\text{diff}}$ derived from all source pairs on each night in table~\ref{table:bulk}. The frequency scaling of the diffractive scale depends on the spectral slope of $P(k)$, $n$. The $n$-values in our analyses typically fall between $2.3$ and $2.7$ which yields a $r_{\text{diff}} \propto f$ scaling\footnote{ Kolmogorov turbulence, with a spectral index of $n=11/6$ in the V15a parameterization scales in a very similar way; $r_{\text{diff}} \propto 6/5 \approx 1$}. Hence, our measurements imply diffractive scales of $\approx 20$\,km and $\approx 10$\,km at 150\,MHz on September 5th and 6th respectively. These values are within the range of typical scales ($5-50$\,km) described in V15a.

\begin{figure}
\includegraphics[width=.48\textwidth]{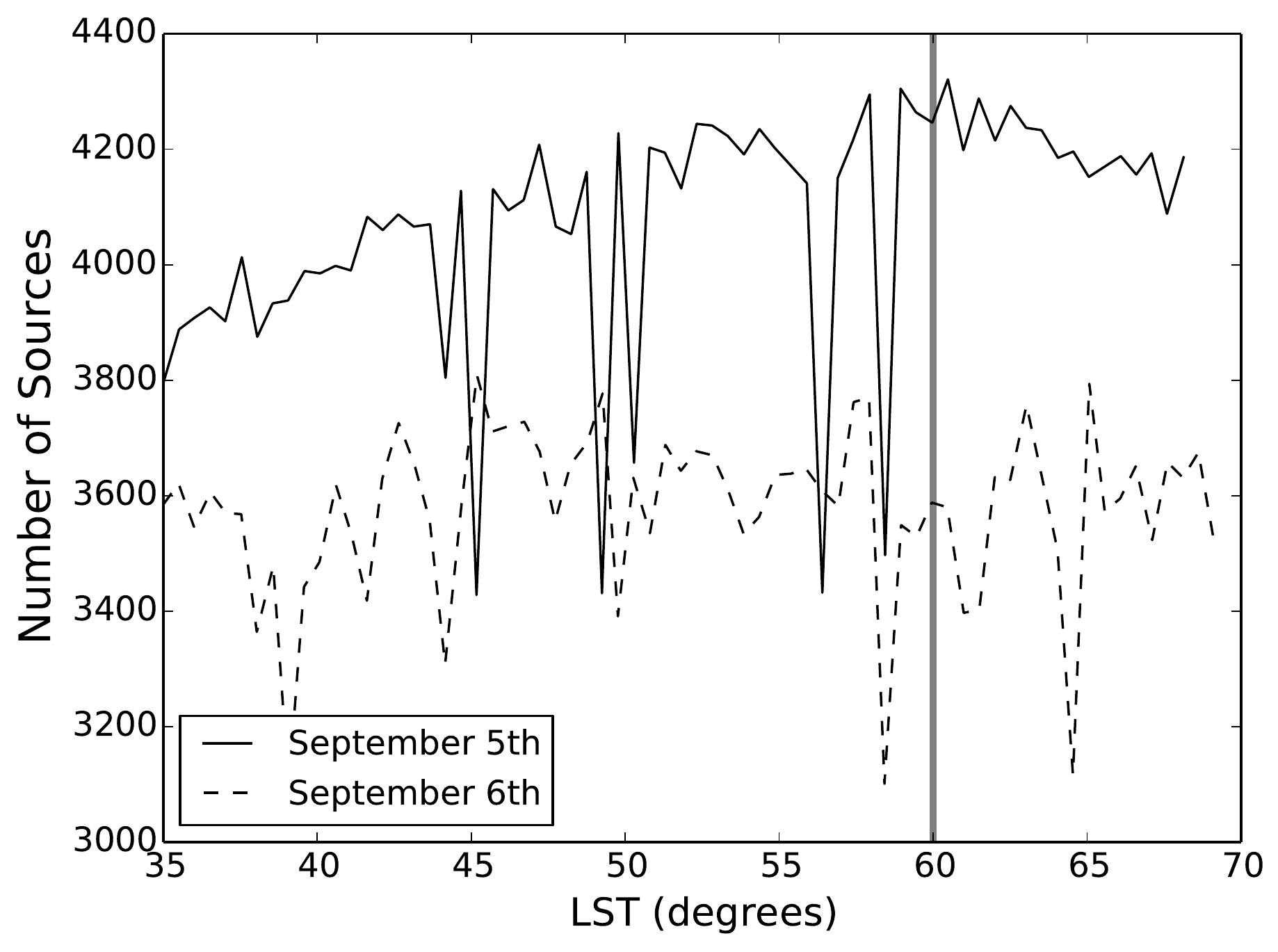}
\caption{The number of sources identified in 112\,s multifrequency synthesis images of Band 1 as a function of time over both nights of observing. On September 5th, the source counts increase with primary beam gain, until transit (vertical gray line
) before decreasing. On September 6th, when more severe ionospheric refraction was observed, the source counts remain significantly lower. Fewer sources were identified in a  handful of September 5th snapshots corresponding to observations in which calibration and flagging anomolies were observed ( Figs.~\ref{fig:cal} and ~\ref{fig:cotter}). We exclude these snapshots from our analysis. }
\label{fig:counts}
\end{figure}

Changes in the refractive index over a source on time scales shorter than the snapshot integration can cause source smearing in image space, resulting in a reduction in the observed peak brightness and reducing the number of source detections \citep{Kassim:2007}. In Fig.~\ref{fig:counts} we compare the number of sources identified in each snapshot over September 5th and 6th. On September 5th, when the level of refraction is significantly lower, we observe that the number of sources increases as the field approaches transit, corresponding to the pointing at which the beam has maximal gain and then turns over. On September 6th, the night in which significantly greater refraction was observed, the number of sources identified stays relatively constant and significantly lower than any of the source counts observed on September 5th. There are also noticeable drops in the source counts on a handful of observations on September 5th which we found to correspond to flagging events in which an entire coarse band was eliminated with significant RFI detections on the edges  (Fig.~\ref{fig:cotter}). We drop these snapshots from the rest of our analysis. 

While refraction varied significantly over both nights and within each night, the bulk solar weather and geomagnetic conditions are nearly identical. In Table \ref{table:bulk} we also list several bulk statistics such as $K_p$ index, 10.7\,cm flux (F10.7)\footnote{$K_p$ and 10.7\,cm flux values were obtained from \url{http://spaceweather.com/archive.php}.}, and mean TEC at the MRO\footnote{Values obtained by averaging global TEC maps downloaded from the MIT Haystack Madrigal database \citep{Rideout:2006}.}. The solar flux at 10.7\,cm is an often used index of solar activity and is known to correlate with the emission of the UV photons responsible for generating the free electrons that impact ionospheric radio propagation \citep{Yeh:1966,Titheridge:1973,daRosa:1973} with values ranging between $50$ and $300$ SFU (1 SFU=$10^{-22}$ W m$^{-2}$).  The $K_p$ index \citep{Bartels:1939}, quantifies the severity of geomagnetic activity and combines many local measurements of the maximal horizontal displacement of the Earth's magnetic field. Values for $K_p$ range from $0$ to $9$ with any values above $5$ indicating a geomagnetic storm.  We find that the bulk values are very similar between nights, indicating that ionospheric indicators derived from the observations themselves are a much better way of assessing data quality. Indeed, \citet{Helmboldt:2012b} also found that the levels of ionospheric turbulence and the incidence of traveling ionospheric disturbances did not appear to correlate strongly with bulk ionosphere and solar weather statistics. A systematic study of ionospheric conditions at the MRO site and correlations with bulk ionosphere statistics at higher frequencies is currently underway \citetext{Loi et al., in preparation}. Studies incorporating information from ancillary probes such as GPS stations are also being carried out \citep{Arora:2015}. 

\begin{table}
\begin{tabular}{|c|c|c|c|c|}\hline
Night & $K_p$  & $\langle$TEC$\rangle$ (TECU) & F10.7 (sfu)  & $\langle r_{\text{diff}} \rangle$ (km) \\ \hline
Sep 5th & 2 & 10.6 & 10.9 & $10.4\pm_{0.2}^{0.4}$ \\ \hline
Sep 6th & 2 & 11.4 & 11.0 & $5.20\pm_{.05}^{.04}$  \\ \hline
\end{tabular}
\caption{Bulk ionsopheric and solar weather properties on the two nights of observing presented here. $\left \langle\text{TEC} \right \rangle$ indicates the mean total electron content over the entire night. We also show the diffractive scale calculated from all source separations on both nights which differ by a factor of two. }
\label{table:bulk}
\end{table}

Having established that one of our observing nights had twice the level of refraction than the other, we can compare power spectra from each night to guage the impact of ionospheric fluctuations on the 21\,cm power spectrum. In \S~\ref{ssec:ionCompare}, we find that the two power spectra appear indistinguishable.

	\subsection{Instrumental Spectral Structure}\label{ssec:reflections}
	At low frequencies, the combination of wider primary beams and intrinsically brighter foregrounds causes leakage into the EoR window to corrupt a wider range of delays which are related to cosmological Fourier modes. In addition, the cosmological modes occupied by a fixed delay depends on redshift, causing features that contaminate low-sensitivity regions of $k$-space at one redshift to contaminate scientifically important regions at another.
		
	When we form a power spectrum from visibilities calibrated by the techniques described in \S~\ref{ssec:initCal}, we are immediately confronted with detections of striped artifacts in the EoR window at discrete delays (see the top left panel of Fig.~\ref{fig:calCompare}). These delays correspond to the round-trip travel times on the lengths of cable that connect the MWA's receivers and beamformers. We can get a rough understanding of how miscalibrated cable reflections affect the power spectrum  by considering the effect of reflections with delay, $\tau_j$ and complex amplitude $\widetilde{r}_j$ on the $j^{th}$ tile. To first order in $| \widetilde{r}_{i/j}|$, the effect of uncalibrated cable reflection is to multiply a visibility by a reflection factor.
	\begin{equation}\label{eq:visreflect}
	V_{ij} \to V_{ij} \left( 1 +\widetilde{r}_i e^{2 \pi i \tau_i f} + \widetilde{r}_j^* e^{-2 \pi i \tau_j f } + \mathcal{O}(\widetilde{r}^2) \right)
	\end{equation}  	
	
	The power spectrum is formed, roughly, by taking a Fourier Transform of the gridded visibilities in frequency and squaring. The square of the Fourier transform of a visibility with uncorrected reflections becomes,
\begin{align}\label{eq:reflections}
	&|\widetilde{V}_{ij}(\tau)|^2 \to |\widetilde{V}_{ij}(\tau)|^2 + \nonumber \\
	& 2 \text{Re} \left[\widetilde{r}_i \widetilde{V}_{ij}(\tau) \widetilde{V}_{ij}(\tau-\tau_i)\right] + \nonumber \\
	& 2\text{Re} \left[ \widetilde{r}_j^*\widetilde{V}_{ij}(\tau + \tau_j) \widetilde{V}_{ij}(\tau) \right]+ \nonumber \\ &2 \text{Re} \left[ \widetilde{r}_i \widetilde{r}_j^* \widetilde{V}_{ij}(\tau-\tau_i) \widetilde{V}_{ij}^*(\tau+\tau_j) \right] + \nonumber \\
	& 2 \text{Re} \left[\widetilde{r}_i \widetilde{r}_j^* \widetilde{V}_{ij}(\tau - \tau_i + \tau_j) \widetilde{V}_{ij}^*(\tau) \right] +  \nonumber \\
	& 2 \text{Re} \left[\widetilde{r}_i^2 \widetilde{V}_{ij}(\tau - 2 \tau_i) \widetilde{V}^*_{ij}(\tau) \right] + \nonumber \\
	& 2 \text{Re} \left[\widetilde{r}_j^{*2} \widetilde{V}_{ij}(\tau + 2 \tau_j) \widetilde{V}^*_{ij}(\tau) \right] + \nonumber \\
	&|\widetilde{r}_i|^2 | \widetilde{V}_{ij}(\tau-\tau_i) |^2 + |\widetilde{r}_j|^2 | \widetilde{V}_{ij}(\tau+\tau_j) | ^2 + \mathcal{O}(\widetilde{r}^3).
\end{align}

 All terms in equation~\ref{eq:reflections} involve the cross multiplications of $\widetilde{V}_{ij}(\tau + \Delta \tau) \widetilde{V}_{ij}^*(\tau + \Delta \tau')$ and a coefficient on the order of $\widetilde{r}^n$.  The cable delays on the MWA are all $\gtrsim 90$\,m, corresponding to round trip delays significantly beyond the wedge for the short baseline lengths considered in our power spectrum analysis. Hence, if $|\widetilde{r}|$ is greater than the ratio between the signal and foreground amplitudes, $R_{fg}$, terms with $\Delta \tau = \Delta \tau'$ dominate equation~\ref{eq:reflections}.  As a consequence, the first and last lines in equation~\ref{eq:reflections} dominate all others. The first ($\mathcal{O}(0)$) is the foregrounds in the absence of reflections and exceeds the signal by a factor of $10^{10}$ but is also contained within the wedge. The last line contaminates $\tau=\tau_i,-\tau_j$ at the level of $\widetilde{r}^2\times10^{10}$ the signal level. All other lines in equation~\ref{eq:reflections} exceed the signal level by $\widetilde{r}^2 10^5$ but also contaminate a greater range of delays: $\tau = \tau_i,-\tau_j,2\tau_i,-2\tau_j,\tau_i-\tau_j$.
 
As we will discuss below, we find that $\widetilde{r} \lesssim 10^{-2}$ (Fig.~\ref{fig:counts}), hence our analysis is only sensitive to the first and last lines arising from first order reflections. The fourth, fifth and sixth terms in equation~\ref{eq:reflections}, which couple second order reflections and beats will be above the level of the cosmological signal twice the round trip delay times and their differences. However, if the $\lesssim 1\%$ reflections can be corrected to be below the level of $\lesssim 10^{-3}$, these cross terms will only appear at the $10^{-3}$ level of the signal and not impede a detection (though they may introduce some bias).

Since the $\mathcal{O}(\widetilde{r}^2)$ terms in the last line of equation~\ref{eq:reflections} dominate the others outside of the wedge, the lowest order effect of a reflection is to  multiply our foregrounds by the reflection coefficient squared and translate them outside of the low-delay region in which they are usually confined (the wedge) to the round-trip delay of our cable which for the MWA is outside of the EoR window. Unless $|\widetilde{r}|$ can be brought below the ratio between the foregrounds and the signal itself ($|\widetilde{r}_i| \lesssim 10^{-5}$), the modes within a wedge translated to $\tau_i$ will remain unusable. Uncorrected reflections at the $10^{-2}$ level will also contaminate higher order harmonics and the differences between the delays.
	
For $\widetilde{r}$ significantly larger than $R_{fg}$ (as is the case here), delays corresponding to higher order reflections will also be contaminated. These higher order terms are well below the sensitivity of this analysis but they can still potentially pose an obstacle to the detection of the signal. Hence, it is worth commenting on the level that $\widetilde{r}^3$ and $\widetilde{r}^4$ terms contaminate our data. 
	First we address $\widetilde{r}^3$. Since every contribution of order $\mathcal{O}(\widetilde{r}^3)$ in equation~\ref{eq:reflections} involves the product of an $\mathcal{O}(\widetilde{r}^3)$ coefficient with delay transformed visibilities translated to two different delays ($\Delta \tau \ne \Delta \tau'$), these terms will contribute at the level of $\widetilde{r}^3 \times 10^5$ the level of the foregrounds. Uncorrected third order terms with $\widetilde{r} \lesssim 10^{-2}$ will contaminate our data at $10^{-1}$ the level of the signal and we do not consider them a serious issue, especially if the reflections are corrected to the $\approx 10^{-3}$ level.

Writing down all $\widetilde{r}^4$ order terms in equation~\ref{eq:reflections} is straightforward but not terribly enlightening.  We can obtain the leading contributions to the $\mathcal{O}(\widetilde{r}^4)$ terms in equation~\ref{eq:reflections} by ignoring the products of delay visibilities translated by different amounts, $\Delta \tau \ne \Delta \tau'$. The $\mathcal{O}(\widetilde{r}^4)$ contributions with $\Delta \tau = \Delta \tau'$ are
\begin{align}
& \ldots + |\widetilde{r}_i|^4 |\widetilde{V}_{ij}(\tau - 2 \tau_i) |^2 + |\widetilde{r}_j|^4 |\widetilde{V}_{ij}(\tau+2 \tau_j)|^2 \nonumber \\
&+ |\widetilde{r}_i|^2 |\widetilde{r}_j|^2 |V_{ij}(\tau - \tau_i + \tau_j)|^2 + \ldots
\end{align}

 We conclude that the sub-percent reflections observed in our data will introduce $\mathcal{O}(\widetilde{r}^4)$ terms to the power spectrum at $\widetilde{r}^4 \times 10^{10}$ times the amplitude of the signal at twice the fundamental delays and their differences. Second order reflections at the $\lesssim 1\%$ level will therefore dominate the signal by two orders of magnitude, impeding a detection. Fortunately, the $\widetilde{r}^4$ dependence of these second order terms greatly amplifies even modest improvements in correcting the reflections. For example, if the reflections are brought to below the $0.1\%$ level, the $\mathcal{O}(\widetilde{r}^4)$ terms will be brought to below $10^{-2}$ the level of the 21\,cm signal. We are able to bring the reflections down to $\approx 0.002$ so they are not a problem in our data. 

What cosmological wave-vectors in our measurements are contaminated by reflections? In cosmological coordinates, a fixed delay corresponds to a line-of-sight wave-number of $k_\|$ (in units of $h$\,Mpc$^{-1}$) of
\begin{equation}\label{eq:eta2kpara}
k_\| \approx \frac{2 \pi H_0 f_{21} E(z)}{c h (1+z)^2} \tau,
\end{equation}
\citep{Morales:2004} where $c$ is the speed of light, $H_0/h=100$\,km\,s$^{-1}$\,Mpc$^{-1}$, $f_{21}$ is the hyperfine emission frequency, and $E(z)=\sqrt{\Omega_M(1+z)^3+\Omega_k(1+z)^2+\Omega_\Lambda}$.  
For constant $\tau$, the $k_\|$ center of the translated delay wedge will decrease with increasing redshift.

In Table \ref{table:reflections} we list the lengths of cable between the MWA receivers and beamformers along with their round-trip delays, and their corresponding $k_\|$ at the center redshift of our two bands along with a redshift typical of an EoR measurement. Because $P(k)$ decreases rapidly with increasing $k$, interferometers are expected to have the highest signal to noise at the smallest delay that is uncontaminated by foregrounds (280\,ns, corresponding to $k_\| \approx 0.1-0.2$\,$h$\,Mpc$^{-1}$ at EoX to EoR frequencies \citep{Pober:2013a}). Assuming that reflections can be corrected to be below the $10^{-3}$ level so that only the last line in equation~\ref{eq:reflections} is above the signal, they should be benign as long as they remain at sufficiently large or small $k_\|$ that they don't leak into the region of maximal sensitivity. Because standing waves translate the entire wedge up to their delay, reflections located at the edge of the wedge will result in excess supra-horizon emission while cable reflections outside of the wedge will contaminate a finite chunk of $k_\|$, not just the delay of the reflection itself. 

\begin{figure*}
\includegraphics[width=\textwidth]{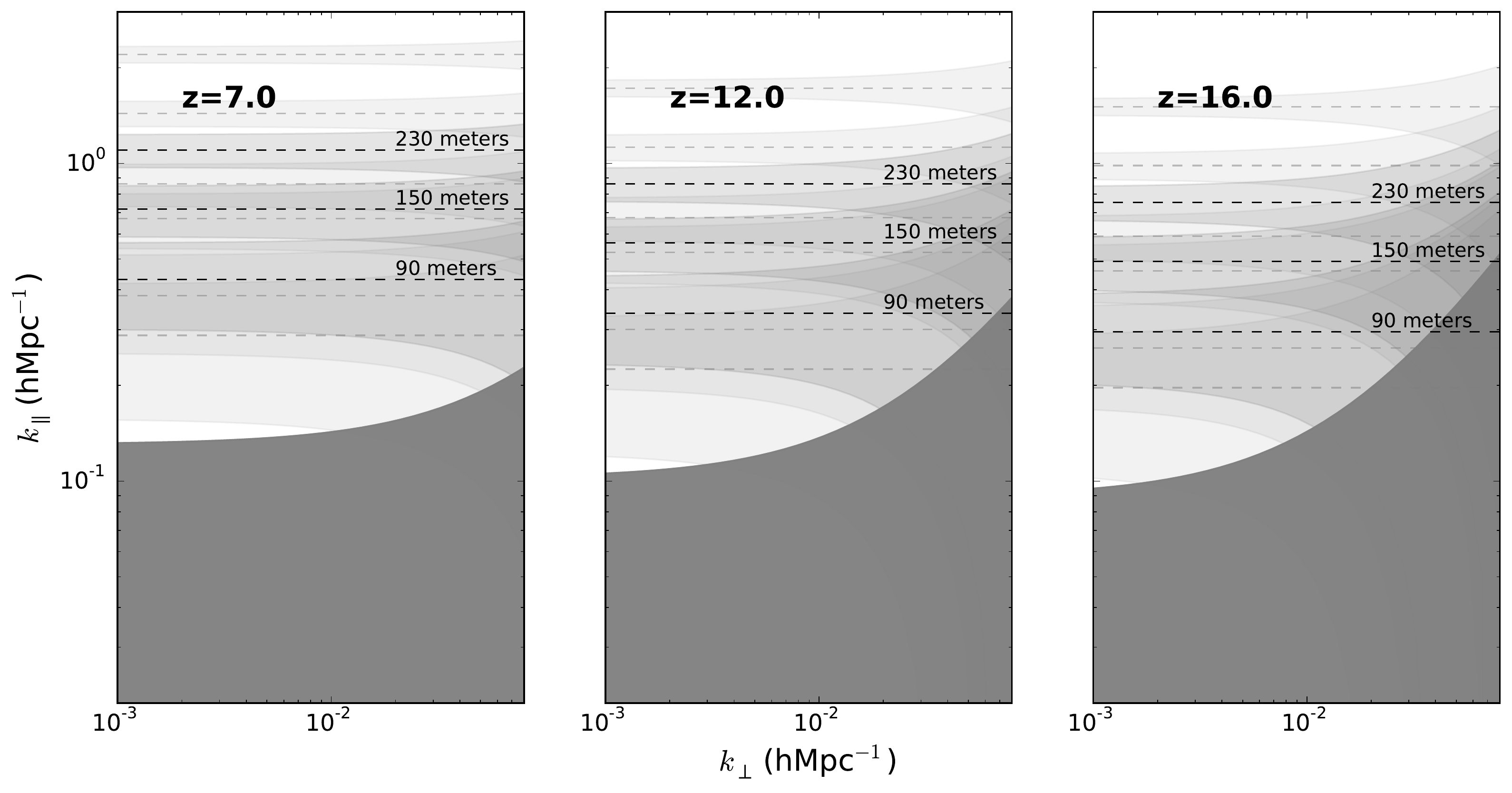}
\caption{The regions of the EoR window contaminated by foregrounds due to uncalibrated cable reflections for several different redshifts.  Dark gray regions denote contamination from first order cable reflections assuming a wedge out to the first null of the primary beam plus the 0.15\,$h$\,Mpc$^{-1}$ at $z=8.5$ buffer observed in \citet{Pober:2013a}. Since the buffer is associated with the intrinsic spectral structure of foregrounds, it lives in delay space. Dark gray regions denote foreground contamination within the wedge which exists even without instrumental spectral structure. At $z=7$, a representative EoR redshift, the contaminated regions remain at relatively high $k_\|$ and have smaller widths due to the smaller primary beam and the scaling of $k_\perp$ and $k_\|$ with $z$. While regions exist between the first order reflections that are somewhat wider at lower redshift, second order reflections can still potentially contaminate nearly all of the EoR window in which interferometers are supposed to be sensitive (light grey shaded regions). Second order reflections are below the sensitivity level of this study but will also pose an obstacle to longer integration unless calibrated out.}

\label{fig:refComparison}
\end{figure*}

We see in Table \ref{table:reflections} that the minimum $k_\|$ associated with a cable ripple on the MWA at $z=7$ is $0.42$\,$h$\,Mpc$^{-1}$ which, even if we allow for a delay width of 280\,ns (the approximate width of the wedge with a supra-horizon emission buffer at $k_\perp =0$), is above the region of maximal sensitivity. However, at $z=16$, the approximate center redshift of our Band 1, this cable ripple lies at $0.27$\,$h$\,Mpc$^{-1}$, which can leak a significant amount of power into the sensitivity sweet spot due to its finite $k_\|$ width. This effect is illustrated in Fig.~\ref{fig:refComparison}. At $z=7$,  the smallest reflection delay on the MWA introduces foreground contamination down to $k_\| \approx$ 0.3\,$h$\,Mpc$^{-1}$, leaving a small foreground-free window in which cosmological measurements may be performed. On the other hand, at  $z=16$, this window becomes smaller with the other reflections filling the EoR window up to $k_\| \sim 0.8$\,$h$\,Mpc$^{-1}$. If the corrected reflections are several times larger than $10^{-3}$, higher order terms below the $90$\,m delay may be comparable to the level of the 21cm signal (shown as light grey regions and centered on light-gray dashed lines). However, only the peak of this foreground power will be near the signal level and the broad wings caused by beam chromaticity and will be well below the signal. 


\begin{table*}
\begin{tabular}{c c c c c c c }\hline
L & N&$\tau_r$ & $f_{30 \lambda } $ & $k_\| (z=7)$ & $k_\| ( z= 12)$ & $k_\| (z=16)$ \\ 
(m) & & ($\mu$s) &  \% & ($h$\,Mpc$^{-1}$) & ($h$\,Mpc$^{-1}$) & ($h$\,Mpc$^{-1}$)  \\ \hline
90 & 19 & .74 & 40 &  .42 & .31 & .27 \\ 
150 & 31 & 1.2  & 80 & .70 & .53 & .50 \\ 
230 & 23 & 1.9 & 29 & 1.1  & .83 & .70 \\ 
320 &8& 2.6 & 0.4 &1.5 & 1.2  & 1.0 \\ 
400 &17& 3.3 &  1.7 & 1.9  & 1.4 & 1.3 \\ 
524 & 30 & 4.3 & 1.9 & 2.5  & 1.9 & 1.7 \\ 
\hline
\end{tabular}
\caption{There are N cables of each length (L) between the MWA receivers and beamformers with associated round-trip delay times $(\tau_r)$. In the $f_{30\lambda}$ column, we list the percentage of baselines within $30\lambda$ at 83\,MHz (where the majority of the MWA's power spectrum sensitivity  lies)  that are formed from at least one tile with the given cable length. We also list the $k_\|$ of each delay given by equation~(\ref{eq:eta2kpara}) for three different redshifts. Cable reflections that are significantly above the $k_\|$ values where we expect to obtain maximum sensitivity to the power spectrum at EoR redshifts $(z \approx 7)$ move into the maximum sensitivity region at EoX redshifts $(z \approx 16)$. Higher order reflections will also contaminate multiples of and differences between the delays and $k_\parallel$ values listed in this table (thought at a lower level).}
\label{table:reflections}
\end{table*}

This effect is purely geometrical in that while the mapping from instrumental delays to k-space varies, the number of measurements in $(u,v,\eta)$ cells which are uncontaminated stays constant. However, as we go to higher redshift the increasing width of the primary beam increases foreground power at supra-horizon delays, effectively reducing the number of usable  $(u,v,\eta)$ cells. As mentioned above, the cosmological power spectrum decreases significantly with increasing $k$ so the fact that smaller $k$ are contaminated by cable reflections at higher redshifts hurts our sensitivity disproportionately. We also note that any partial reflections from kinks and bends within the cable itself can lead to contamination of additional delays below the round-trip travel time on the cable.

We make a first attempt to remove these reflections by fitting reflection functions (equation \ref{eq:reflection}) to our self calibration solutions divided by the smooth fit in equation~\ref{eq:smooth}. Since our per snapshot calibration solutions are too noisy to obtain good fits for the beamformer-receiver reflection, we fit these on calibration solutions that are averaged over each night of observing. We find that this method is of limited efficacy in removing the receiver to beamformer ripples (Fig.~\ref{fig:calCompare}, top right).

Since, to avoid cosmological signal removal and spurious frequency structure due to mis-modeled foregrounds, we are attempting to model our bandpass with a small number of parameters we are unable to capture the full spectral structure of the cable reflections. As we see in Fig.~\ref{fig:rHist} the reflection parameters are frequency dependent and we do not have a clear picture of their precise evolution. Calibration exploiting redundant baselines might be able to make headway on the problem since it is not sensitive to unmodeled signal except in for deriving an overall phase and amplitude scaling that averages over all baselines \citep{Wieringa:1992,Liu:2010,Zheng:2014}. The MWA's baselines are designed for very low redundancy, making the technique unusable here. Future upgrades to the MWA are expected to include a significant number of redundant baselines \citetext{Tingay, private communication}.

\subsection{Calibration with Autocorrelations}\label{ssec:autocal}

Confronted with the problem of reflections contaminating the power spectrum, we apply an alternative approach that uses tile autocorrelations to obtain calibration amplitude gains with sub-percent level accuracy.

What information is encoded in the autocorrelations? Let $I({\bf s},f)$ be the brightness distribution on the sky at frequency $f$ and direction ${\bf s}$. Consider the sky signal entering the antenna and traveling through a signal chain in which the $m^{th}$ successive element applies a multiplicative complex gain $g_{j,m}$ and adds a zero mean noise component with variance $N^2_{j,m}$. Correlating the output at the $M^{th}$ gain element with itself to form an autocorrelation yields
\begin{equation}\label{eq:autoCorr}
V_{jj}(f) = | g_j |^2 \left[ \sum_{m=0}^M \frac{N^2_{j,m}(f)}{\Pi_{n=0}^{m} |g_{j,n}|^2}+ \int d\Omega A_j({\bf s},f) I({\bf s},f) \right].
\end{equation}
Here, $g_j = \Pi_{m=0}^M g_{j,m}$ is the net gain and $A_j({\bf s},f)$ is the antenna beam. Using simulations of diffuse emission and an analytic model of the MWA primary beam, we find that $\int A_j I({\bf s},f) d\Omega$ is fit at the $10^{-5}$-$10^{-6}$ level by a third order polynomial while $N_{j,m}^2(f)$, which is due to noise in analogue electronics should also vary smoothly with frequency. Hence $\sqrt{V_{jj}}$ is well approximated by the product of $| g_j |$ multiplied by a smooth factor which may be modeled by a low-order polynomial. 

To remove this multiplicative factor, we need a model for the ratio of the square root of each autocorrelation to the amplitude of the calibration solution. We use the product of a third order polynomial and a 7\,m cable reflection (modeling the LNA-beamformer cables) and fit it to the ratio, averaged over 112\,s intervals, using our noisy initial calibration solutions.

\begin{figure}
\includegraphics[width=.48\textwidth]{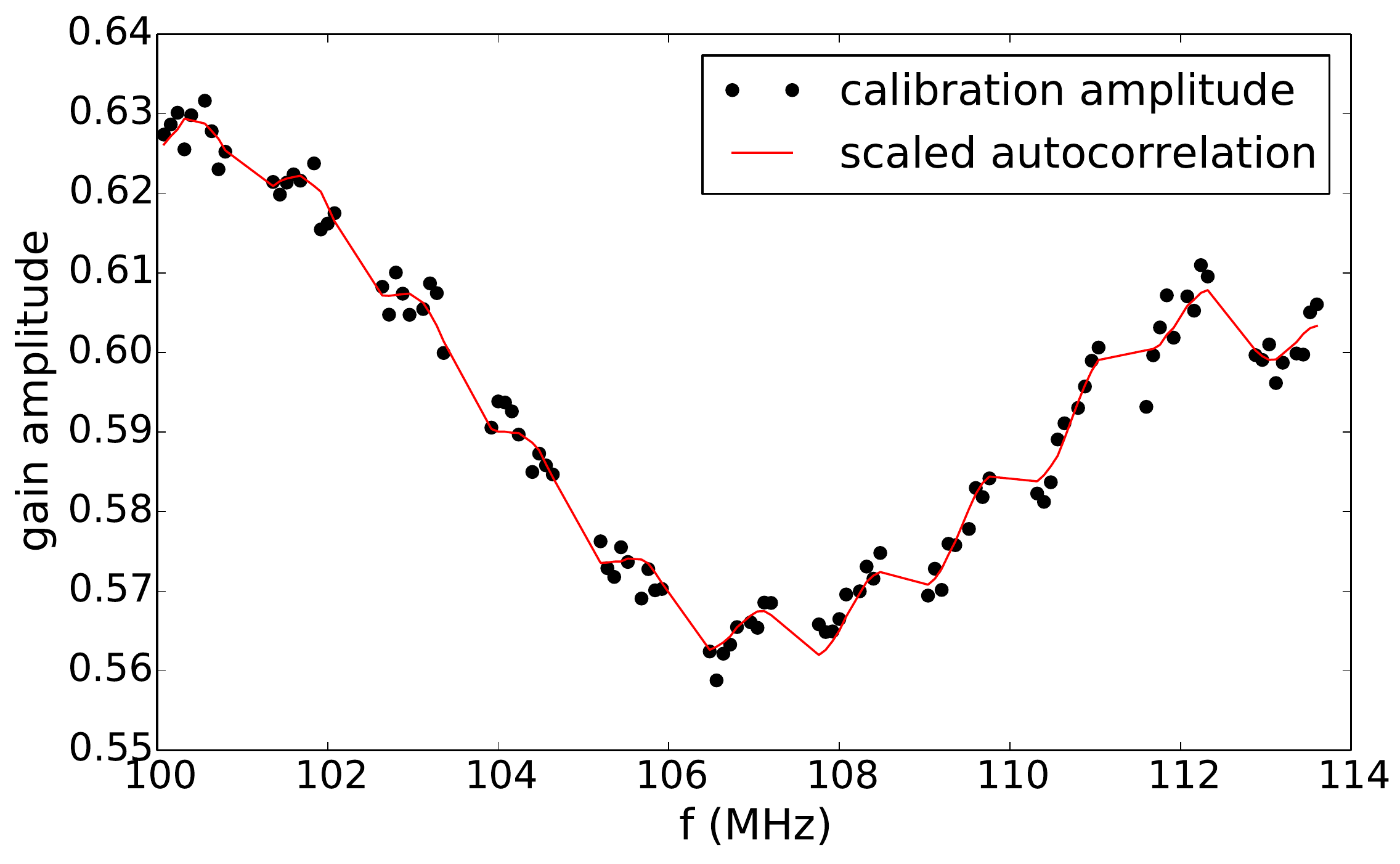}
\caption{We show the amplitude of a calibration gain averaged over a fifteen minute pointing (black circles) along with the square root of our  autocorrelations which have been scaled by a third order polynomial and a single seven meter reflection to match the calibration solution (red line). After multiplying the autocorrelations by a smooth function, they are brought into good agreement with the calibration gains.}
\label{fig:autoCompare}
\end{figure}

We then use the square root of the autocorrelation divided by this polynomial as our calibration amplitude. In Fig.~\ref{fig:autoCompare} we demonstrate the validity of this technique by comparing the smoothly corrected autocorrelations for a single snapshot with a calibration amplitude that has been averaged over a single pointing and see that they are in very good agreement.

\begin{figure*}
\includegraphics[width=\textwidth]{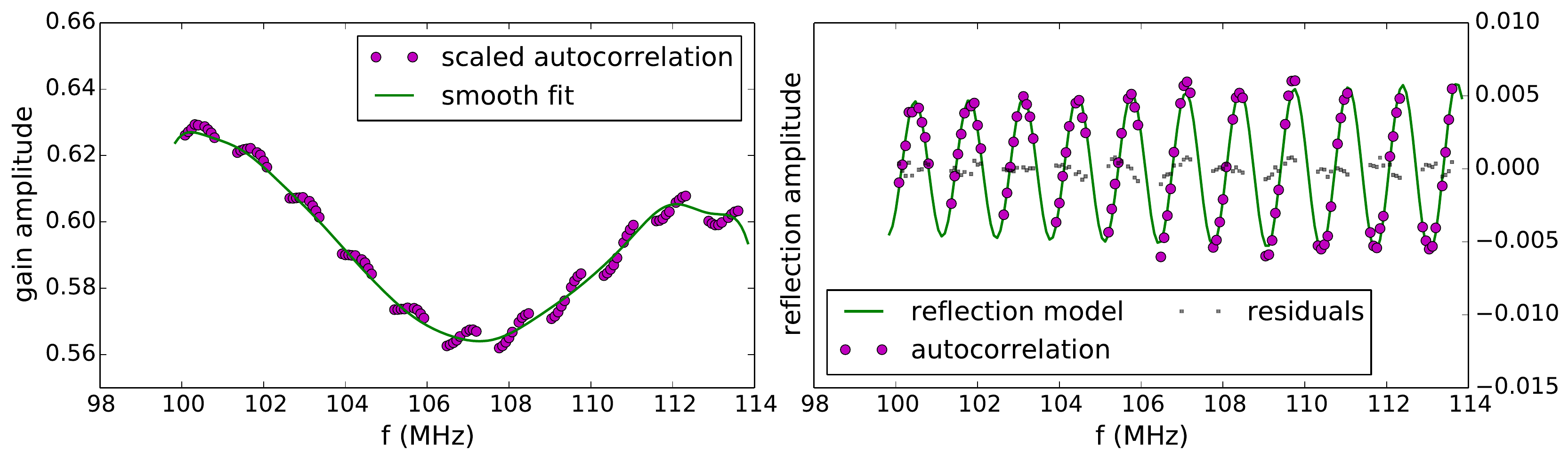}
\caption{Left: In order to obtain reflection parameters, we divide our scaled autocorrelation (magenta circles) by a smooth fuction consisting of a third order polynomial and large scale reflections (green line). Right: We fit this ratio (magenta circles) to a reflection function (green line) and are left with $\sim 10\%$ residuals (grey points).}
\label{fig:autoResid}
\end{figure*}

While the autocorrelations can be used for our amplitude calibration, there still remains the problem of adding the reflection ripple to phase calibration. We can use autocorrelations to predict this ripple in the phases. We obtain its parameters following the same fitting procedure described in \S \ref{ssec:initCal} except this time we fit the scaled autocorrelations, with a smooth polynomial divided out, to the amplitude of the reflection term in equation~\ref{eq:reflection},
\begin{equation}\label{eq:refAmp}
|R_j(f)|= \frac{1}{\sqrt{1-2 r_j \cos ( 2 \pi f \tau_{L(j)}+ \phi_j)+r_j^2}}.
\end{equation}
In Fig.~\ref{fig:autoResid} we illustrate the fitting procedure by showing the autocorrelations divided by the smooth fit along with the best fit model reflection. One can see that the residuals in the reflection fit tend to be on the order of 10\% hence there is some fine scale structure at the $10^{-3}$ level that we are still unable to model. Since our model includes the impact of higher order reflections, we think that these residuals arise from unmodeled frequency dependence in the reflection coefficients, sub-reflections in the cables, and digital artifacts present in the autocorrelations. Recalling our discussion in \S~\ref{ssec:reflections}, $10^{-3}$-level residuals will leave contamination in our power spectra at the level of $10^4$ times the signal level due the $|\widetilde{r}|^2$ terms in the last line of equation~\ref{eq:reflections} but suppress all higher order reflections to below the signal level.  

From $\tau_{L(j)}$, $\phi_j$, and $r_j$, we add the reflection's additive contribution to the gain phase
\begin{multline}\label{eq:refPhase}
\text{Arg}(g_j) \to \text{Arg}(g_j') \\ =\text{Arg}(g_j)+\tan^{-1} \left [  \frac{-r_j \sin (2 \pi f \tau_{L(j)} + \phi_j)}{1 - r_j \cos ( 2 \pi f \tau_{L(j)} + \phi_j)} \right].
\end{multline}

\begin{figure}
\includegraphics[width=.48\textwidth]{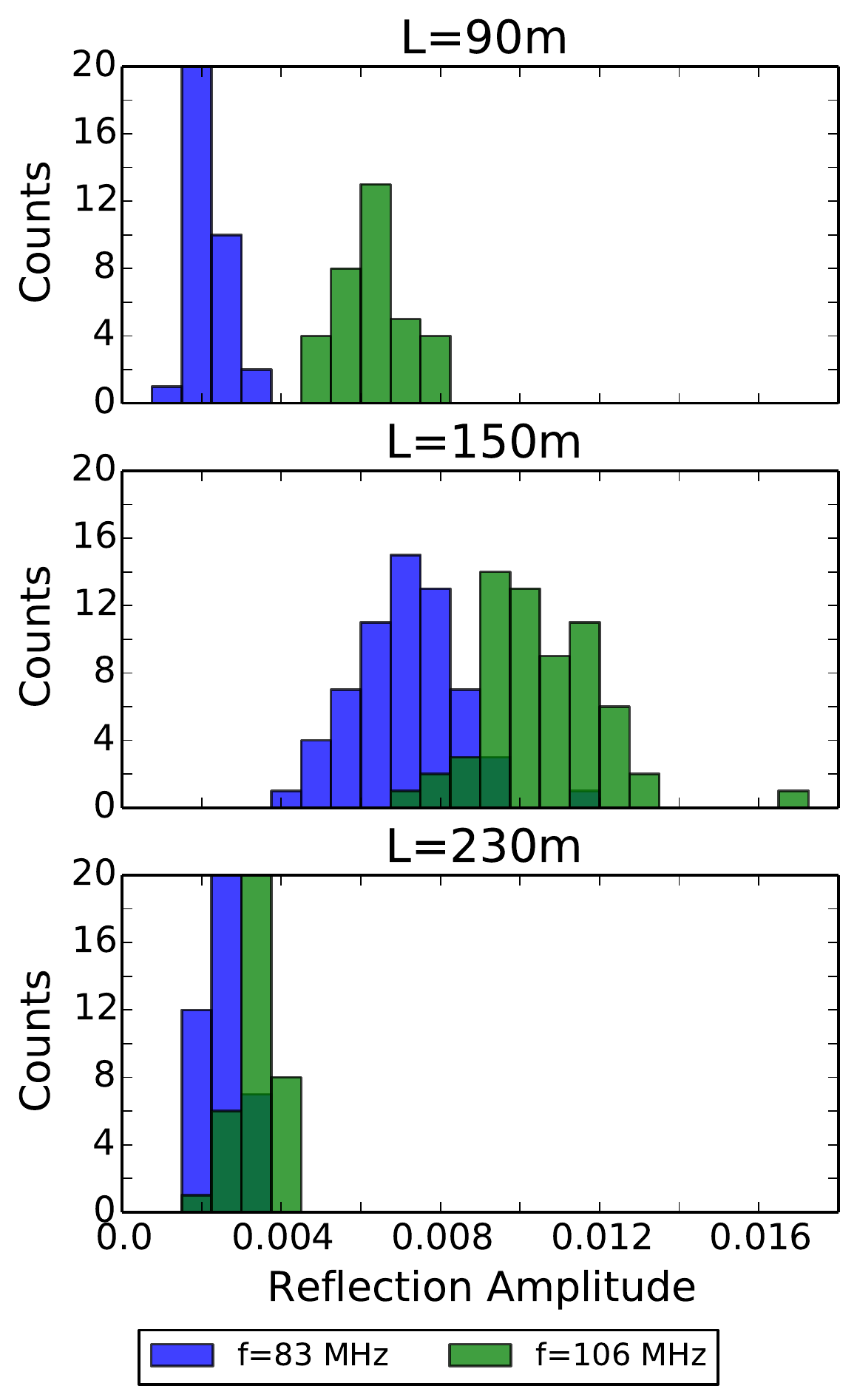}
\caption{Histograms of fitted cable reflection amplitudes for Band 1 (blue) and Band 2 (green) obtained from fits to autocorrelations for three different cable lengths between MWA receivers and tiles. The reflection amplitudes range from $0.2-1\%$ making them difficult to fit using the noisy self calibration solutions. Reflection amplitudes in Band 2 are systematically larger than Band 1 for all cable lengths, indicative of non-trivial frequency evolution in the reflection parameters.}
\label{fig:rHist}
\end{figure}

Since tiles with 320, 400, and 524\,m cables only contribute to  $\sim 4 \%$ of our sensitive baselines, we discard them entirely. In Fig.~\ref{fig:rHist} we show the distributions of the reflection amplitudes fitted from autocorrelations (averaged over the night of September 5th) inferred for our 90, 150, and 230\,m cables for both the high and low bands. One can see that the reflection coefficients are on the order of fractions of a percent and vary significantly from cable length to cable length. This is reasonable since the cable impedance, which determines the reflection amplitude, is a function of both its geometry and dielectric properties (with equal length cables likely formed from cable batches of similar dielectric properties). In addition, frequency evolution of the reflection amplitude is apparent by comparing the fit in Bands 1 and 2 implying that a single delay standing wave is not quite the correct model to use in our phases. 

\begin{figure*}
\includegraphics[width=\textwidth]{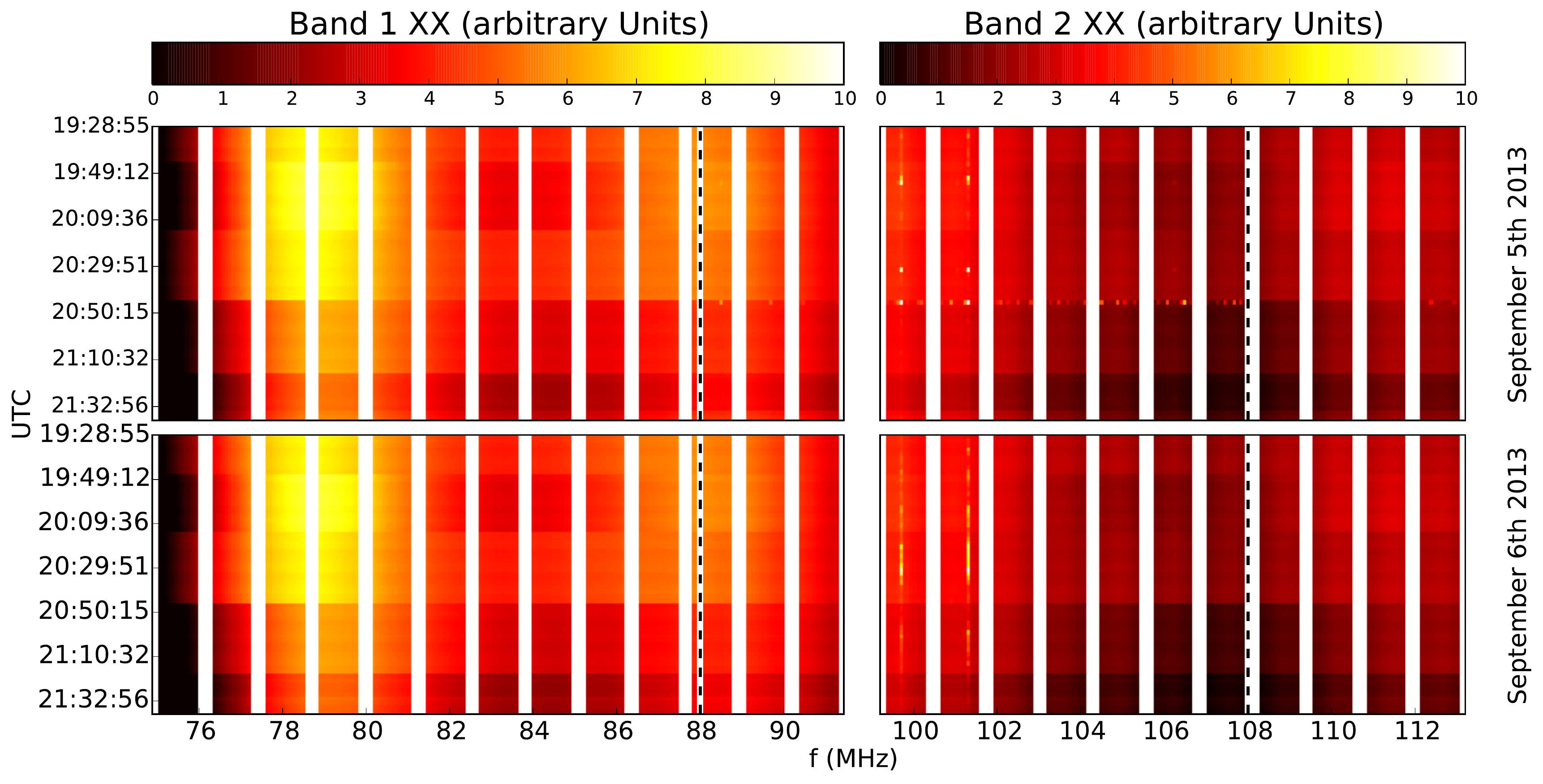}
\caption{Dynamic spectra of the square root of a representative tile autocorrelation. Note the different color bars for the two frequency bands since Band 1 evolves more steeply in frequency than Band 2. The autocorrelations exhibit repetitive structure in time from night to night with smooth time variations occuring as the sky rotates overhead and steep transitions occuring every $\approx 30$\,minutes due to changes in the analogue beamformer settings as the antennas track the sky. Limited RFI is plainly visible within the FM band, especially in Band 2, and the events are consistent with the flagging events identified by {\tt cotter} shown in Fig.~\ref{fig:cotter}.}
\label{fig:autoWaterfall}
\end{figure*}

Autocorrelations are particularly susceptible to RFI and potential contamination due to cross talk and other artifacts. In Fig.~\ref{fig:autoCompare}, we saw that after flagging the channel edges, the spectral structure in the autocorrelations was consistent with our calibration solutions up to a smooth polynomial factor. In Fig.~\ref{fig:autoWaterfall}, we inspect for artifacts and RFI in a typical tile autocorrelation as a function of time. We see that RFI is present at similar times that were flagged in autocal (Fig.~\ref{fig:cotter}). We also see that the time evolution of each autocorrelation is consistent between the two nights with rapid $10\%$ transitions occuring at $\approx 30$\,minute intervals when the analogue beamformer settings are changed to track the sky. Ripples in frequency are also visible, corresponding to the structure in the standing wave reflections. It is difficult to pick out small artifacts in this dynamic spectrum view unless more large-scale smooth structure is fitted out, as is done in our calibration procedure. The residuals after this fitting give a better picture of what fine spectral features exist in the dynamic spectra of the autocorrelations which we discuss in the next section.

\subsection{The time dependence of residual structure.}
We noted in \S~\ref{ssec:autocal} that our fits to reflections tended to have $\sim 10\,\%$ residuals. Since we rely on these fits to predict the reflections in our gain phases, we expect residuals of these fits that are also present in the phases to contribute reflection power at a similar level. Our residuals could arise from thermal noise in the autocorrelations and calibration solutions. If this were the case we might expect them to average down with time. On the other hand, these residuals might also arise from mismodeling of the reflections themselves and would not average down with time. The result would be a systematic floor which can only be overcome by finding the correct model of the reflections or removing them from the signal path. 

\begin{figure*}
\includegraphics[width=\textwidth]{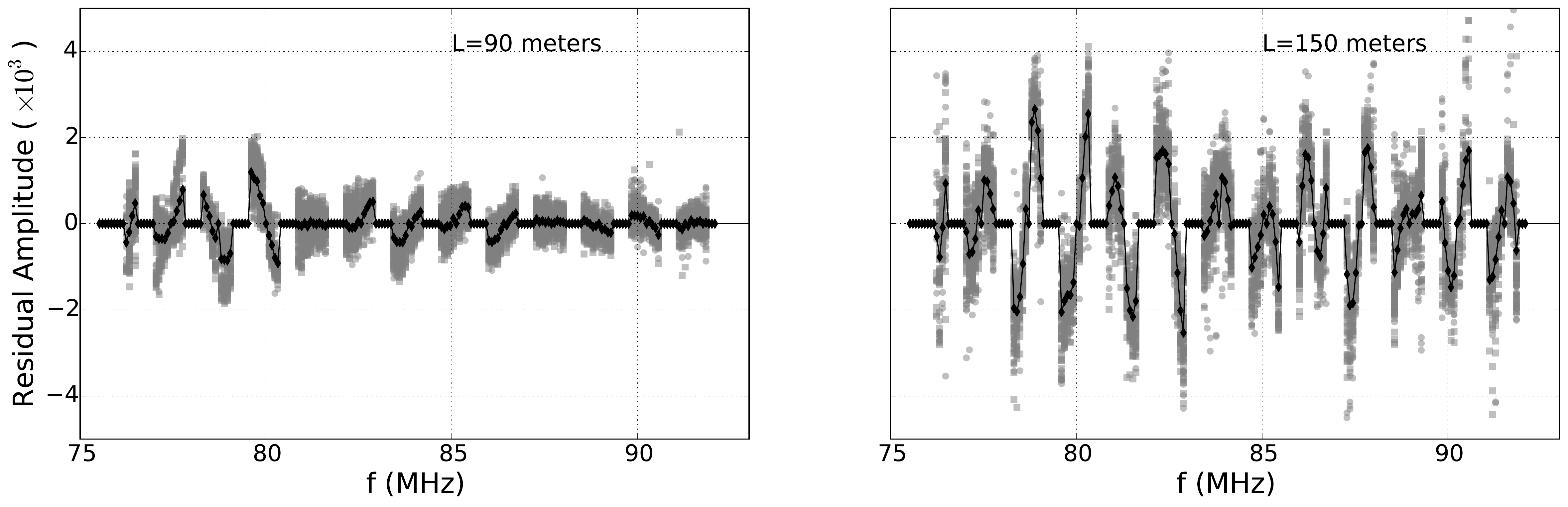}
\caption{Left: The residuals to fitting reflection functions in our autocorrelations for all two minute time-steps in our analysis for a representative tile with a 90\,meter beamformer to receiver connection (light grey points). While some scatter exists in the residuals due to fitting noise, they average to non-zero values on the order of $\sim 10^{-3}$ (black dots). These residuals are due to mismodeling the reflections and at a lower level potentially arise from digital artifacts. Right: The same as the left but for a 150\,meter cable whose reflection coefficient is $\sim \times 2$ as large as the 90\,meter cable, leading to larger residuals due to mismodeling.}
\label{fig:residTimeDependence}
\end{figure*}

Plotting the fit residuals for two representative 90\,m and 150\,m tiles over the low band (Fig.~\ref{fig:residTimeDependence}), we find them to be at the $\sim 10^{-3}$ level. While there is some scatter in these residuals due to fitting noise, their frequency dependent shape is relatively constant. As a consequence, the residuals average to a spectrum with frequency structure. These residuals are likely due to mismodeling of the frequency dependent amplitude, phase, and period of the reflections but at a lower level may have some contributions from digital artifacts and cross-talk present in the autocorrelations. Because the component in these residuals that is sourced by reflections is also present in the phases which we are trying to model, there remains an uncorrected component to the gains that we are not calibrating out and does not average down with time. 

While calibration with autocorrelations still appears to be limited by fine frequency artifacts arising from reflections, the high SNR of the reflections in the autocorrelations does offer significant improvement over fitting the reflections in the calibration solutions themselves. We provide a more quantitative look at the improvement achieve using calibration with auto-correlations in \S~\ref{ssec:calComparison}.

\section{Power Spectrum Results}\label{sec:results}

We can now present our power spectrum results and the first upper limits on the Epoch of X-ray heating power spectrum. We form cross power spectra of the even and odd timestep data cubes through the empirical covariance modeling technique developed in \citep{Dillon:2015b} (D15). 
In this procedure, the foreground residual model used in the inverse-covariance weighted quadratic power spectrum estimates and in the associated error statistics is derived from the data. It assumes that foreground residuals are correlated in frequency but uncorrelated in the $uv$ plane and depend only on frequency and $|\mathbf{u}|$. 
We refer the reader to D15 and its predecessors \citep{Tegmark:1997a,Liu:2011,Dillon:2013,Dillon:2014} for a thorough discussion of how this technique works. Along with estimates of the power spectrum amplitude, our pipeline outputs error bars and window functions which describe the mixing of the true power spectrum values into each estimate. We form 1d power spectra by binning our 2d power spectra using the optimal estimator formalism of \citet{Dillon:2014} with the weights of all modes lying outside of the EoR window or with $k_\|$ values showing consistent cable reflection contamination set to zero (D15).

First we will examine our two dimensional power spectra for Bands 1 and 2, derived from $\approx 15$\,MHz of bandwidth each, and comment on systematics (\S \ref{ssec:ps2d}) and how well our calibration techniques mitigate them (\S~\ref{ssec:calComparison}). We finish by presenting our spherically binned 1d power spectra, our most sensitive data product. We use our 1d power spectra to compare foreground contamination from ionospheric systematics on both nights (\S~\ref{ssec:ionCompare}) and determine our best upper limits (\S~\ref{ssec:upper}).

\subsection{Systematics in the 2d Power Spectrum.}\label{ssec:ps2d}

\begin{figure*}
\includegraphics[width=.48 \textwidth]{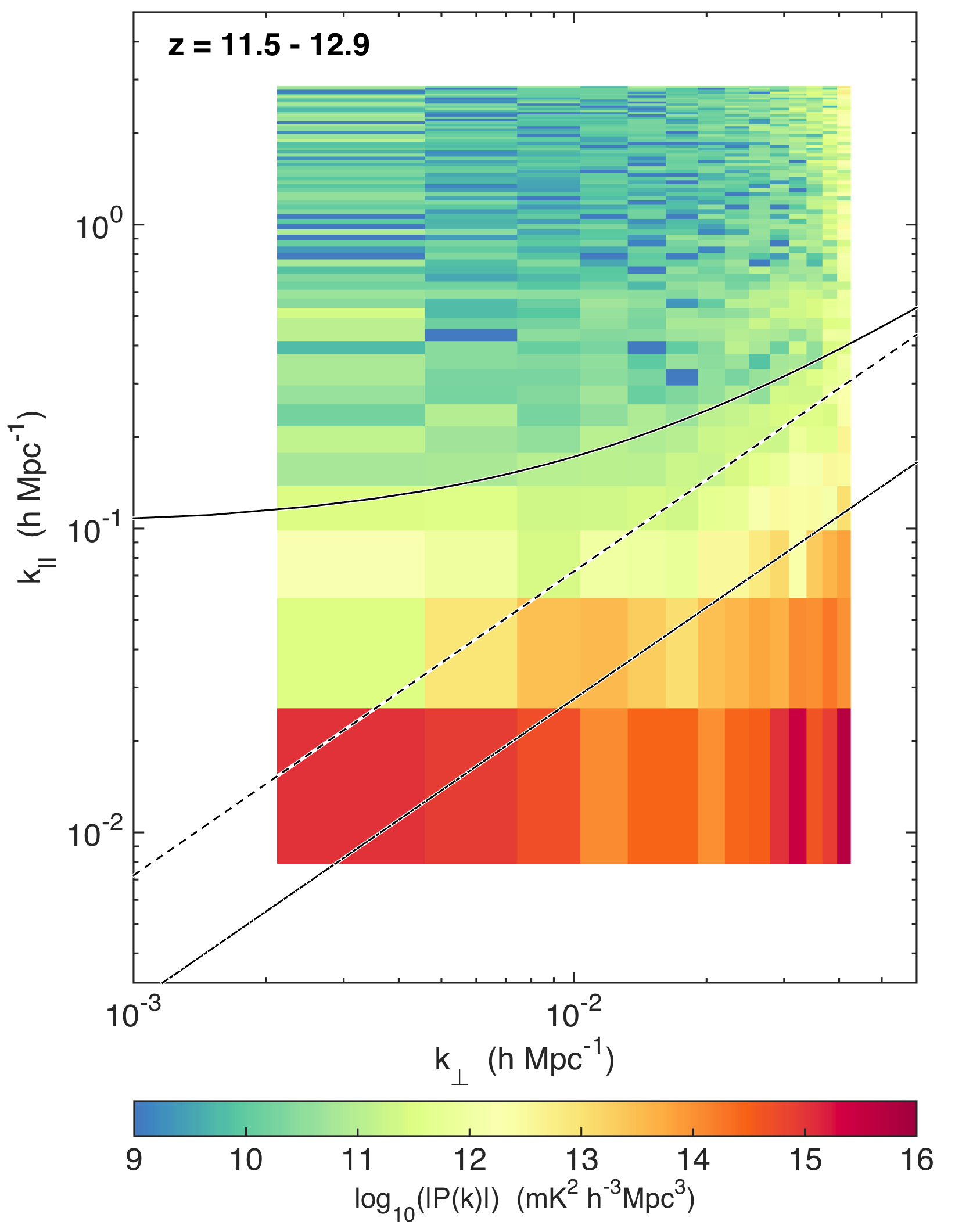}
\includegraphics[width=.48 \textwidth]{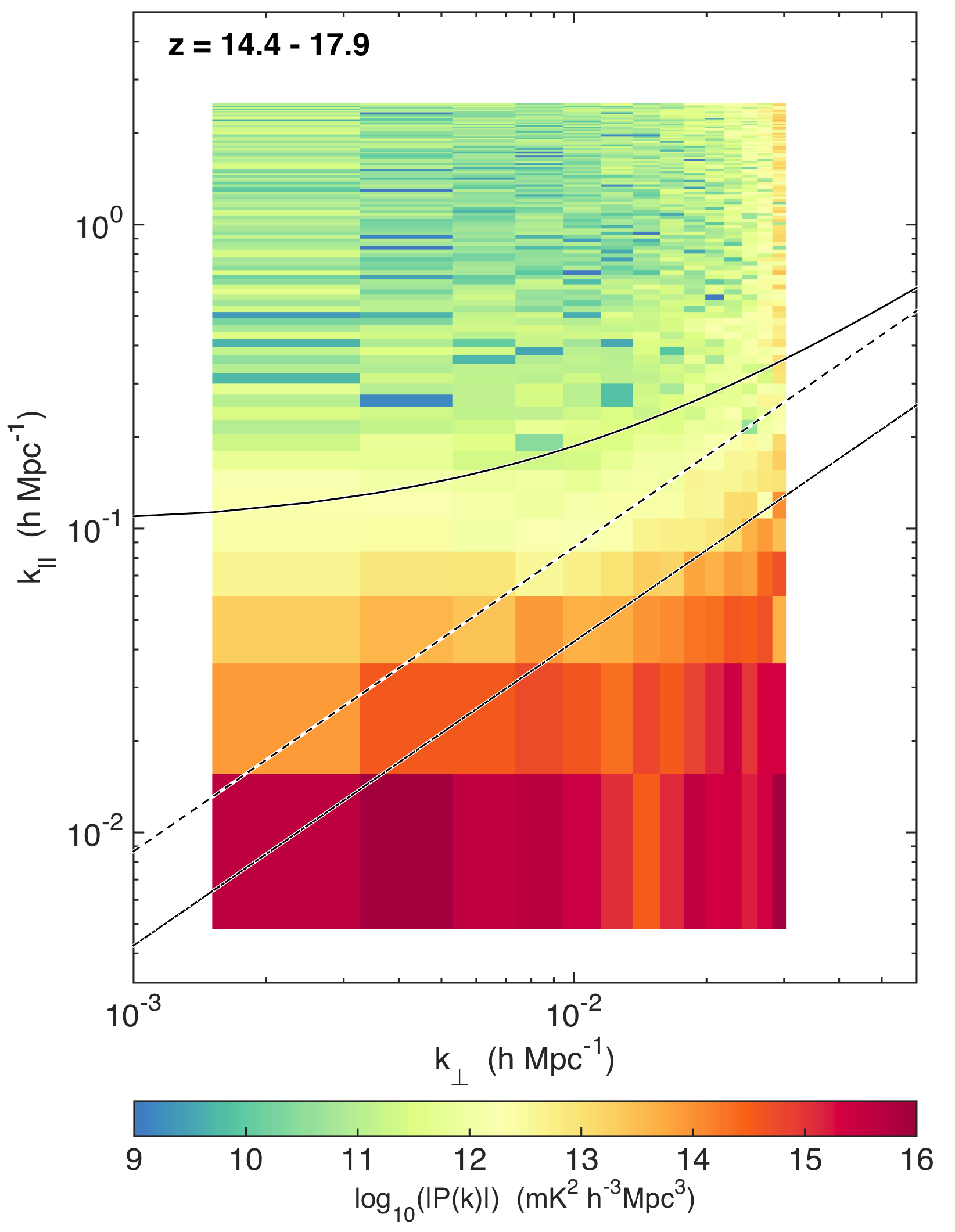}
\caption{The absolute value of our cylindrical power spectrum estimate from our three nights of observing on Band 2 (left) and Band 1 (right). We overplot the locations of the primary beam (dash-dotted), horizon (dashed), and horizon plus a 0.1\,$h$\,Mpc$^{-1}$ buffer (solid black) wedges. We see that the foregrounds are primarily contained within the wedge and that the EoR window is, for the most part, noise-like. There is some low SNR structure below $k_\| \approx 0.5$\,$h$\,Mpc$^{-1}$, corresponding to $k_\|$ modes  contaminated by cable reflections. The amplitude in power rises very quickly due to an increase in thermal noise which rises very quickly at large $k_\|$ due to a rapid falloff in $uv$ coverage beyond $k_\perp \sim 0.2$\,$h$Mpc$^{-1}$.}
\label{fig:pK2d}
\end{figure*}

The absolute values of our two dimensional power spectrum estimates using data calibrated with auto-correlations are shown in Fig.~\ref{fig:pK2d} for both bands. The distinctive ``wedge" confines the majority of our foreground power with some supra-horizon emission clearly present out to $\sim 0.1$\,$h$\,Mpc$^{-1}$ as was found in observations of foreground contamination with a similarly large PAPER primary beam \citep{Pober:2013a}. Smooth frequency calibration errors, arising from foreground mismodeling, may also contribute to the supra-horizon emission along with intrinsic chromaticity in the primary beam itself. As we expected, the level of foregrounds and thermal noise is noticeably higher in our measurement of Band 1. We note that at the edge of our $k_\perp$ range, there is a significant increase in power which is due to a rapid increase in thermal noise from the drop-off in the $uv$-coverage of our instrument. Though somewhat hard to see by eye, there are signs of coherent non-noise-like structures in both bands below $k_\| \approx 0.5$\,$h$\,Mpc$^{-1}$.

\begin{figure*}
\includegraphics[width=.48 \textwidth]{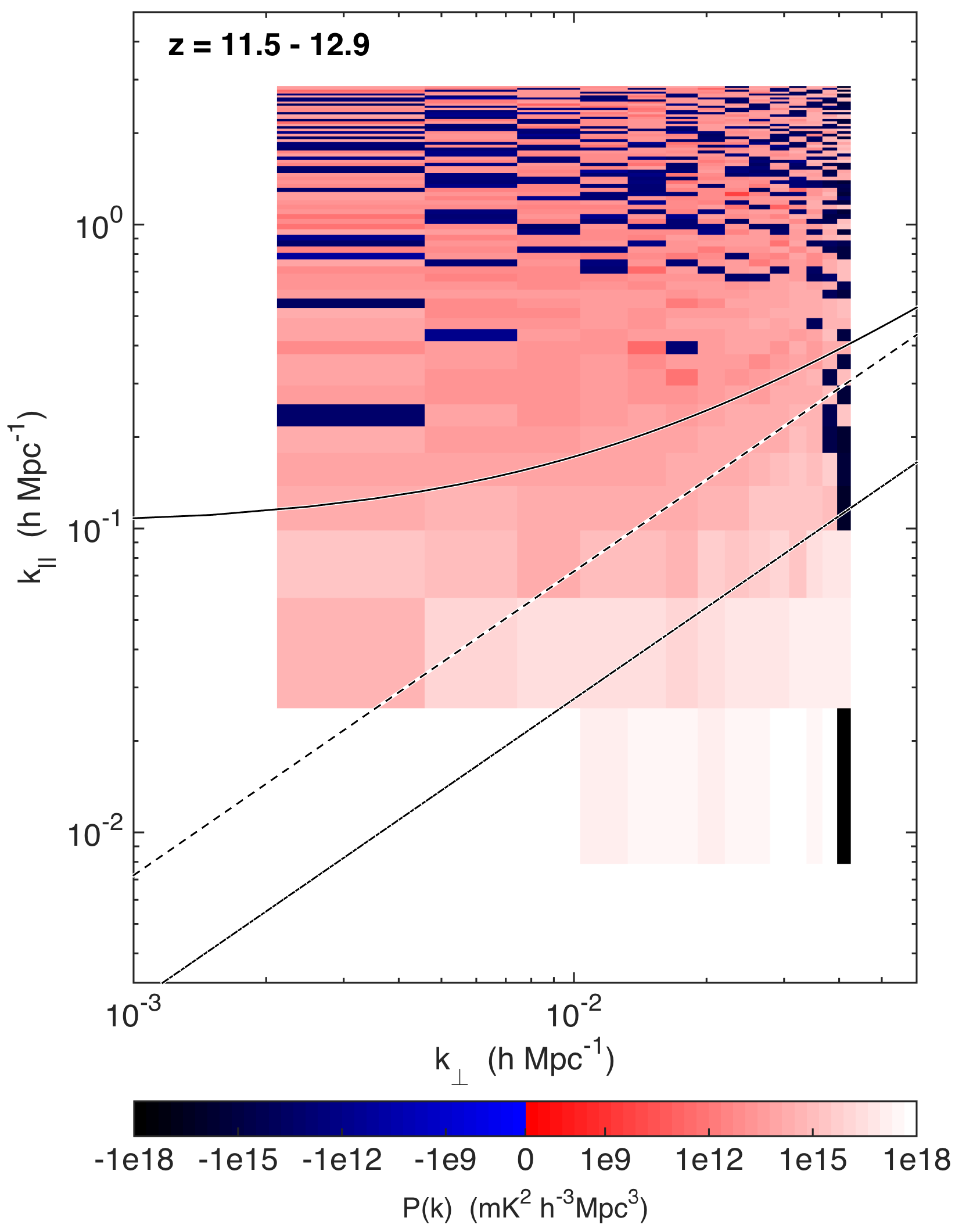}
\includegraphics[width=.48 \textwidth]{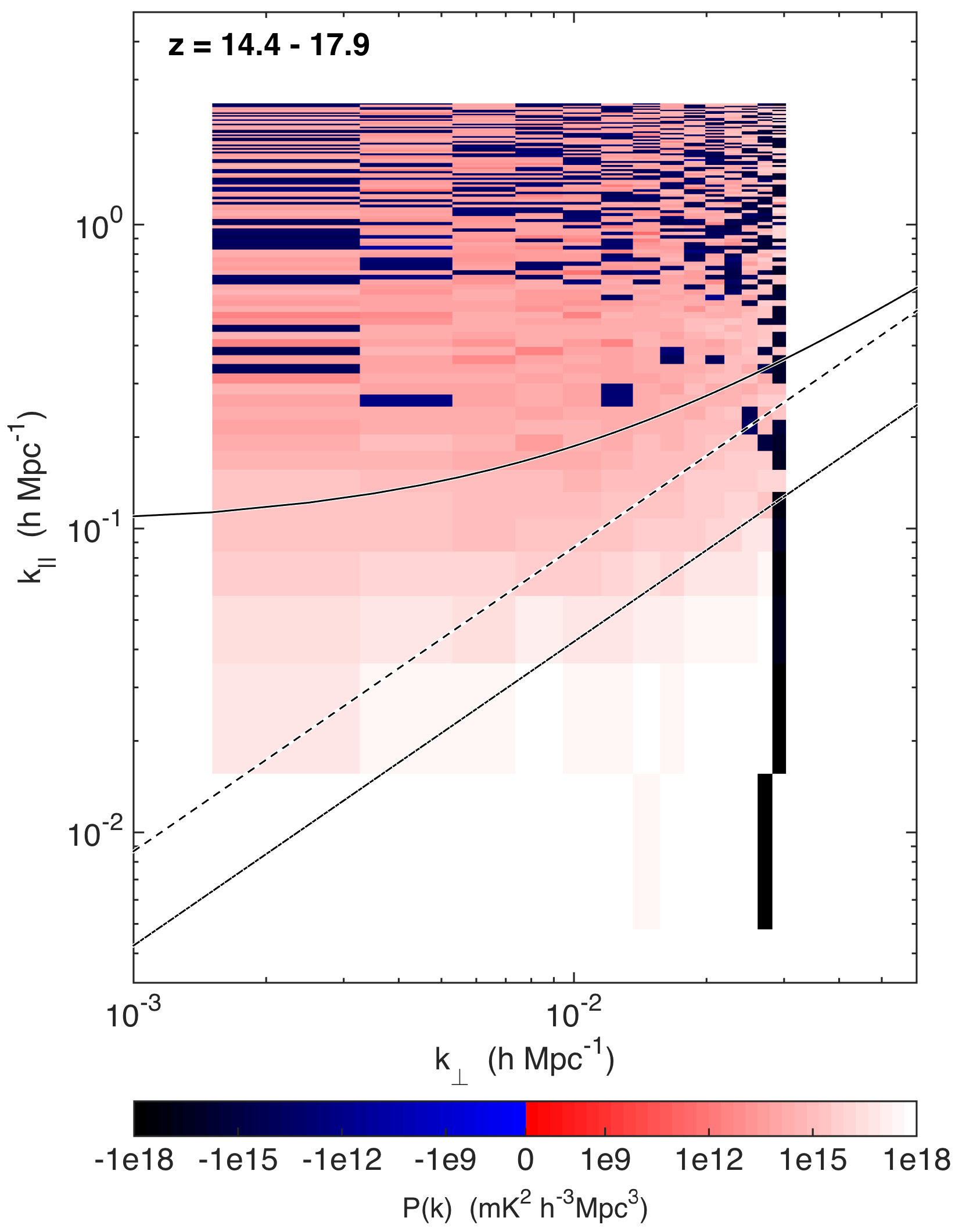}
\caption{$P(k)$ over Band 2 (left) and Band 1 (right)  with a color scale that highlights cells with positive or negative values. We expect regions that are thermal noise dominated to contain an equal number of positive and negative estimates and regions that are dominated by foreground leakage to be entirely positive. We observe significant foreground contamination outside of the wedge up to $k_\| \approx 0.5$\,$h$\,Mpc$^{-1}$ in both bands.}
\label{fig:pSpecSign}
\end{figure*}

We confirm these faint $k_\| \lesssim 0.5$\,$h$\,Mpc$^{-1}$ structures as systematic contamination by inspecting the sign of our power spectrum estimate over the $k_\perp$-$k_\|$ plane. While the expected value of the even/odd cross power spectrum is always positive, $k$-bins that are dominated by noise have an equal probability of being positive or negative. Regions in which band powers are predominately positive are detections of foregrounds or systematics. In Fig.~\ref{fig:pSpecSign} we show $P(k)$ from data calibrated with auto-correlations for both bands with an inverse hyperbolic sine color scale to highlight regions of k-space that have positive or negative values. It is clear that the region of with $k_\| \lesssim 0.5$\,$h$\,Mpc$^{-1}$ is not well described by thermal noise.

\begin{figure*}
\includegraphics[width=.48 \textwidth]{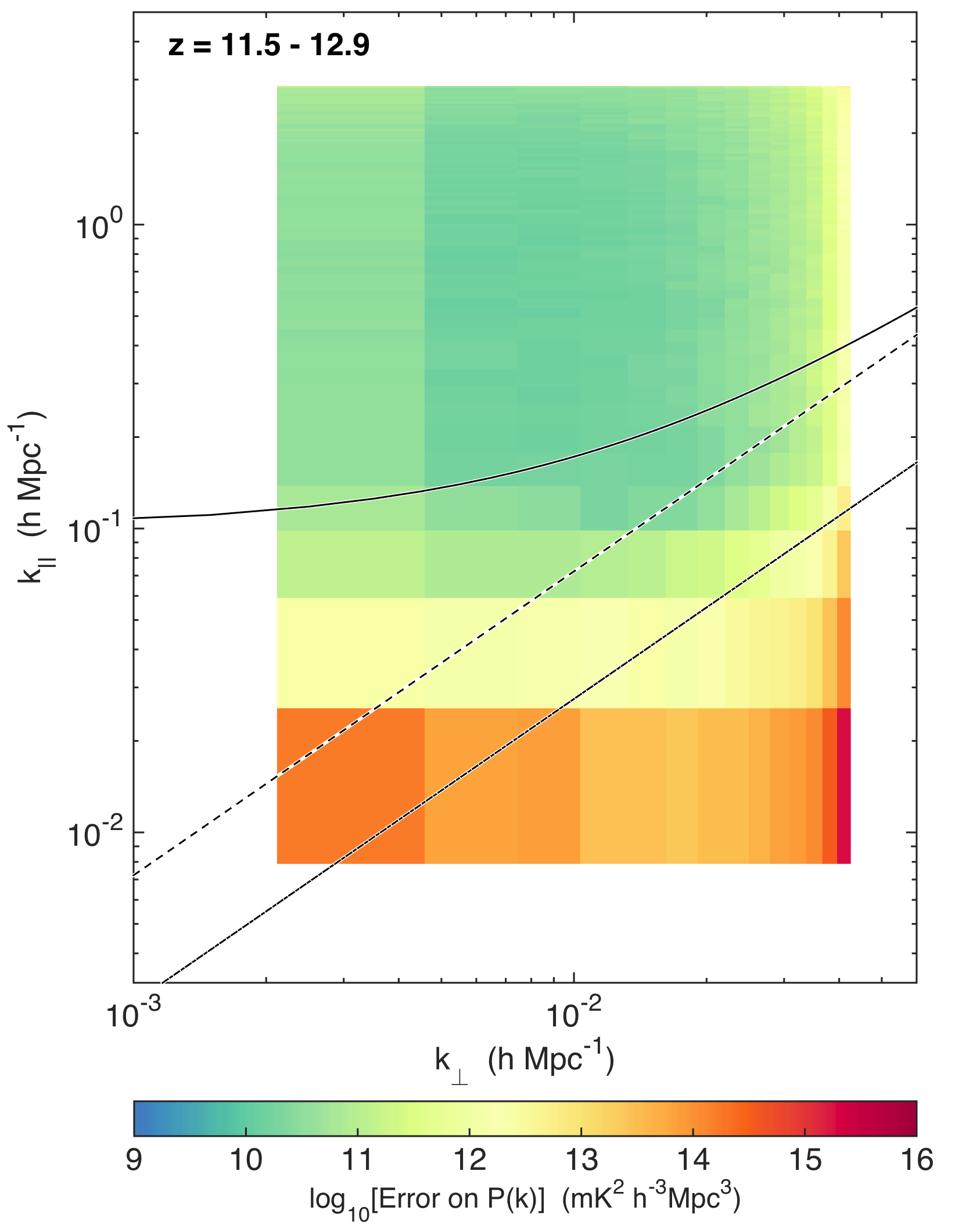}
\includegraphics[width=.48 \textwidth]{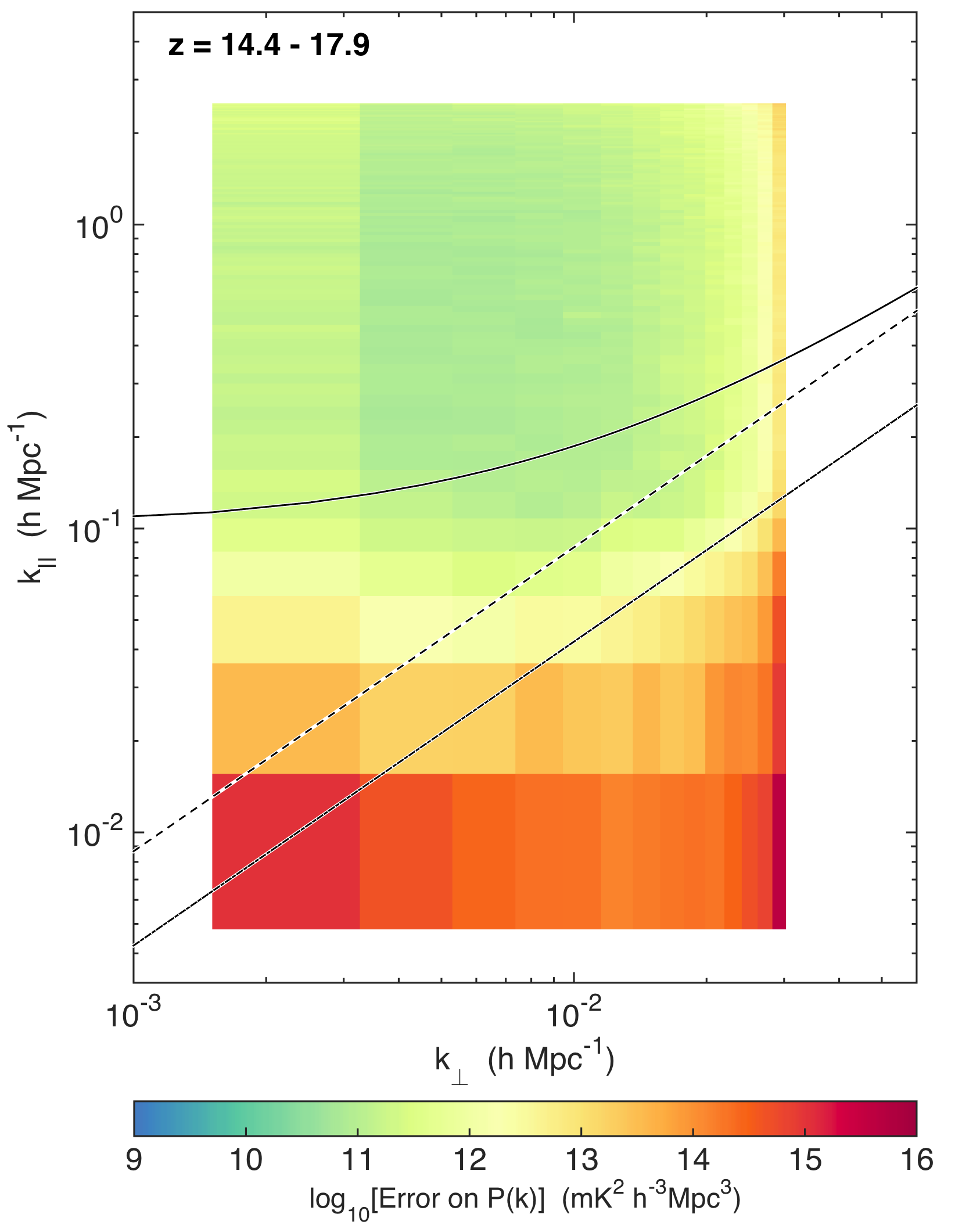}
\caption{The errors on $\boldsymbol{\widehat{p}}$ arising from residual foregrounds and thermal noise are determined by looking at even/odd difference cubes and foreground-subtracted residual cubes using the method of \citet{Dillon:2015b}. We show the error bars on our cylindrical power spectrum here, seeing that errors arising from foregrounds are contained within the wedge. These foreground errors are maximized at the smallest  and largest $k_\perp$ arising from large power in diffuse emission and increasing thermal noise from a dropoff in baseline density respectively. }
\label{fig:2dErrors}
\end{figure*}

\begin{figure*}
\includegraphics[width=.48 \textwidth]{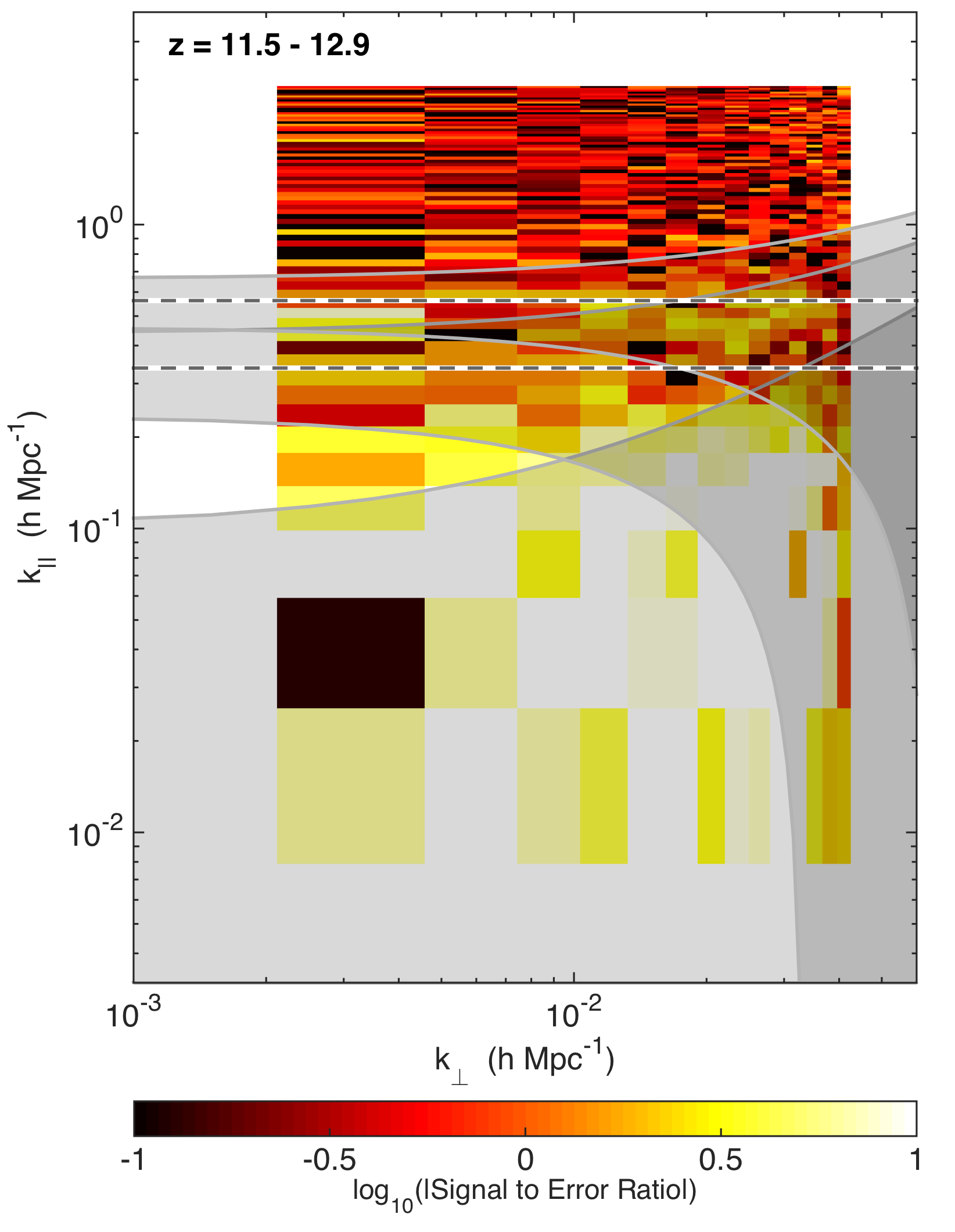}
\includegraphics[width=.48 \textwidth]{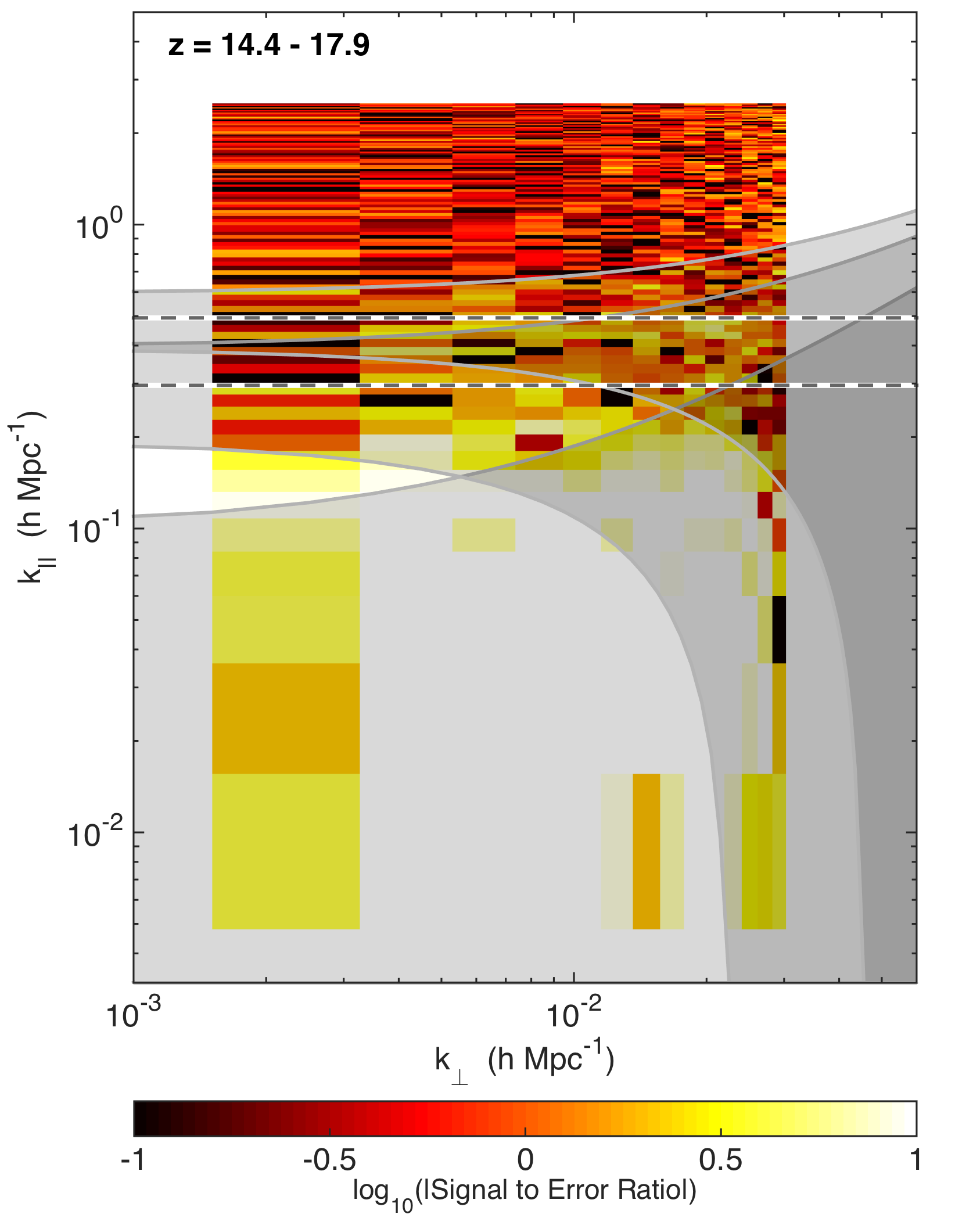}
\caption{The foreground contamination within the wedge along with residual detections due to miscalibrated fine frequency features in the bandpass are especially clear in plots of the ratio between power and the error bars estimated by the empirical covariance method of D15. We overplot the wedge with a $0.1$\,$h$\,Mpc$^{-1}$ buffer along with the wedge translated to cable reflection delays of our $90$ and $150$\,m receiver to beamformer cables to highlight the effect of this systematic.}
\label{fig:2dSNR}
\end{figure*}

Detections of foregrounds and systematics are especially visible in the ratio between the power spectrum and error bars predicted by our empirical covariance method (Fig.~\ref{fig:2dErrors}). In Fig.~\ref{fig:2dSNR} we observe excess power at the $\sim 2 \sigma$ level. While this is not a significant excess on a per cell basis, we detect this same power at high significance when we average in bins of constant $k\equiv \sqrt{k_\perp + k_\|}$.

Because this excess power is present at similar levels over both of our observing subbands, one of which has a significantly greater overlap with the FM, we cannot attribute this excess to RFI. In our 1d power spectra we also find that excess power is detected in our highest redshift bin which is outside of the FM entirely (Fig.~\ref{fig:1dPower}, right panel). The best explanation we have for this leakage is the residual structure in the MWA's bandpass caused by standing wave reflections on the beamformer to receiver cables. To demonstrate the plausibility of this explanation, we overlay the wedge translated to the $k_\|$ modes corresponding to the delays of our 90 and 150 meter cables. For clarity, we do not show the 230 meter cable reflections in this overlay since their amplitudes and the number of tiles affected is comparatively small. We also observe this reflection in the 1d power spectrum (Fig.~\ref{fig:1dPower}) which has higher signal to noise. We find that the region where one might expect contamination from a cable reflection is in good agreement with the observed excess power.

\subsection{Comparing Calibration Techniques}\label{ssec:calComparison}

\begin{figure*}
\includegraphics[width=\textwidth]{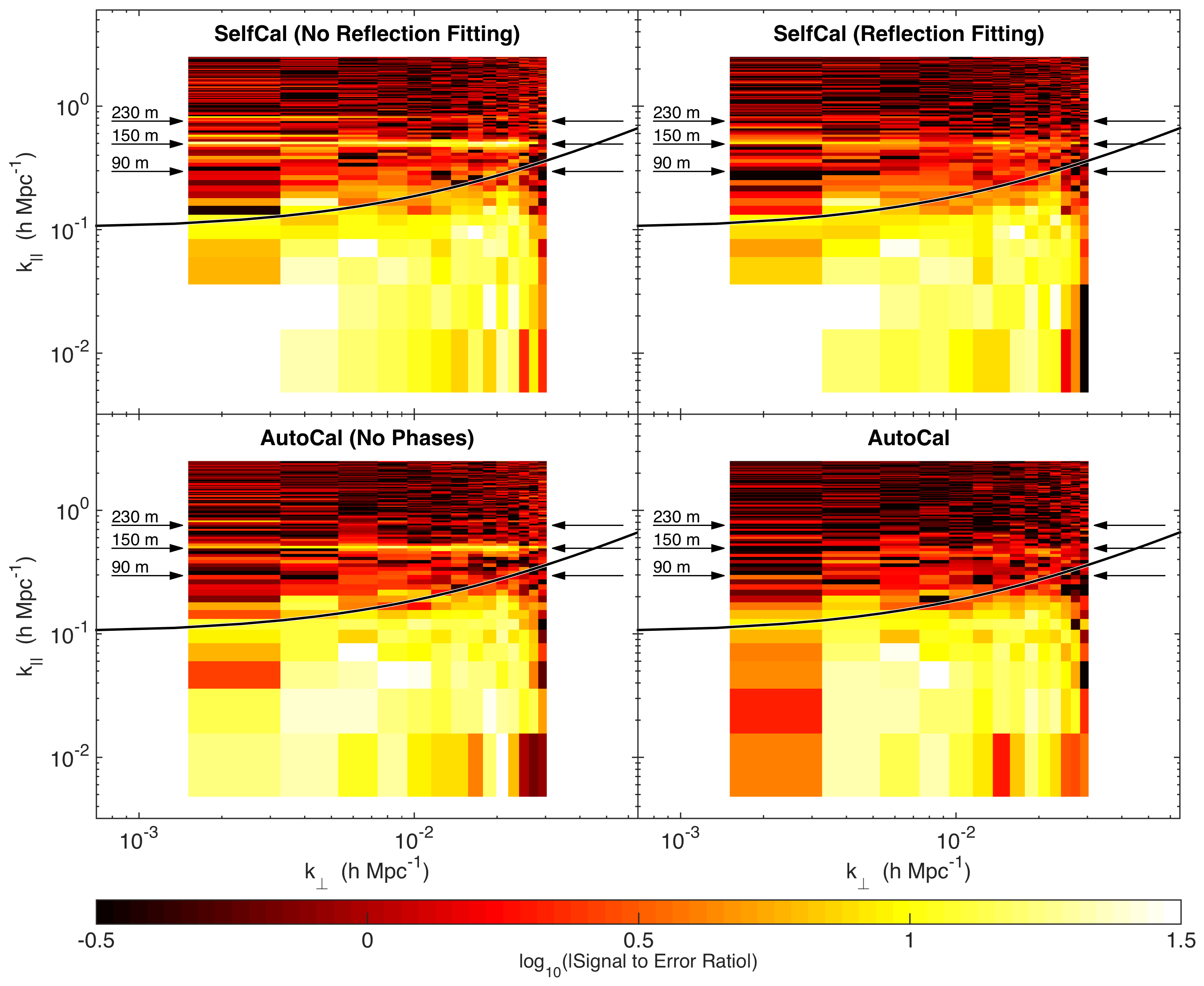}
\caption{Here we show the ratio between our 2D power spectrum and the error bars estimated by the emprical covariance method of \citet{Dillon:2015b}. On the top left, we show our data calibrated using our initial calibration (see \S~\ref{ssec:initCal}) with no attempt made to correct for standing wave structure in the MWA bandpass. Bright, band-like structures are clearly visible at the delays associated with reflections. On the top right, we show a first attempt to correct for cable reflections by fitting a sinusoidal model to rather noisy calibration solutions that had been integrated over a night of observing (1.5 hours each night). While the bands appear weaker, they are still quiete visible above the noise. In the bottom right panel, we show the same plot with calibration solutions using scaled autocorrelations described in \S~\ref{ssec:autocal}. In the lower left panel we show a power spectrum with calibration solutions using autocorrelations for the amplitudes but without any attempt to correct reflections in the phase solutions. Pronounced reflection features are visible in this power spectrum, indicating that any mismodeled reflection structure in the phases will contaminate our measurement.}
\label{fig:calCompare}
\end{figure*}

Having formed 2d power spectra and estimates of the vertical error bars, we are in a position to asses the performance of our calibration solution in removing systematics. By inspecting the signal to error ratio in the EoR window, we compare our different calibration techniques.  In Fig.~\ref{fig:calCompare} we show the ratio of  $P(k)$, binned over annuli, to the error bars in Band 1 for the calibration techniques discussed in this work. For all calibration methods, the majority of foreground detections are contained within the wedge with a $\sim 0.1$\,$h$\,Mpc$^{-1}$ buffer, indicating that all perform at a similar level in removing smooth gain structure within the wedge.

We first inspect a power spectrum derived from data calibrated using the initial method described in \S~\ref{ssec:initCal} in which coarse band structure is removed by averaging over tiles, the per tile amplitudes and phases of each antennas are fit to smooth polynomials, and no attempt is made to model the beamformer-receiver reflections (top left corner). Significant foreground power is visible beyond the wedge to $k \sim 0.5$\,$h$\,Mpc$^{-1}$ and is especially bright at the delays corresponding to the $k_\|$ values of the cable reflections in Table~\ref{table:reflections}. The fact that the 150\,m delay dominates the others stems from the fact that most of our short baselines are formed from 150\,m cables and that the amplitude of the reflections in the 150\,m cables is larger compared to the 90 and 230\,m cable lengths (Fig.~\ref{fig:rHist}). We next show a first attempt to fit out the reflections by averaging all calibration amplitudes in a night, dividing out a polynomial, and fitting equation~\ref{eq:refAmp}. While the power in the bands is reduced significantly, residuals remain at the $2$-$10\sigma$ level, especially in the reflection bands. Since our initial calibration solutions are so noisy, it makes sense that they are difficult to fit.

We finally inspect results from calibrations derived from the autocorrelations described in \S \ref{ssec:autocal} (lower right). While there is significant reduction compared to the amplitude on the top right corner, there still exist residuals outside of the window at the $\sim 1-2 \sigma$ level. We think that these residuals arise from imperfect modeling of the reflection coefficients in the autocorrelation amplitudes, which will leave some reflection structure in the visibility phases. To demonstrate the impact of unmodeled reflection structure in the phases, we leave the phases of our auto-calibration solutions uncorrected for any fitted reflection coefficients (lower left) and find that significant power is reintroduced into the window. 

\begin{figure*}
\includegraphics[width=\textwidth]{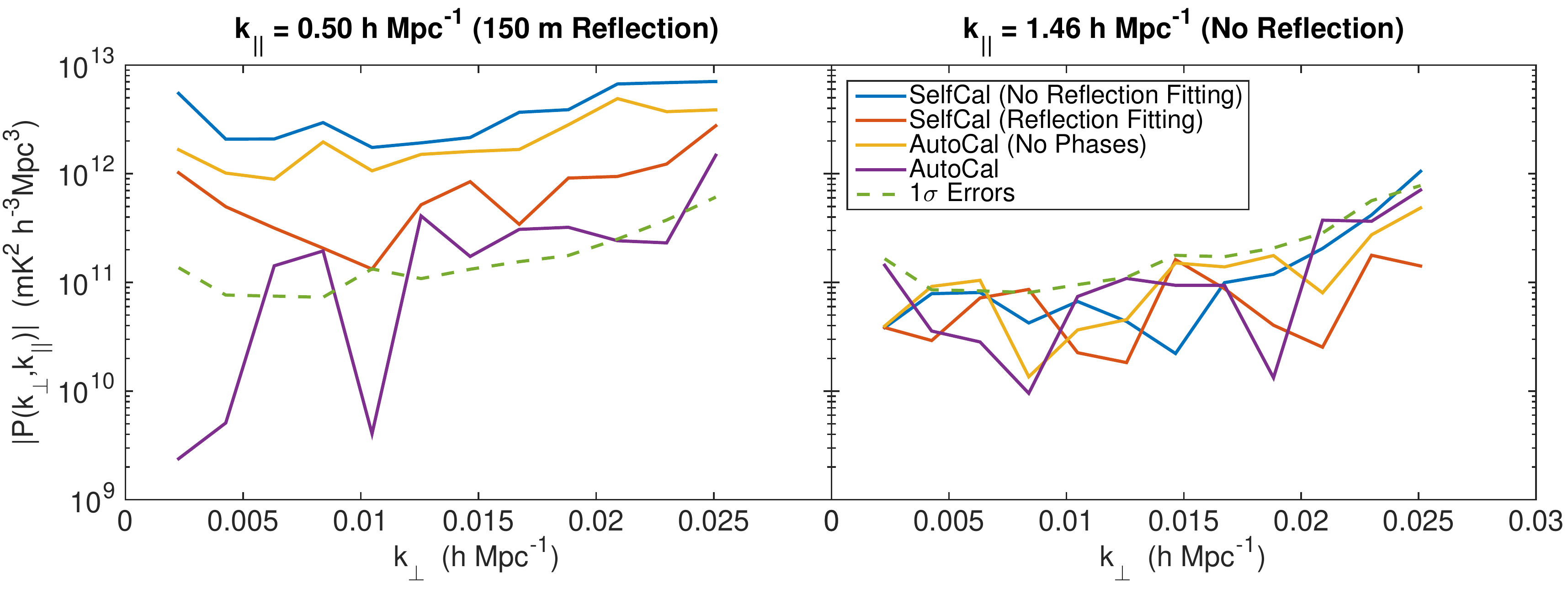}
\caption{The level of power at a fixed $k_\|$ corresponding to the delay of reflections from our 150\,m cable (left), and comparing it to a value of $k_\|$ unaffected by cable reflections (right). The blue line shows the power spectrum level for calibration in which the bandpass is modeled as a polynomial with no attempt to correct fine frequency scale reflections. We see that power is on the order of $\sim 50$ times the thermal noise level (green-dashed line). Attempting to fit the reflections to calibration solutions integrated over each night gives us an improvement in the power level by roughly an order of magnitude (orange solid line). Using calibration solutions derived from autocorrelations brings down the reflection power by another factor of a few (purple solid line) but is still unable to bring the majority of measurements below the $\sim 1 \sigma$ level. While we think that the autocorrelations accurately capture the fine frequency structure of the gains, we are still forced to model this fine frequency structure and predict it in the phases. Residual power is likely due to inaccuracies in this modeling. The right hand panel shows all data below the stimated noise level. This is due to the fact that in \citep{Dillon:2014} it is shown that the method for calculating error bars layed-out in \citet{Liu:2011,Dillon:2013} slightly over-estimates the noise.}
\label{fig:calCompare2}
\end{figure*}

We can get a more quantitative view of how much autocorrelations can improve calibration by taking a slice through the cylindrical power spectrum at the $k_\|$ of our 150\,m cable reflection (Fig.~\ref{fig:calCompare2}) where we see that fitting the calibration solutions was able to remove roughly an order of magnitude of the power in the reflection while AutoCal removes a factor of $\sim 20$. Since the power spectrum is proportional to the square of the visibilities which are primarily contaminated by first order reflection contributions, this corresponds to an accuracy of $\approx$ 20$\%$ in removing the reflections in the visibilities and is consistent with the residuals observed in Fig.~\ref{fig:residTimeDependence}. Such inaccuracy likely arises from our inability to model the precise frequency dependence of the reflection parameters in the phases and is on a similar order to the residuals observed in Fig.~\ref{fig:autoResid}. Since the reflections are removed to this accuracy in the visibility, we can briefly comment on how the relative contribution of second order reflections (which are below our noise floor even without any calibration). Since the second order reflections appear in the data at the $\widetilde{r}^{4}$ level and we have reduce their amplitude in the data from $\lesssim 0.01$ to $\lesssim 0.003$, they will enter the power spectrum at the level of $\lesssim 10\times10^{-1}$ the level of the 21\,cm signal.

We attempted to better model the reflections by allowing for frequency evolution of the amplitudes but found little improvement in the power spectrum. We also found that we are able to obtain better fits of the autocorrelations by adding additional smooth reflections terms to equation \ref{eq:smooth} which could be important if unmodeled large scale structures bias our fits of small scale ones. However, using more complicated fits of the large scale structure, we did not observe significant improvement in power spectrum contamination. The solutions that we ultimately settled on in this analysis allow for a power law evolution of the reflection amplitude and add an additional small delay reflection term to equation \ref{eq:smooth} which lies well within the wedge. In most of our autocorrelation fits, residuals remained at the $\sim 10\%$ level, some of which may arise from secondary reflections in bent or kinked cables. While these residuals were clearly present at high SNR in the autocorrelations, we have not found a way to sufficiently model the contribution of these low level structures to our phases.

	While using the autocorrelations has allowed us to characterize and subtract the fine spectral structure in the instrument better, it may not be a viable long term solution, even in the regions of the EoR window that currently appear foreground free. RFI contamination and digital artifacts are known to contaminate autocorrelations and likely exist below our current noise level.

\subsection{Power Spectra Comparison Between Nights of Varying Ionospheric Activity}\label{ssec:ionCompare}

\begin{figure}
\includegraphics[width=.48 \textwidth]{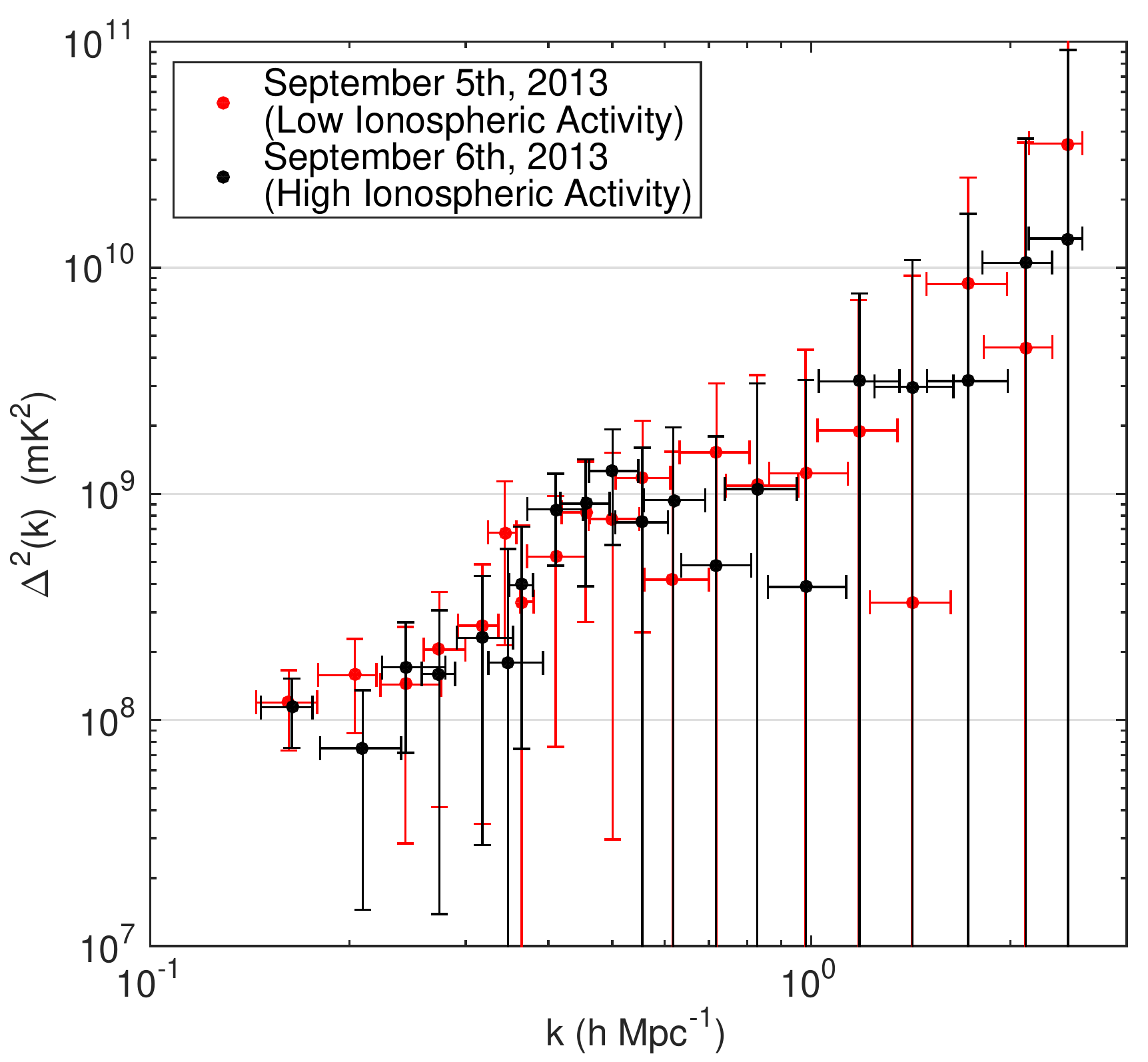}
\caption{The Band 1, 1d power spectra from our two nights of observing: September 5th, 2013 (black) and September 6th, 2013 (red). We saw in Fig.~\ref{fig:refraction}, that the magnitude of refractions on September 6th were on average twice as severe. The two power spectra nearly indistinguishable (within error bars) despite the significant differences in conditions, indicating that ionospheric systematics do not have a significant effect after three hours of integration, even at these low frequencies.}
\label{fig:ionCompare}
\end{figure}

An open question is whether or not the ionosphere will significantly hamper measurements of the power spectrum. The fact that the severity of ionospheric effects increase with $\lambda^2$ makes the question especially pertinent at low frequency. Changes in foreground emission induced by ionospheric effects can enter the power spectrum in two ways: through calibration and through the foreground residuals themselves. We check whether either of these potential error sources have an observable effect on our 1d power spectrum in Fig.~\ref{fig:ionCompare} by comparing power spectra derived from $1.4$ hours of Band 1 data on September 5th, over which ionospheric activity was comparatively mild to the same number of hours of Band 1 data on September 6th where differential refraction was approximately twice as severe. 

We find that the power spectra, which are estimated from data outside of the wedge, are consistent with each other. This result confirms the intuitive idea that since ionospheric errors in the foreground model are spectrally smooth (evolving as $\sim \lambda^2$), they should be contained within the wedge. We also extended our 1d power spectrum estimation into the wedge to see whether the foreground detections appeared to be significantly different and find that they are not. This suggests that the random errors induced by the ionosphere average down with time. It is important to keep in mind that the spatial scales being probed in our analysis are relatively large, on the order of $\gtrsim 2.5^\circ$, while ionospheric refraction at these frequencies effects sub-arcminute scales. Hence the contamination that we might expect from ionospheric refraction should be small. Amplitude scintillation effects are prominant on short baselines (V15a) and likely dominate any contamination, however their spectral coherence still constrains them to be predominantly within the wedge (V15b). 

\subsection{First Upper Limits on the 21\,cm Power Spectrum During the Pre-Reionization Epoch}\label{ssec:upper}

\begin{figure*}
\includegraphics[width=\textwidth]{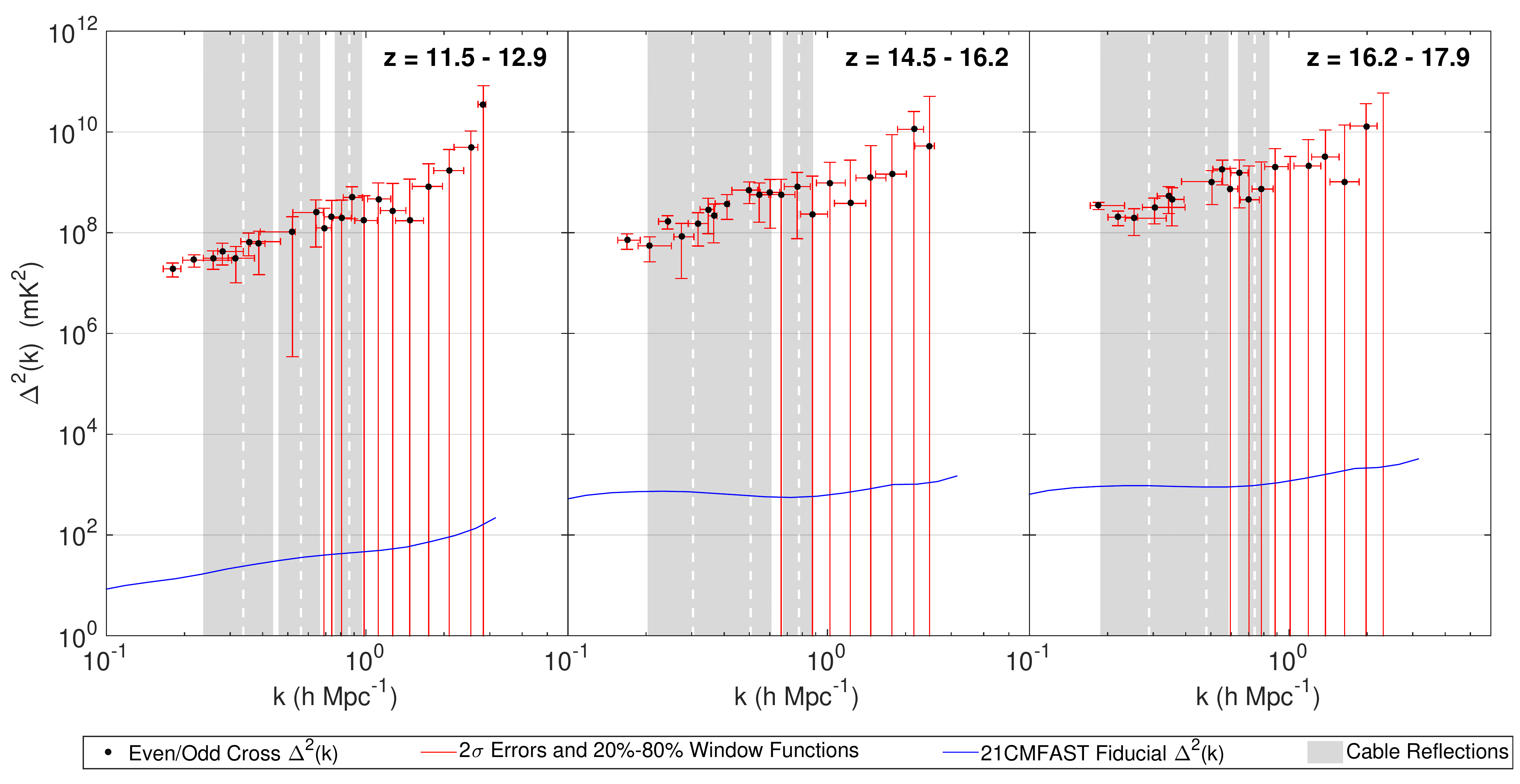}
\caption{Dimensionless 1d power spectra derived by Integrating spherical shells excluding the foreground contaminated wedge region with a $0.1$\,$h$\,Mpc$^{-1}$ buffer. Black dots indicate the mean estimated from the weighted average in each bin. Vertical error bars denote the $2\sigma$ uncertainties while horizontal error bars indicate the width of window functions. We also shade regions of k-space that we expect to have some level of foreground contamination due to uncalibrated cable reflection structure. Gray shaded regions clearly correspond to regions in which our power spectrum measurements are not consistent with thermal noise. We note that where our upper limits do agree with thermal noise, the power spectrum is on the order of $\sim 100$ times larger than the upper limits set with the MWA at $\approx 180$~MHz(D15). This factor is reasonable given that the sky noise (noise power spectrum) scales with $\sim f^{-2.6}$ ($f^{-5.2}$) and the primary beam solid angle increases as $\sim f^2$} 
\label{fig:1dPower}
\end{figure*}

We limit our 1d power spectra to redshift widths of $\Delta z \sim 1.5$ to minimize effects from cosmic evolution. A redshift interval of $\Delta z \sim 0.5$ is the range most cited in the literature over which the statistics of the brightness temperature field are expected to be stationary \citep{Mao:2008}. However, at higher redshift, the frequency range corresponding to $\Delta z \approx 0.5$ decreases as $(1+z)^{-2}$ with $\Delta z =0.5$ corresponding to a bandwidth of only $2.45$~MHz by $z=16$. Reducing our bandwidth to such a small interval leads to poor $k$ resolution which we prefer to maintain for assessing  systematics. Since we are far from a detection, we opt for a larger redshift interval than we would otherwise use if we were actually observing the cosmological signal. In Fig.~\ref{fig:1dPower} we show 1d power spectra derived from our three hours of observing. Vertical error bars give $2 \sigma$ uncertainties and the horizontal error bars give the width of our window functions. The amplitudes of our power spectrum values are consistent with thermal noise except for the regions of k-space below $k_\| \lesssim 0.5$\,$h$\,Mpc$^{-1}$. At $k\approx 1$\,$h$\,Mpc$^{-1}$, where our measurements are well described by thermal noise, our upper limits are on the order of $100$ times higher than the results presented in D15 in which a similar three hour upper limit was established at $\approx 180$~MHz. This factor of $\approx 100$ is consistent with what we expect from equation~\ref{eq:var}. The sky temperature increases with decreasing frequency as $f^{-2.6}$ leading to a factor of $\approx 30$ from $T_{sys}^2$ while $\lambda^4/A_e^2$ introduces an additional factor of 4-10. 

The detections at small $k$ are many orders of magnitude larger than the expected cosmological signal from a {\tt 21cmFAST} simulation \citep{Mesinger:2011} (blue solid lines) so they cannot possibly originate from the redshifted HI emission. Instead, these detections are most likely the miscalibrated reflection structure observed in our 2D power spectra. We shade out regions of the $k$~axis in which we expect contamination given the reflections discussed above and find that they correspond to the same modes where detections are observed. These systematic detections occupy the regions of Fourier space where our interferometer has the greatest sensitivity to the cosmic signal.  Since we do not expect the systematics dominated regions to integrate down, a detection with the MWA in its current state using the techniques presented in this work would have to take place at $k \gtrsim 0.5$\,$h$\,Mpc$^{-1}$, requiring over $10^5$ hours of integration---a rather infeasible time scale. Thus, in order to probe the pre-reionization epoch, improvements in calibration and/or changes in the hardware of the MWA will have to be implemented. We note that at lower redshifts, the primary beam is smaller and the k-modes occupied by reflections are farther away from the sensitivity sweet spot, so it is less likely that this problem will prevent the MWA from detecting the EoR power spectrum. 

Our best upper limits fall within the region of Fourier space with systematic errors and, while we do not expect them to integrate down with more observing time, we can infer that $\Delta^2(k)$ is less than $2.5 \times 10^7$\,mK$^2$ at $k=0.18$\,$h$\,Mpc$^{-1}$ and $z=12.2$, $8.3 \times 10^7$\,mK$^2$ at $k=0.21$\,$h$\,Mpc$^{-1}$ and $z=15.35$, and $2.7 \times 10^8$\,mK$^2$ at  $k=0.22$\,$h$\,Mpc$^{-1}$ and $z=17.05$, all at 95\% confidence.

\subsection{The outlook for EoR Measurements on the MWA.}\label{ssec:EoR}
A pertinent question arising from our analysis is how much the observed reflections impact or limit observations with the MWA of the Epoch of Reionization power spectrum at higher frequencies. The answer depends significantly on the calibration and reduction approach and as of now, several different efforts using alternative calibration and reduction schemes are being undertaken \citep{Jacobs:2016}. The analyses in \citet{Dillon:2015b} and \citet{Beardsley:2016} are calibrated in a way similar to this work, employing limited calibration parameters to avoid detrimental modeling errors. After an integration time of $\approx 3$\,hours, \citet{Dillon:2015b} also observe the cable reflections above the thermal noise level, however the smallest $k$ mode occupied by the shortest 90\,m cable lies at $k \approx 0.4$\,$h$Mpc$^{-1}$ while the delay width is narrower due to the smaller primary beam, causing the wedge to occupy fewer $k_\parallel$ modes. As a result, there are regions of k-space below the shortest reflection that are still consistent with noise. Such miscalibrated structure should be highly detectable after $\approx 10-30$ hours of integration but the results of such an analysis are still forthcoming \citep{Beardsley:2016}. If the reflections can be corrected to the $\sim 10^{-3}$ level as was done in this analysis, the region below the first reflection should remain free of contamination from the beam-former to receiver reflections.

Additional calibration pipelines, which include far greater degrees of freedom, such as the Real Time System (RTS) \citep{Mitchell:2008,Ord:2010} and the reduction pipeline discussed in \citet{Offringa:2016} include direction dependent calibration, ionospheric phase fitting, and greater frequency resolution, are also being applied to MWA data sets. A recent upper limit at 180\,MHz derived from RTS calibrated data and the CHIPS power spectrum estimator did not show evidence of the cable reflections being present \citep{Trott:2016a}. It is likely that the enhanced degrees of freedom allowed by RTS calibration did a better job at removing the structure from reflections but it is difficult to tell given that the error bars due to thermal noise at the comoving scales relevant to these reflections are an order of magnitude larger than those in \citet{Dillon:2015b} because of the shorter integration time. Ultimately, the increase in the number of fitting parameters may enhance the removal of instrumental chromaticity in the EoR window, however simulations by \citet{Barry:2016} show that small errors in ones calibration model will introduce power into the window in excess of the 21\,cm signal unless the intrinsic bandpass is smooth enough to be modeled by a small number of parameters or a source model exists with an accuracy significantly beyond what is currently available. 
Ultimately, the increase in fitting parameters may enhance the removal of instrumental chromaticity in the EoR window. Whether they can be introduced without adding power into the EoR window in excess of the signal, due to small errors in source modeling is still an open question that is currently being investigated.

\section{Conclusions and Future Experimental Considerations}\label{sec:conclusion}

In this paper, we have presented low frequency radio observations with the MWA at unprecedentedly high redshifts between 11.6 and 17.9. Our goals in conducting these observations were to place upper limits on the 21\,cm power spectrum during the Epoch of X-ray heating and to assess the levels of systematics which are expected to be generally worse than at EoR frequencies. These systematics include ionospheric effects, RFI (due to the FM band) and increased thermal noise. We need to control these systematics if we are to learn the detailed properties of the sources that heated the IGM; be they the first generation of stellar mass black holes, the hot interstellar medium left over from the first supernovae explosions in the universe, or dark matter annihilation. 

With regards to RFI, we have found after three hours of integration that existing algorithms are sufficient to flag RFI below the FM  band. Within the FM band, we have found that only a handful of channels are contaminated continuously and that after discarding them our power spectra do not show any evidence of RFI contamination. This bodes well for future planned 21\,cm experiments at the MRO such as the SKA-low which is expected to make high signal to noise detections of the power spectrum \citep{Koopmans:2015}. However, we are still many orders of magnitude above the level of a detection and reducing the thermal noise through longer integrations may reveal lower level RFI.

Over two nights of observing, we encounter different ionospheric conditions, observed quantitatively using the differential refraction metric described in \citet{Cohen:2009}. We establish that ionospheric fluctuations are the source of observed position shifts by comparing the level of refraction in our two observing bands and find that they exhibit the expected $\lambda^2$ evolution. Diffractive scales on the second night of $\approx 5$\,km are a factor of two shorter than the first night, indicative of more severe ionospheric activity. When we compare the 1d power spectra derived from an equal amount of data on each night, we find that they are very similar to each other, lending support to the idea that since ionospheric effects on calibration and foreground residuals are spectrally smooth, they should not contaminate the EoR window.

While the majority of foreground power is contained within the wedge, we find high-significance foreground detections within the EoR window out to a $k_\| \lesssim 0.5$\,$h$\,Mpc$^{-1}$. These contaminated regions are consistent with  miscalibrated cable reflections. We are able to obtain an order of magnitude improvement on removing the worst of these features using fits to autocorrelations, however they still limit our sensitivity at the 2-5 $\sigma$ level. In addition, since auto-correlations are generally contaminated by RFI and digital artifacts, it is likely that in reducing the dominant obstacle in our data, we have introduced additional features that are below the noise level of this analysis. Since the reflections occupy the regions of $k$-space where we would otherwise expect the greatest cosmological sensitivity, our best upper limits are a factor of a few larger than the limits we would obtain if we were thermal noise limited. Cable reflections are especially pernicious at higher redshifts because the increasing primary beam width adds foreground power to delays ever closer to the horizon. While supra horizon emission off of the wedge moves up in $k_\|$, the modes occupied by cable reflections move down, increasing in width. The EoR window is crushed between the shortest reflection mode and the top of the wedge.

While our observations on the MWA will not integrate down below $\approx 10^8$\,mK$^2$ at $k\lesssim 0.5 h$Mpc$^{-1}$ and is limited by the intrinsic spectral structure of the instrument, the systematics encountered in this analysis do not prevent 21\,cm observations at high redshift in general. A robust source catalog, that includes emission all the way down to the horizon along with precise models of the primary beam will lead to less foreground power bleeding from the edge of the wedge, \citep{Thyagarajan:2015a,Thyagarajan:2015b,Pober:2016} and potentially open up a foreground free region under the first cable reflection. Resolving the question of cosmological signal loss and mixing of foreground spectral structure from large to short baselines may enable us to calibrate with more free parameters, better capturing the spectral structure of the bandpass. More robust calibration of these features may also be obtainable with a redundant array \citep{Wieringa:1992,Liu:2010,Zheng:2014}. The 128-tile MWA has very little redundancy by design, however an additional 128 tile expansion is expected to introduce two highly redundant, hex-packed, subarrays \citetext{Tingay, private commmunication}. The final plan for HERA, which is currently under construction, is dominated by 331 hexagonally packed dishes. Its layout is designed to take advantage of redundant calibration as well \citep{Pober:2014}. Finally, calibration using injected signals \citep{Patra:2015} can also be employed to make high precision measurements of the bandpass. 

The most sure way of eliminating reflection features is to remove them in hardware either by ensuring better impedance matching on the cable connections, changing the cable lengths to move reflections out of the window, or early digitization. The current HERA design employs cables no longer than 35\,m in length, translating to $k_\| = 0.09$\,$h$\,Mpc$^{-1}$ at $z=16$  and ensures that reflections within the dish are below an acceptable level \citep{EwallWice:2016a,Patra:2016,Thyagarajan:2016}, while the planned MWA phase III upgrade and the SKA are considering digitization at the beamformers \citetext{Tingay, private communication}, eliminating reflections altogether. 

While measurements of the 21\,cm line at EoR frequencies can teach us about the nature of UV photon sources and constrain cool thermal histories, a significant number of scenarios predict saturation of heating's contribution to brightness temperature fluctuations during reionization. In order to learn of the detailed properties of the sources that heated the IGM and to exploit the full potential of the 21\,cm line as a cosmological and astrophysical probe, we will invariably want to extend our search to as low a frequency as possible. In this work we have obtained a first look at the systematics facing us in this high redshift realm and have found that most of them are navigable. As of now, our primary limitation lies in the design of our instrument and calibration, both of which can be dramatically improved on relatively short time-scales. Ultimately, we expect a combination of improvements in instrumental design including shorter/no cables to keep reflections inside of the wedge and redundant baseline layouts allowing for more robust calibration to allow for much deeper integrations in the near future.

\section*{Acknowledgements}
We would like to thank the referees for their helpful comments. We extend our gratitude to Jeff Zhang and Adrian Liu for useful discussions. This work was supported by NSF Grants AST-0457585, AST-0821321, AST-1105835, AST-1410719, AST-1410484, AST-1411622, and AST-1440343, by the MIT School of Science, by the Marble Astrophysics Fund, and by generous donations from Jonathan Rothberg and an anonymous donor. AEW acknowledges support from the National Science Foundation Graduate Research Fellowship under Grant No. 1122374. AM acknowledges support from the European Research Council (ERC) under the European Union’s Horizon 2020 research and innovation program (grant agreement No 638809 - AIDA).

This scientific work makes use of the Murchison Radio-astronomy Observatory, operated by CSIRO. We acknowledge the Wajarri Yamatji people as the traditional owners of the Observatory site. Support for the MWA comes from the U.S. National Science Foundation (grants AST-0457585, PHY-0835713, CAREER-0847753, and AST-0908884), the Australian Research Council (LIEF grants LE0775621 and LE0882938), the U.S. Air Force Office of Scientific Research (grant FA9550-0510247), and the Centre for All-sky Astrophysics (an Australian Research Council Centre of Excellence funded by grant CE110001020). Support is also provided by the Smithsonian Astrophysical Observatory, the Raman Research Institute, the Australian National University, and the Victoria University of Wellington (via grant MED-E1799 from the New Zealand Ministry of Economic Development and an IBM Shared University Research Grant). The Australian Federal government provides additional support via the Commonwealth Scientific and Industrial Research Organisation (CSIRO), National Collaborative Research Infrastructure Strategy, Education Investment Fund, and the Australia India Strategic Research Fund, and Astronomy Australia Limited, under contract to Curtin University. We acknowledge the iVEC Petabyte Data Store, the Initiative in Innovative Computing and the CUDA Center for Excellence sponsored by NVIDIA at Harvard University, and the International Centre for Radio Astronomy Research (ICRAR), a Joint Venture of Curtin University and The University of Western Australia, funded by the Western Australian State government.

\bibliography{ms}

\appendix
\section{The Effect of Cable Reflections on Tile Gains}\label{app:reflections}
Throughout this work, we reference several expressions dealing with standing waves on cables that arise from mismatched impedances at their connections. In this section we derive these expressions for the reader's convenience. Discussions of this problem can be found in most elementary electricity and magnetism texts. 

An voltage signal, $A(x,t)$ incident on the end of a transmission line with impedance $Z_0$ and length $L$ that is terminated by some resistance $R_L$ will be partially reflected $B(x,t)$ and transmitted $C(x,t)$. The amplitudes of the reflected and transmitted components can be found by enforcing continuity in the voltage across the connection and are given by\begin{equation}
B(L,t) = \frac{Z_0-R_L}{Z_0+R_L}A(L,t)  \equiv \widetilde{R} A(L,t)
\end{equation}
\begin{equation}
C(L,t)= \frac{2 Z_0}{Z_0 + R_L} A(L,T) \equiv \widetilde{T} A(L,t)
\end{equation}
The impedance of a length $L$ coaxial line is given by
\begin{equation} \label{eq:imp}
Z_0 = R_0 + i\left( 2 \pi  f \ell_0 L - \frac{1}{2 \pi f c_0 L} \right) 
\end{equation}
where $c_0$ is the capacitance per unit length and $\ell_0$ is the inductance per unit length. A ubiquitous undergraduate electricity and magnetism exercise involves finding these quantities for a coaxial cable filled with a dielectric of permittivity $\epsilon$ and permeability $\mu$ \citep{Griffiths:2013}, yielding
\begin{equation}
c_0 = \frac{2 \pi \epsilon}{\ln \frac{d_o}{d_i}}
\end{equation}
and
\begin{equation}
\ell_0 = \frac{\mu}{2 \pi} \ln \frac{d_o}{d_i}.
\end{equation}
Here $d_i$ is the radius of the inner wire of the coaxial cable and $d_o$ is the radius of the outer shell. It is clear from equation \ref{eq:imp} that the reflection coefficients are dependent on frequency in a way that is influenced by the cable geometry and dielectric properties. 

Now we consider the coaxial cable terminated on both ends with reflection coefficients $\widetilde{R}_0$ and $\widetilde{R}_L$. A monochromatic voltage signal with frequency $f$ entering the cable at $x=0$ with amplitude $s(f)$ will travel to the end of the cable $(x=L)$ where part of it will be transmitted and the other part reflected. The complex amplitude of the transmitted component is $\widetilde{T}_L(f) s(f)e^{ \pi i \tau f }$  while the reflected component has complex amplitude $\widetilde{R_L} s(f) e^{\pi i \tau f}$, where $\tau$ is the time it takes for the signal to propagate down the length of the cable and back. The reflected component will travel back down to $x=0$ and be re-reflected and transmitted with an amplitude of $\widetilde{T}_L(f) \widetilde{R}_0 \widetilde{R}_L s(f) e^{3 \pi i \tau f }$. We may compute the total output at $x=L$ as a series of  transmitted waves where the $n^{th}$ summand has gone through $n$ partial reflections,
\begin{align}
s_{eff}(f) &= \widetilde{T}_0 e^{\pi i \tau f} \sum_{n=0}^{\infty} \left( \widetilde{R}_0 \widetilde{R}_L  e^{2 \pi i \tau f} \right)^n \\
& = s_{eff}(f) \widetilde{T}_0 e^{\pi i \tau} \frac{1}{1 - \widetilde{R}_0 \widetilde {R}_L e^{2 \pi i \tau f}}.
\end{align}
The term $\widetilde{T}_0 e^{\pi i \tau f}$ has a phase and amplitude that evolves gradually with frequency so we may treat it as part of a smooth complex gain $g(f)$ which will include the contributions from all other steps in the signal path. The gain of the tile in the presence of reflections becomes
\begin{equation}
g(f) \to g(f)' = g(f) \frac{1}{1 - r e^{i(2 \pi \tau + \phi)}}
\end{equation} 
where $re^{i \phi} = \widetilde{R}_0 \widetilde{R}_L$, both terms potentially evolving with frequency. 

\section{The Power Spectrum of ionospheric phase fluctuations from measurements of Differential Refraction.}\label{app:Refraction}
In this section, we derive the relationship between the structure function of source offsets and the underlying power spectrum of ionospheric phase fluctuations. We will addopt the common assumption that the TEC above the MWA, and hence the phases added to transiting electromagnetic waves are described approximately by a Gaussian random field \citep{Rufenach:1972,Singleton:1974} whose power spectrum we denote as $P({\bf k})$. In \S~\ref{ssec:ion}, we measure the differential refraction of source positions which we may express in terms of the gradients of the phase screen.
\begin{align}
D(\theta) &= 2 \left(\frac{c}{2 \pi f} \right)^2  \left \langle \nabla \phi({\bf r}_0) \cdot \nabla \phi({\bf r}_0) -  \nabla \phi({\bf r}_0) \cdot \nabla \phi^*({\bf r}_0 +  {\bf r}) \right \rangle \nonumber \\
&= 2 \left( \frac{c}{2 \pi f} \right)^2 \left[ \rho_\nabla(0) - \rho_\nabla({\bf r}) \right],
\end{align}
where $\rho_\nabla({\bf r})$ is the correlation function of the ionospheric gradients. We can write $\rho_\nabla({\bf r})$ in terms of the power spectrum by expanding $\nabla \phi({\bf r})$ in terms of its Fourier components
\begin{equation}
\nabla \phi ({\bf r}) = \frac{i}{(2\pi)^2} \int d^2{\bf k} \widetilde{\phi}({\bf k}){\bf k}e^{ i {\bf k} \cdot {\bf r}}
\end{equation}
Hence,
\begin{equation}
\rho_\nabla({\bf r}) = \frac{1}{(2 \pi)^2} \int d^2{\bf k} k^2 e^{- i {\bf k} \cdot {\bf r}} P({\bf k}), 
\end{equation}
where we have used the definition of the power spectrum,
\begin{equation}
\left \langle \widetilde{\phi}({\bf k}) \widetilde{\phi}({\bf k}') \right \rangle = (2 \pi)^2 P({\bf k}) \delta_D^{(2)}({\bf k} - {\bf k}').
\end{equation}
If we assume isotropy of the field, we have
\begin{equation}
\rho_\nabla(r) = \frac{1}{2 \pi} \int dk k^3 P(k) J_0(k r).
\end{equation}
Thus, by measuring the structure function of source offsets, we effectively measure the power spectrum of the ionospheric fluctuations.

\section{The Amplitude of Scintillation Noise in MWA Visibilities}\label{app:Scintillation}
In this section, we estimate the amplitude of scintillation noise present in each of the two second time steps that we interleave to estimate the system temperature. The time between the interleaved steps used to compute our system temperature is smaller than the coherence time given in V15a. However,computing the amplitude of scintillation noise, assuming that it is entirely decorrelated between our two-second time steps allows us to place an upper limit on what systematic bias in $T_{\text{sys}}$ that might arise. We estimate the level of scintillation noise a baseline with length ${\bf b}$ arising from a source population with a number density per solid angle and intrinsic flux bin given by,
\begin{equation}\label{eq:counts}
\frac{d^2N(S_t,f)}{dS_t d \Omega} = C S_t^{-\alpha} f^{-\beta},
\end{equation}
using equation (2.7) in V15b
\begin{align}\label{eq:scintNoise}
\sigma^2_{\text{scint}}[V({\bf b})] &= 4 S_{eff}^2  \int d^2{\bf k}P({\bf k}) \sin^2(\pi \lambda h {\bf k}^2 - \pi {\bf b \cdot q}) \nonumber \\
S_{eff}^2 &\approx \frac{C B_{eff}(f) f^{-\beta}}{3-\alpha} S_{max}^{3-\alpha}.
\end{align}
Here $P({\bf k})$ is the power spectrum of ionospheric phase fluctuations and $S_{max}$ is the maximal apparent source flux for which ionospheric effects have not been calibrated out. $B_{eff}$ is the effective primary beam of the instrument and can be computed from the equation 
\begin{equation}
B_{eff}(f) = \int d\Omega B^{\alpha -1}(f,\boldsymbol{\ell})
\end{equation}
where $B(f,{\bf \ell})$ is the antenna primary beam. This equation is derived assuming a small field of view $\lesssim 10^\circ$. However, V15b find that it is accurate to within $\approx 10$\% for substantially larger fields such as the MWA's. 

Since we do not attempt to calibrate out the fluctuations on $2$\,s intervals, this source flux can be obtained by setting the number of sources in the field of view of the instrument with fluxes equal to $S_{max}$ to one (V15a),
\begin{equation}
S_{max}(f) = \left( \frac{\alpha-1}{C f^{-\beta} B_{eff}(f) }\right)^{1/(1-\alpha)}s
\end{equation}
For our source population, we use fits by \citet{DiMatteo:2002} to the source counts observed in the 6C survey \citep{Hales:1988} at $151$\,MHz,
\begin{equation}
\frac{d^2N}{dS_t d\Omega}(S_t,f_0=150\,\text{MHz}) = k \left(\frac{S_t}{.880 \text{Jy}} \right)^{-\gamma} 
\end{equation}
where $k = 4000 \, \text{sr}^{-1} \text{Jy}^{-1}$ and $\gamma = 2.5$. Assuming that all of the sources have a spectral index close to the observed mean of $\delta=0.8$, we determine the frequency dependence of the source counts by setting the number of sources with fluxes above flux $S_t$ at $f_0 = 150$\,MHz equal to the number of sources at $f$ with fluxes greater than $S_t'=S_t(f/f_0)^{-\delta}$. Doing this, we obtain
\begin{equation}
\frac{d^2N(S_t,f)}{dS_t d\Omega} = k \left(\frac{f}{f_0}\right)^{\delta (1-\gamma)} S_t^{-\gamma}
\end{equation}
which is similar to the expression in \citet{Trott:2015} except for an order-unity difference in the frequency power law which was neglected in that work since $f/f_0 \approx 1$ and here, where $f/f_0 \approx 1/2$ accounts for an $\approx 25\%$ enhancement in the source counts. 

We substitute $\beta = \delta(\gamma-1)=1.2$, $C=k f_0^{-\delta(1-\gamma)} = 1.6\times10^6 \text{Jy}^{-1} \text{sr}^{-1} \text{MHz}^{-2.5}$, $\alpha = \gamma=2.5$, and an effective beam area of $ \ref{eq:counts}$. Using the short dipole model of the MWA beam, we compute a $B_{eff}(83\,\text{MHz})$ of $0.33$\,sr. From these numbers, we obtain $S_{eff}=212.3$\,Jy. 

The final ingredient is $P(\bf k)$ which we compute from our fits of our differential refraction measurements described in appendix~\ref{app:Refraction} and using the functional form in equation~\ref{eq:structurePS}. Applying equation~\ref{eq:scintNoise}, we obtain values for $\sigma_{\text{scint}}[V({\bf b})]$ between $\approx 4-6$\,Jy on the 30 minute intervals on September 5th and $\approx 1-2$\,Jy on the 30 minute intervals on September 6th at 83\,MHz. 

We estimate the noise on a single antenna for each two second interleaved time interval is given by \citep{Morales:2004}

\begin{equation}
\sigma_v = \frac{k_b T_{\text{sys}}}{A_e \sqrt{2 d f \tau }}
\end{equation}

where $df$ is the channel width, $A_e$ is the effective area of the tile, and $T_{\text{sys}}\approx T_{\text{sky}} = 60 (\lambda/\text{meter})^{-2.6}\,\text{K}$ \citep{Rogers:2008,Fixsen:2011} which dominates the MWA's system temperature at lower frequency. Using $df=80$\,kHz and $\tau=2$\,s, we obtain $\sigma_v \approx 315$\,K. Hence, on a single two second integration for each of our visibilities, scintillation noise contributes at the level of $\lesssim 2$\% relative to the system noise during the most severe times and $\lesssim 0.3$\% during the calmest intervals.

\end{document}